\documentclass[superscriptaddress,onecolumn,pre,nofootinbib]{revtex4}

\usepackage{amsmath}
\usepackage{graphicx,color}
\usepackage{amssymb}
\usepackage{amsfonts}
\usepackage{bm}

\newcommand{\be}{\begin{equation}}
\newcommand{\ee}{\end{equation}}
\newcommand{\bea}{\begin{eqnarray}}
\newcommand{\eea}{\end{eqnarray}}

\newcommand{\erf}{\textrm{erf}}

\begin{document}
\sloppy


\title{Dissipative self-gravitating Bose-Einstein condensates with
arbitrary nonlinearity \\
as a model of dark matter halos}

\author{Pierre-Henri Chavanis}
\affiliation{Laboratoire de Physique Th\'eorique, 
Universit\'e Paul Sabatier, 118 route de Narbonne 31062 
Toulouse, France}

\begin{abstract}
We develop a general formalism applying to Newtonian self-gravitating
Bose-Einstein
condensates. This formalism may find
application in the
context of dark matter halos. We introduce a generalized Gross-Pitaevskii
equation including a source of dissipation (damping) and an arbitrary
nonlinearity. Using the Madelung transformation, we derive the hydrodynamic
representation of this generalized Gross-Pitaevskii equation and obtain a damped
quantum Euler equation 
involving a friction force proportional and opposite to the velocity and
a pressure force associated with an equation of
state determined by the nonlinearity present in the generalized Gross-Pitaevskii
equation.
In the strong friction limit, we obtain a  quantum Smoluchowski equation.
These equations satisfy an $H$-theorem for a free energy functional constructed
with a generalized entropy. We specifically consider the Boltzmann and
Tsallis entropies associated with isothermal and polytropic equations of state.
We also consider the entropy associated with the logotropic equation of
state. We derive the virial theorem corresponding to the generalized
Gross-Pitaevskii equation, damped quantum Euler equation, and quantum
Smoluchowski equation. Using a Gaussian ansatz, we obtain a simple equation
governing the dynamical evolution of the size of the condensate.
We develop a mechanical
analogy associated with this gross dynamics. We highlight a specific model of dark matter halos
corresponding to a generalized Gross-Pitaevskii equation with a logarithmic
nonlinearity and a cubic
nonlinearity. It corresponds to a damped quantum Euler equation associated
with a mixed entropy combining the Boltzmann and Tsallis
entropies. It leads to dark matter halos with an equation
of state $P=\rho k_B T_{\rm eff}/m+2\pi
a_s\hbar^2\rho^{2}/m^3$ presenting a condensed core (BEC/soliton) and an
isothermal
halo with an effective temperature $T_{\rm eff}$. We propose that this model
provides an effective coarse-grained parametrization of dark matter halos
experiencing
gravitational cooling.  Specific applications of our formalism to dark
matter halos will be developed in future papers.

\end{abstract}

\maketitle


\section{Introduction}
\label{sec_introduction}

Bose-Einstein condensates (BECs) play an important role in condensed matter
physics \cite{bec}. Recently, it has been suggested that they
could also play a
major role in astrophysics and cosmology
(see \cite{revueabril,revueshapiro,bookspringer} for recent reviews). Indeed,
dark
matter halos could be giant self-gravitating BECs at zero temperature
\cite{baldeschi,khlopov,membrado,sin,jisin,leekoh,schunckpreprint,
matosguzman,sahni,
guzmanmatos,hu,peebles,goodman,mu,arbey1,silverman1,matosall,silverman,
lesgourgues,arbey,fm1,bohmer,fm2,bmn,fm3,sikivie,mvm,lee09,ch1,lee,prd1,prd2,
prd3,briscese,
harkocosmo,harko,abrilMNRAS,aacosmo,velten,pires,park,rmbec,rindler,lora2,
abrilJCAP,mhh,lensing,glgr1,ch2,ch3,shapiro,bettoni,lora,mlbec,madarassy,
abrilph,playa,stiff,guth,souza,freitas,alexandre,schroven,pop,eby,cembranos,
braaten,davidson,schwabe,fan,calabrese,bectcoll,chavmatos,hui,zhang}. At the
scale
of galaxies, Newtonian gravity can be used  so the
evolution of the wave function is governed by the Gross-Pitaevskii-Poisson (GPP)
equations (see, e.g., \cite{prd1,prd2,prd3}).
Using the Madelung transformation
\cite{madelung}, the GPP equations can be written under the form
of hydrodynamic equations, the so-called quantum Euler-Poisson (EP)
equations. These equations are similar to the equations of
cold dark matter (CDM) except that they include an anisotropic quantum pressure
(or quantum potential) accounting for the Heisenberg uncertainty principle and
an isotropic pressure due to the self-interaction (scattering). For usual BECs,
described by the Gross-Pitaevskii (GP) equation with a cubic nonlinearity, the
equation of state is
quadratic, $P=2\pi a_s\hbar^2\rho^{2}/m^3$, where $m$ is the mass
of the bosons and $a_s$ is their scattering length \cite{revuebec}. At large
(cosmological) scales, quantum
effects are negligible and one recovers the classical hydrodynamic
equations of CDM which are remarkably
successful in explaining the large-scale structure of the universe
\cite{ratra}. At small (galactic) scales, the BEC model differs from the CDM
model because of the pressure due to quantum mechanics. Gravitational collapse
is prevented by the pressure arising from the Heisenberg uncertainty
principle or by the pressure arising from the repulsive scattering of the
bosons. Dark matter halos can then reach an equilibrium state
with a smooth core
density.
On the
other hand,
the BEC model has a finite Jeans length that provides a sharp small-scale
cut-off in the matter power spectrum.  Therefore, quantum mechanics  may be a
way to solve the problems of the CDM model such as the cusp problem and the
missing satellite problem.\footnote{The CDM model encounters many problems at
the scale of galactic or sub-galactic structures. Indeed, CDM simulations lead
to $r^{-1}$ cuspy density profiles \cite{nfw} at galactic centers (in the scales
of the order of $1$ kpc and smaller) while most rotation curves indicate a
smooth core density \cite{burkert}. On the other hand, the predicted number of
satellite galaxies around each galactic halo is far beyond what we see around
the Milky Way \cite{satellites}. These problems might be solved,
without altering the virtues of the CDM model, if the dark matter is composed of
BECs \cite{baldeschi,khlopov,membrado,sin,jisin,leekoh,schunckpreprint,
matosguzman,sahni,
guzmanmatos,hu,peebles,goodman,mu,arbey1,silverman1,matosall,silverman,
lesgourgues,arbey,fm1,bohmer,fm2,bmn,fm3,sikivie,mvm,lee09,ch1,lee,prd1,prd2,
prd3,briscese,
harkocosmo,harko,abrilMNRAS,aacosmo,velten,pires,park,rmbec,rindler,lora2,
abrilJCAP,mhh,lensing,glgr1,ch2,ch3,shapiro,bettoni,lora,mlbec,madarassy,
abrilph,playa,stiff,guth,souza,freitas,alexandre,schroven,pop,eby,cembranos,
braaten,davidson,schwabe,fan,calabrese,bectcoll,chavmatos,hui,zhang}. The wave
properties of the dark matter may stabilize the system against gravitational
collapse providing halo cores instead of cuspy density profiles in agreement
with observations. The resulting  coherent configuration may be understood as
the ground state of some gigantic bosonic atom where the boson particles are
condensed in a single macroscopic quantum state $\psi({\bf r})$. In the BEC
model, the formation of dark matter structures at small scales is suppressed by
quantum mechanics.} For that reason, the BEC model is a good candidate to
describe dark matter.\footnote{The good  properties of the BEC model are
nevertheless not sufficient to vindicate that model. Other dark
matter models based, e.g.,  on a fermionic
\cite{vega3,urbano,rar,kingclassique,kingfermionic} or on a logotropic
\cite{delog,delogplb}
equation of state  also give relevant results.}

However, the self-gravitating
BEC model faces apparent difficulties.
First of all, the GPP equations are
conservative (dissipationless) equations, so it is not clear at first sight {\it
how} they can
relax towards a steady state representing a dark matter halo. If we ignore this
difficulty for a moment and consider stable steady state solutions of the GPP
equations,\footnote{They correspond to the ground state of the
GPP equations for which the wave function has no node. Excited states for which
the wave
function has  nodes are unstable.} we can
construct models of dark matter halos and determine their
mass-radius relationship. This has been done in Refs. \cite{prd1,prd2}
for an arbitrary value of the scattering length $a_s$ connecting the
non-interacting limit ($a_s=0$) to the Thomas-Fermi (TF)
limit valid when $GM^2ma_s/\hbar^2\gg 1$. However, these results are not
consistent with the observations of large dark matter halos. When the bosons are
noninteracting, one finds that the radius of the halos should decrease with
their mass as $R\propto M^{-1}$ and when the bosons are self-interacting, one
finds that the halos should have the same radius, independently of their mass.
These results
are in contradiction with the  observations that reveal that the radius of dark
matter halos increases with their mass as $R\propto M^{1/2}$ corresponding to a
constant surface density
(see \cite{vega3,kingclassique,kingfermionic}). This apparent
paradox can be solved (see Appendix F of \cite{kingfermionic}) by considering
that the stable steady state solution
of the
GPP equations (soliton) describes only the {\it core} of the
halos\footnote{The profile of the solitonic core, which is
static solution of the GPP equations, a has been computed in
\cite{rb,membrado,gul0,gul,prd2,ch2,ch3,pop} for
noninteracting BECs. The
case self-interacting BECs has been considered in \cite{gul,prd2}.}
and that
this core is surrounded by an envelope in which the 
 density profile decays approximately as
$r^{-3}$ at large distances like the Navarro-Frank-White (NFW)
\cite{nfw} and Burkert \cite{burkert} density profiles. This
core-halo structure has been evidenced in the numerical simulations of
\cite{ch2,ch3,schwabe}. In conclusion, in order to solve the apparent
difficulties of the self-gravitating BEC model, we need to find a source of
dissipation leading to a relaxation mechanism, and
understand the formation of an
envelope surrounding the solitonic core.

A solution is provided by the concept of
gravitational cooling that was introduced by Seidel and Suen \cite{seidel94} in
the context of boson stars. This is a
dissipationless mechanism similar in some respect to the concept of violent
relaxation  introduced by  Lynden-Bell \cite{lb} for collisionless
self-gravitating systems\footnote{Collisionless self-gravitating systems such as
elliptical galaxies \cite{lb} and dark matter halos made of massive neutrinos
\cite{kingclassique,kingfermionic} are described by the Vlasov equation which is
a conservative equation. A spatially homogeneous collisionless self-gravitating
system undergoes gravitational collapse (Jeans instability) and
forms regions of overdensity. When the density has sufficiently grown, these
regions collapse under their own gravity at first in free fall. Then, as
nonlinear gravitational effects become important at higher densities, they
undergo damped oscillations due to an exchange of kinetic and potential energy.
They heat up and
finally settle into a quasi stationary state (QSS) with a core-halo structure on
a coarse-grained scale (virialization). The system is able to form a dense core
by sending some of the particles (stars or neutrinos depending on the context)
at large distances. This process is related to phase mixing and
nonlinear Landau damping. The resulting Lynden-Bell distribution is similar to
the
Fermi-Dirac distribution. Elliptical galaxies are nondegenerate \cite{lb}.
Fermionic dark matter halos may be degenerate
\cite{kingclassique,kingfermionic}. In that case, the QSS has a core-halo
structure with a completely
degenerate core at $T=0$ (fermion ball) and an isothermal atmosphere with an
effective temperature $T_{\rm eff}$. The density is finite in the core and
decreases as $r^{-2}$ in the halo leading to flat rotation curves. Actually,
the
halo cannot be exactly isothermal otherwise it would have an infinite mass. The
density rather decreases as $r^{-3}$ at large distances like in the
NFW \cite{nfw} and Burkert \cite{burkert} profiles. This
steeper decay may be due to incomplete relaxation \cite{lb}, tidal
effects from
nearby galaxies such as those accounted for in the fermionic King model
\cite{kingclassique,kingfermionic}, or external stochastic perturbations.
Violent relaxation \cite{lb} explains how
collisionless self-gravitating systems can rapidly thermalize  and reach a
statistical equilibrium state with a large
effective temperature $T_{\rm eff}$ even if the initial temperature is low.} 
but ending on a unique 
final state independent of the initial condition. A self-gravitating BEC at
$T=0$, described by the GPP equations, that is not initially in a steady state
undergoes gravitational collapse (Jeans instability), displays damped
oscillations, and finally settles into a QSS (virialization) by radiating part
of the scalar
field \cite{seidel94,gul0,gul}. For example, if the BEC is initially in an
excited state
(that is unstable), it spontaneously evolves towards the ground state (that is
stable) by ejecting scalar radiation. This cooling mechanism allows the halo to
get rid of its excess kinetic energy necessary to form a compact bosonic core.
This process may also be at work during hierarchical
clustering.\footnote{Hierarchical clustering is the mechanism by which small
dark matter halos merge and form larger halos in a bottom-up structure
formation scenario.
It is believed that dark matter halos acquire a NFW profile in the
envelope as a result of successive mergings.} As a result of gravitational
cooling, dark matter halos take a core-halo structure with a condensed
core (soliton/BEC) with an equation of state $P=2\pi a_s\hbar^2\rho^{2}/m^3$,
which is a stable stationary solution of the GPP equations at
$T=0$ (ground state), surrounded by a halo of scalar radiation (waves) that is
approximately isothermal like in the process of violent collisionless
relaxation \cite{lb}. The
halo of scalar radiation is similar to an isothermal envelope with
an equation of state $P=\rho k_B T_{\rm eff}/m$ involving an effective
temperature $T_{\rm eff}$. In that case, the density decreases as $r^{-2}$ at
large distances
leading to flat rotation curves. Gravitational cooling explains
how self-gravitating
bosons can rapidly thermalize and acquire a large
effective temperature $T_{\rm eff}$ even if $T=0$ fundamentally. Therefore, although
the true thermodynamic temperature is $T=0$, everything happens {\it as if} the
system had a core-halo structure with a core at $T=0$ (BEC/soliton) and a halo with
an effective temperature $T_{\rm eff}\neq 0$. We emphasize that $T_{\rm
eff}$ is an effective temperature, not a thermodynamic temperature. Bosonic
dark matter halos  are fundamentally described by the GPP equations at $T=0$
(it is shown in Appendix F of \cite{kingfermionic} that the
temperature $T$ of the halos is always much smaller than the condensation
temperature $T_c$ whatever their size).
However, we propose that,
because
of gravitational cooling, bosonic dark matter halos acquire an 
envelope of scalar radiation 
that is similar to an
isothermal atmosphere with an effective temperature $T_{\rm eff}$. In
the analogy with the process of violent relaxation of collisionless self-gravitating  systems, the bosonic core (BEC/soliton) corresponds to the
fermion ball and the halo made of scalar radiation corresponds to the isothermal
halo predicted by Lynden-Bell's theory. Actually, the halo cannot be exactly
isothermal for the reason given in footnote 5. In reality, the density in the
halo decreases as $r^{-3}$, similarly to the NFW \cite{nfw} and Burkert
\cite{burkert} profiles, instead of $r^{-2}$ (isothermal sphere \cite{bt}). This
extra-confinement may be due
to incomplete relaxation, tidal effects, stochastic
perturbations... Several types of halos
are possible depending on their size. Dwarf
dark
matter halos are compact objects that are completely condensed without
an atmosphere. Therefore, their size is equal to the size of the BEC/soliton. By
contrast, large dark matter halos are extended objects with a core-halo
structure. They have a condensed core (BEC/soliton) surrounded by an extended
atmosphere made of scalar radiation with a density profile decaying as
$r^{-3}$ at large distances like the  NFW and Burkert profiles. It is the
 atmosphere that fixes their proper
size.
The atmosphere can be much larger than the size of the soliton. The presence of
the halo of scalar radiation explains why the radius of the dark matter halos
increases with their mass. In this way, there is no paradox with the BEC model
at $T=0$.

Because of gravitational cooling, a self-gravitating BEC at $T=0$ reaches a steady state
with a condensed core (soliton) and an approximately isothermal atmosphere made
of scalar
radiation. In this paper, we propose to heuristically model the process of
gravitational cooling by a generalized GP equation including a source of dissipation (damping)
and an arbitrary nonlinearity. This equation may be viewed as an
{\it effective} description of the system's dynamics on a coarse-grained
scale. In other words, it provides a
coarse-grained parametrization of the (fined-grained) GP equation
at $T=0$.\footnote{In
this sense, the generalized GP equation represents the counterpart of the
relaxation equation for the
coarse-grained distribution function introduced in Refs.
\cite{csr,mnras,dubrovnik}
in the context of Lynden-Bell's theory of violent relaxation \cite{lb}. This
relaxation equation provides a
coarse-grained parametrization of the (fined-grained) Vlasov equation.}  By
using
Madelung's transformation, we show that the generalized GP equation is
equivalent to a damped quantum Euler equation involving a friction force
proportional and opposite to the velocity and a pressure force associated with
an equation of state related to the nonlinearity present in the
generalized GP equation. In the strong friction limit, we obtain a 
quantum
Smoluchowski equation. These equations satisfy an $H$-theorem for a
free energy functional associated with a generalized entropy. A logarithmic
nonlinearity $\ln|\psi|^2\, \psi$ in the GP equation leads to
an isothermal equation of state associated with the Boltzmann entropy. A power
law nonlinearity $|\psi|^{2(\gamma-1)}\, \psi$ leads to a polytropic
equation of state associated with the Tsallis entropy. We also consider an
hyperbolic
nonlinearity $\psi/|\psi|^2$ leading to the logotropic equation of state
associated with a logarithmic entropy. We highlight a specific model of dark
matter halos
corresponding to the generalized GPP equations with a logarithmic nonlinearity
and a cubic
nonlinearity. It corresponds to the damped quantum isothermal-polytropic EP
equations associated 
with a mixed entropy combining the Boltzmann and Tsallis
entropies. We propose that this model provides an effective coarse-grained model
of dark matter halos experiencing
gravitational cooling. It leads to dark matter halos with an equation of state
$P=\rho k_B T_{\rm eff}/m+2\pi a_s\hbar^2\rho^2/m^3$ presenting a condensed core
(soliton) and an isothermal halo. This model is not perfect since the density
in the halo should decrease as $r^{-3}$ (NFW/Burkert) instead of $r^{-2}$
(isothermal), but it can be interesting in a first approach. The
$r^{-3}$ (NFW/Burkert) 
profile
may result from a more complicated physics such as incomplete relaxation, tidal
effects, stochastic forcing... Tidal effects can be taken into account
heuristically by introducing a confining potential in the GP
equation. Alternatively, the $r^{-3}$ (NFW/Burkert) profile could be
accounted for by
using a more complicated nonlinearity in the generalized GP equation (or,
equivalently, a more complicated equation of state in the orresponding Euler
equation).

The paper is organized as follows. In section \ref{sec_qsp}, we introduce the
generalized GPP equations, derive their hydrodynamic representation, and
establish
the general condition of hydrostatic equilibrium. In section \ref{sec_thermo},
we develop a generalized thermodynamic formalism and derive an $H$-theorem.
In section \ref{sec_virial}, we derive the virial theorem. In section
\ref{sec_eos}, we consider particular forms of generalized GPP and
quantum EP equations, and determine their associated equations
of state, generalized entropies, and equilibrium distributions. In section
\ref{sec_standmod}, we highlight a specific model of dark matter halos
corresponding to the generalized GPP equations with a logarithmic nonlinearity
and a
cubic nonlinearity, equivalent  to  the damped quantum isothermal-polytropic EP
equations. In section \ref{sec_pi}, we compare our results with previous
works. In section \ref{sec_ansatz}, we make a Gaussian ansatz and obtain a
simplified equation governing the dynamical evolution of the size of the
condensate. We develop a mechanical analogy associated with this gross dynamics
and  obtain a general analytical expression for the mass-radius relation of dark
matter halos. We study their stability and determine their pulsation period.
This paper introduces a general formalism appropriate to Newtonian
self-gravitating
BECs. This formalism covers a great variety of situations. Specific applications
to dark matter halos will be developed in future works (in preparation).
However, the generalized GPP equations that we study in this paper are
interesting in their own right and may find applications for other systems
beyond the context of dark matter halos.

\section{Self-gravitating Bose-Einstein condensates}
\label{sec_qsp}

\subsection{The Gross-Pitaevskii-Poisson equations}
\label{sec_mfgp}

We consider a system of $N$ bosons with mass $m$ interacting via a
binary potential $u(|{\bf r}-{\bf r}'|)$ \cite{bogoliubov}. At $T=0$, all
the bosons condense into the same quantum ground state and the system is
described by one order parameter $\psi({\bf r},t)$ called the condensate wave
function.\footnote{The condensation of the bosons takes place when their thermal
(de Broglie) wavelengths $\lambda_{T}=(2\pi\hbar^2/mk_BT)^{1/2}$ overlap, that is, when the
 thermal wavelength is greater than the mean
inter-particle distance  $l=n^{-1/3}$ ($n$ is the number density of the bosons).
This
leads to the inequality $\lambda_T>l$, $n\lambda _{T}^{3}>1 $ or $T<T_c$ where  $T_c\sim 2\pi
\hbar^2 n^{2/3}/m k_B$ is the  critical condensation temperature (up to a
numerical proportionality factor of order unity).} In the mean-field approximation, this gas of
interacting BECs is governed by the time-dependent self-consistent field
equations (or mean-field Schr\"odinger equation)
\cite{bogoliubov,gross1,gross2,gross3,pitaevskii2}:
\begin{eqnarray}
\label{mfgp1}
i\hbar \frac{\partial\psi}{\partial t}({\bf r},t)=-\frac{\hbar^2}{2m}\Delta\psi({\bf r},t)+m\Phi_{\rm tot}({\bf r},t)\psi({\bf r},t),
\end{eqnarray}
\begin{eqnarray}
\label{mfgp2}
\Phi_{\rm tot}({\bf r},t)=\int \rho({\bf r}',t) u(|{\bf r}-{\bf r}'|)\, d{\bf r}',
\end{eqnarray}
\begin{eqnarray}
\label{mfgp3}
\rho({\bf r},t)=|\psi({\bf r},t)|^2,
\end{eqnarray}
\begin{eqnarray}
\label{mfgp3b}
\int |\psi({\bf r},t)|^2\, d{\bf r}=M=Nm.
\end{eqnarray}
Equation (\ref{mfgp3b}) is the normalization condition, Eq. (\ref{mfgp3})
gives the density of the BEC, Eq. (\ref{mfgp2}) determines the associated
potential, and Eq. (\ref{mfgp1}) determines the evolution of the wave function.
We assume that the potential of interaction can be written as $u=u_{\rm LR}+u_{\rm SR}$,
where $u_{\rm LR}$ refers to long-range interactions and $u_{\rm SR}$ to short-range
interactions. For self-gravitating BECs, the potential of long-range
interactions is the gravitational potential  $u_{\rm LR}=-G/|{\bf r}-{\bf r}'|$,
where $G$ is the constant of gravity. On the other hand, following
Gross \cite{gross1,gross2,gross3} and
Pitaevskii \cite{pitaevskii2}, we
assume that the
short-range
interactions
correspond to binary collisions that can be modeled by the effective potential
$u_{\rm SR}=g\delta({\bf r}-{\bf r}')$ \cite{hy,lhy}, where the coupling
constant (or
pseudo-potential) $g$ is related to the s-wave scattering length $a_s$ of the bosons through
$g={4\pi a_s\hbar^2}/{m^3}$ \cite{revuebec}.
For the sake of generality, we allow $a_s$ to be positive or negative ($a_s>0$
corresponds to a short-range repulsion and $a_s<0$ corresponds to a short-range
attraction). Under these conditions, the total potential can be written as
$\Phi_{\rm tot}=\Phi+h(\rho)$ where $\Phi({\bf r},t)=-G\int \rho({\bf r}',t)/|{\bf r}-{\bf r}'| \, d{\bf r}'$ is the gravitational potential that is the solution of the Poisson equation $\Delta\Phi=4\pi  G\rho$ and
$h(\rho)=g\rho=g|\psi|^2$
is an effective potential modelling short-range interactions.
Regrouping
these results, we obtain the Gross-Pitaevskii-Poisson (GPP) equations
\begin{eqnarray}
\label{mfgp13}
i\hbar \frac{\partial\psi}{\partial t}=-\frac{\hbar^2}{2m}\Delta\psi+m\Phi\psi+ \frac{4\pi a_s\hbar^2}{m^2}|\psi|^{2}\psi,
\end{eqnarray}
\begin{equation}
\label{mfgp14a}
\Delta\Phi=4\pi G |\psi|^2.
\end{equation}
We note that the GP equation (\ref{mfgp13}) involves a cubic
nonlinearity. As mentioned in the Introduction, dark matter
halos could be self-gravitating BECs described by the GPP equations
(\ref{mfgp13}) and (\ref{mfgp14a}). 

\subsection{The generalized Gross-Pitaevskii-Poisson equations}
\label{sec_ggpp}

In this paper, we consider the generalized GPP equations
\begin{eqnarray}
\label{mfgp9}
i\hbar \frac{\partial\psi}{\partial t}=-\frac{\hbar^2}{2m}\Delta\psi
+m\lbrack\Phi
+h(|\psi|^2)+\Phi_{\rm ext}\rbrack\psi
-i\frac{\hbar}{2}\xi\left\lbrack \ln\left (\frac{\psi}{\psi^*}\right )-\left\langle \ln\left (\frac{\psi}{\psi^*}\right )\right\rangle\right\rbrack\psi,
\end{eqnarray}
\begin{equation}
\label{mfgp14}
\Delta\Phi=S_d G |\psi|^2.
\end{equation}
A physical interpretation of these equations will be given in Sec. \ref{sec_pi}.
These equations generalize the GPP equations
(\ref{mfgp13}) and (\ref{mfgp14a}) in several ways:

(i) We have written these equations in a space of dimension $d$ (the dimension
$d=3$ corresponds to spherical halos, the dimension $d=2$ corresponds to
filaments, and the dimension $d=1$ corresponds to sheets). In that case, the
Poisson equation is written as $\Delta\Phi=S_d G\rho$, where
$S_d=2\pi^{d/2}/\Gamma(d/2)$ is the surface of a hypersphere of unit radius in a
$d$-dimensional space (the gravitational constant $G$ depends on the
dimension of space $d$ but, for convenience, we shall not write this dependence
explicitly). We recall that $S_3=4\pi$ in $d=3$, $S_2=2\pi$ in $d=2$, and
$S_1=2$ in $d=1$. The gravitational potential can be explicitly written as
\begin{eqnarray}
\label{pot1}
\Phi({\bf r},t)=-\frac{G}{d-2}\int \frac{\rho({\bf r}',t)}{|{\bf r}-{\bf r}'|^{d-2}} \, d{\bf r}'\qquad (d\neq 2),
\end{eqnarray}
\begin{eqnarray}
\label{pot2}
\Phi({\bf r},t)={G}\int {\rho({\bf r}',t)}\ln{|{\bf r}-{\bf r}'|} \, d{\bf r}'\qquad (d=2).
\end{eqnarray}

(ii) The last term in Eq. (\ref{mfgp9}) represents a source of dissipation measured by the friction
coefficient $\xi$ (this interpretation will become clear in Sec. \ref{sec_mad}
where we introduce a hydrodynamic representation of the generalized GPP
equations). 
The brackets denote spatial averaging: $\langle X\rangle=\frac{1}{M}\int \rho
X\, d{\bf r}$.

(iii) We have introduced an arbitrary nonlinearity determined by the effective potential $h(|\psi|^2)$ instead of the
usual quadratic potential $h(|\psi|^2)=g|\psi|^2$ describing pair contact
interactions \cite{gross1,gross2,gross3,pitaevskii2}. In
this way, we can describe a larger class of systems. We recall
that the GP equation can be
derived from the Klein-Gordon (KG) equation 
\begin{eqnarray}
\label{kg}
\square\varphi+\frac{m^2c^2}{\hbar^2}\varphi+2\frac{dV}{d|\varphi|^2}\varphi
-i\frac{\xi m}{\hbar}\left\lbrack \ln\left
(\frac{\varphi}{\varphi^*}\right )-\left\langle \ln\left
(\frac{\varphi}{\varphi^*}\right )\right\rangle\right\rbrack\varphi=0
\end{eqnarray}
in the nonrelativistic limit
$c\rightarrow +\infty$ (we have generalized the KG equation by introducing
dissipative effects). In that case, the
effective potential $h(|\psi|^2)$ in the GP
equation is related to the self-interaction potential $V(|\phi|^2)$ in the
KG equation by  (see \cite{playa} and Appendix C of \cite{delog}):
\begin{eqnarray}
\label{kggp}
h(|\psi|^2)=\frac{dV}{d|\psi|^2}, \qquad {\rm i.e.}\qquad h(\rho)=V'(\rho).
\end{eqnarray}
As a result,
we can rewrite the generalized GP equation (\ref{mfgp9}) as 
\begin{eqnarray}
\label{mfgp9new}
i\hbar \frac{\partial\psi}{\partial t}=-\frac{\hbar^2}{2m}\Delta\psi
+m\left\lbrack\Phi
+\frac{dV}{d|\psi|^2}+\Phi_{\rm ext}\right\rbrack\psi
-i\frac{\hbar}{2}\xi\left\lbrack \ln\left (\frac{\psi}{\psi^*}\right )-\left\langle \ln\left (\frac{\psi}{\psi^*}\right )\right\rangle\right\rbrack\psi.
\end{eqnarray}

(iv) We have assumed that the particles are subjected to an external potential $\Phi_{\rm ext}({\bf r})$.  For illustration, we shall consider the harmonic potential
\begin{equation}
\label{mfgp8}
\Phi_{\rm ext}=\frac{1}{2}\omega_0^2 r^2.
\end{equation}
When $\omega_0^2=-\Omega^2<0$, this potential mimics the effect of a solid-body 
rotation of the system (this analogy is exact in $d=2$). When $\omega_0^2>0$,
this potential mimics the effect of a confining trap. In astrophysics, where the
GPP equations describe dark matter halos, this trapping potential could account
for tidal interactions arising from neighboring galaxies.

\subsection{The Madelung transformation}
\label{sec_mad}

We use the Madelung \cite{madelung} transformation to rewrite the
generalized GP
equation (\ref{mfgp9}) under the form of hydrodynamic equations. We write the
wave
function as
\begin{equation}
\label{mad1}
\psi({\bf r},t)=\sqrt{{\rho({\bf r},t)}} e^{iS({\bf r},t)/\hbar},
\end{equation}
where  $\rho({\bf r},t)$ is the density and $S({\bf r},t)$ is the real action.
We have
\begin{equation}
\label{mad2}
\rho=|\psi|^2\qquad {\rm and}\qquad S=-i\frac{\hbar}{2}\ln\left
(\frac{\psi}{\psi^*}\right ).
\end{equation}
We note that the dissipative term in the GP equation (\ref{mfgp9}) can be written as $\xi (S-\langle S\rangle)\psi$. Following Madelung, we introduce the velocity field
\begin{equation}
\label{mad5}
{\bf u}=\frac{\nabla S}{m}.
\end{equation}
Since the velocity is potential, the flow is irrotational: $\nabla\times {\bf u}={\bf 0}$. Substituting Eq. (\ref{mad1}) into Eq. (\ref{mfgp9}) and separating real and imaginary parts, we obtain
\begin{equation}
\label{mad6}
\frac{\partial\rho}{\partial t}+\nabla\cdot (\rho {\bf u})=0,
\end{equation}
\begin{equation}
\label{mad7}
\frac{\partial S}{\partial t}+\frac{1}{2m}(\nabla S)^2+m \left\lbrack \Phi+h(\rho)+\Phi_{\rm ext}\right\rbrack+Q+\xi (S-\langle S\rangle)=0,
\end{equation}
where
\begin{equation}
\label{mad8}
Q=-\frac{\hbar^2}{2m}\frac{\Delta \sqrt{\rho}}{\sqrt{\rho}}=-\frac{\hbar^2}{4m}\left\lbrack \frac{\Delta\rho}{\rho}-\frac{1}{2}\frac{(\nabla\rho)^2}{\rho^2}\right\rbrack
\end{equation}
is the quantum potential which takes into account the Heisenberg uncertainty
principle.  The first equation is similar to the equation of continuity in
hydrodynamics. It accounts for the local conservation of mass $M=\int \rho\,
d{\bf r}$. The second equation has a form similar to the classical
Hamilton-Jacobi equation with an additional quantum term and a source of
dissipation. It  can also be interpreted as a generalized Bernoulli equation for
a potential flow. Taking 
the gradient of Eq. (\ref{mad7}), and using the well-known  identity of
vector analysis $({\bf u}\cdot \nabla){\bf u}=\nabla ({{\bf u}^2}/{2})-{\bf
u}\times (\nabla\times {\bf u})$ which reduces to $({\bf u}\cdot \nabla){\bf
u}=\nabla ({{\bf u}^2}/{2})$ for an irrotational flow, we obtain an equation
similar to the Euler equation with a linear friction and a quantum force
\begin{equation}
\label{mad9}
\frac{\partial {\bf u}}{\partial t}+({\bf u}\cdot \nabla){\bf u}=-\nabla h-\nabla\Phi-\nabla \Phi_{\rm ext}-\frac{1}{m}\nabla Q-\xi {\bf u}.
\end{equation}
This equation shows that the effective potential $h$ appearing in the GP equation can be
interpreted as an enthalpy in the hydrodynamic equations.  We can also write Eq.
(\ref{mad9}) under the form
\begin{equation}
\label{mad10}
\frac{\partial {\bf u}}{\partial t}+({\bf u}\cdot \nabla){\bf
u}=-\frac{1}{\rho}\nabla P-\nabla\Phi-\nabla \Phi_{\rm ext}-\frac{1}{m}\nabla
Q-\xi {\bf u},
\end{equation}
where $P({\bf r},t)$ is a pressure.  Since $h({\bf
r},t)=h\lbrack \rho({\bf
r},t)\rbrack$, the pressure $P({\bf r},t)=P\lbrack \rho({\bf r},t)\rbrack$ is a
function of the density, i.e., the flow is barotropic. The equation of state
$P(\rho)$ is determined by the potential $h(\rho)$ through the relation
\begin{equation}
\label{mad11}
h'(\rho)=\frac{P'(\rho)}{\rho},
\end{equation}
which can be viewed as a Gibbs-Duhem relation $dP=\rho dh$ (we
shall see later that $h$ is one component of the chemical potential). Equation
(\ref{mad11}) can be integrated into
\begin{equation}
\label{mad11b}
P(\rho)=\rho h(\rho)-V(\rho)=\rho V'(\rho)-V(\rho),
\end{equation}
where $V$ is a primitive of $h$ (this notation is consistent with
Eq. (\ref{kggp}) where $V$ represents the potential of self-interaction in
the KG equation that reduces to the GP equation in the nonrelativistic limit
$c\rightarrow +\infty$). The speed of sound is $c_s^2=P'(\rho)=\rho
V''(\rho)$. In conclusion, the generalized GPP equations are equivalent to
the
hydrodynamic equations
\begin{equation}
\label{mad12}
\frac{\partial\rho}{\partial t}+\nabla\cdot (\rho {\bf u})=0,
\end{equation}
\begin{equation}
\label{mad13}
\frac{\partial {\bf u}}{\partial t}+({\bf u}\cdot \nabla){\bf
u}=-\frac{1}{\rho}\nabla P-\nabla\Phi-\nabla \Phi_{\rm ext}-\frac{1}{m}\nabla
Q-\xi{\bf u},
\end{equation}
\begin{equation}
\label{mad14}
\Delta\Phi=S_d G\rho.
\end{equation}
For the harmonic potential defined by Eq. (\ref{mfgp8}), we have $\nabla\Phi_{\rm ext}=\omega_0^2{\bf r}$. Using the continuity equation (\ref{mad12}), the Euler equation (\ref{mad13}) can be rewritten as
\begin{eqnarray}
\label{mad14b}
\frac{\partial}{\partial t}(\rho {\bf u})+\nabla(\rho {\bf u}\otimes {\bf u})
=-\nabla P-\rho\nabla\Phi-\rho\nabla \Phi_{\rm ext}-\frac{\rho}{m}\nabla
Q-\xi\rho {\bf u}.
\end{eqnarray}
From that equation, one can introduce a momentum tensor (see Appendix
\ref{sec_stress}). We shall refer to these equations as the damped quantum
barotropic EP equations. We note that the hydrodynamic
equations
(\ref{mad12})-(\ref{mad14}) do {\it not} involve viscous terms since they
are equivalent to the GP equations at $T=0$. As a result, they describe a
superfluid.  When the quantum
potential can be neglected, we recover the classical damped barotropic EP
equations. For dissipationless systems ($\xi=0$), they reduce to the
quantum and classical barotropic EP equations. On the
other hand, in the overdamped limit $\xi\rightarrow +\infty$, we can formally
neglect the inertia of the particles in Eq. (\ref{mad13}) so that
\begin{equation}
\label{mad15}
\xi{\bf u}\simeq -\frac{1}{\rho}\nabla P-\nabla\Phi-\nabla \Phi_{\rm
ext}-\frac{1}{m}\nabla Q.
\end{equation}
Substituting this relation into the continuity equation (\ref{mad12}), we obtain
the quantum barotropic Smoluchowski-Poisson (SP) equations:\footnote{Because of
the complex nature of the wave function, it is
not possible to take the strong friction limit directly in the generalized
GP
equation (\ref{mfgp9}). We must necessarily split this equation into its real
and imaginary parts (which can be done by means of the Madelung transformation)
and take the limit $\xi\rightarrow +\infty$ in the damped Euler equation
(\ref{mad13}) which corresponds to the gradient of the real part of the
generalized GP
equation
(\ref{mfgp9}).}
\begin{equation}
\label{mad16}
\xi\frac{\partial\rho}{\partial t}=\nabla\cdot\left (\nabla
P+\rho\nabla\Phi+\rho\nabla \Phi_{\rm ext}+\frac{\rho}{m}\nabla Q\right ),
\end{equation}
\begin{equation}
\label{mad17}
\Delta\Phi=S_d G \rho.
\end{equation}
When the quantum potential can be neglected, we obtain the classical barotropic
SP equations.\footnote{The classical barotropic 
SP equations also describe a gas of self-gravitating  Brownian
particles in the overdamped limit \cite{sgb1,sgb2}. These particles experience a
random force in
addition to the gravitational interaction. In that context, the barotropic
Smoluchowski equation can be interpreted as a nonlinear Fokker-Planck (NFP)
equation associated with stochastic processes \cite{gen,entropy}. It is
interesting to note that overdamped BECs and overdamped Brownian particles are
described by similar equations. However, we emphasize that (besides the presence
of the quantum potential) their physical
interpretation is different. For example, in the case of Brownian particles, the
pressure term in the Smoluchowski equation leading to (nonlinear) diffusion
comes from a (multiplicative) random force. By contrast, in the case of BECs, it
comes from the effective potential $h(|\psi|^2)$ accounting for short-range
interactions between the particles. As we show in Secs. \ref{sec_bol}
and \ref{sec_st}, a logarithmic
effective potential  leads to  normal (linear)
diffusion while a pair contact potential leads to anomalous (quadratic)
diffusion. We also note that the damped quantum Euler equation
(\ref{mad13}) can be rigorously justified for dissipative BECs while the damped
 Euler equation is not rigorously justified for
Brownian particles except in the strong friction limit
$\xi\rightarrow +\infty$ where it reduces to the barotropic Smoluchowski
equation (see the discussion in
\cite{just1,just2}). Indeed, the
hydrodynamic equations describing dissipative BECs do not involve viscous
terms (because of their superfluid nature) while the hydrodynamic equations
describing 
Brownian particles generally do.}
Finally, if we neglect the advection term $\nabla(\rho {\bf u}\otimes {\bf u})$
in Eq. (\ref{mad14b}), but retain the term $\partial (\rho {\bf
u})/\partial t$, and
combine the resulting equation with the continuity equation (\ref{mad12}), we
obtain the quantum barotropic telegraphic equation
\begin{equation}
\label{tmad12}
\frac{\partial^2\rho}{\partial t^2}+\xi\frac{\partial\rho}{\partial
t}=\nabla\cdot\left (\nabla P+\rho\nabla\Phi+\rho\nabla \Phi_{\rm
ext}+\frac{\rho}{m}\nabla Q\right ).
\end{equation}
It can be seen as a generalization of the quantum barotropic 
Smoluchowski equation (\ref{mad16}) taking inertial (or memory) effects into
account.

\subsection{The energy}

If we define the energy  by
\begin{equation}
\label{enxia}
E=-\left (\frac{\partial S}{\partial t}\right )_{\xi=0}
\end{equation}
and use the Hamilton-Jacobi equation (\ref{mad7}), we obtain
\begin{equation}
\label{enxi}
E({\bf r},t)=\frac{1}{2}m{\bf u}^2+m\left\lbrack \Phi+h(\rho)+\Phi_{\rm
ext}\right\rbrack+Q.
\end{equation}
This is the sum of the kinetic energy, the gravitational potential, the
enthalpy, the external potential, and the quantum potential. We note that the
damped quantum  barotropic Euler equation (\ref{mad13}) can be written as
\begin{equation}
\label{enxidf}
\frac{\partial {\bf u}}{\partial t}=-\frac{\nabla E}{m}-\xi{\bf u}.
\end{equation}

\subsection{The quantum force}

The quantum potential (\ref{mad8}) first appeared in the work of  Madelung
\cite{madelung} and was rediscovered by Bohm \cite{bohm}. For that reason, it is
sometimes called
``the Bohm potential''.\footnote{A relativistic version of the
quantum potential appears in the works of de
Broglie \cite{broglie1927a,broglie1927b,broglie1927c} and London \cite{london}
who developed a hydrodynamic representation of the KG equation independently
from Madelung \cite{madelung}.} The ``quantum
force'' by unit of mass writes
\begin{equation}
\label{ntmad13a}
{\bf F}_Q=-\frac{1}{m}\nabla Q.
\end{equation}
We note the identity
\begin{equation}
\label{mad19}
(F_Q)_i=-\frac{1}{m}\partial_i Q= -\frac{1}{\rho}\partial_j P_{ij},
\end{equation}
where $P_{ij}$ is the quantum stress (or pressure) tensor defined by
\begin{equation}
\label{mad20}
P_{ij}^{(1)}=-\frac{\hbar^2}{4m^2}\rho\, \partial_i\partial_j\ln\rho\qquad {\rm or}\qquad
P_{ij}^{(2)}=\frac{\hbar^2}{4m^2}\left (\frac{1}{\rho}\partial_i\rho\partial_j\rho-\delta_{ij}\Delta\rho\right ).
\end{equation}
This tensor is manifestly symmetric: $P_{ij}=P_{ji}$. The identity (\ref{mad19})
shows that the quantum force $-\nabla Q$ is equivalent to the
force produced by an anisotropic
pressure tensor $P_{ij}$. In comparison, the effective potential
$h(\rho)$ is equivalent to an isotropic pressure $P(\rho)$.  The tensors defined
by Eq. (\ref{mad20}) are related to each other by
\begin{equation}
\label{tmad16}
P_{ij}^{(1)}=P_{ij}^{(2)}+\frac{\hbar^2}{4m^2}(\delta_{ij}\Delta\rho-\partial_i\partial_j\rho).
\end{equation}
They differ by a tensor
$\chi_{ij}=\delta_{ij}\Delta\rho-\partial_i\partial_j\rho$ satisfying 
$\partial_j\chi_{ij}=0$. Contracting the indices, we obtain
\begin{equation}
\label{mad20b}
P_{ii}^{(1)}=-\frac{\hbar^2}{4m^2}\rho\, \Delta\ln\rho,\qquad P_{ii}^{(2)}=\frac{\hbar^2}{4m^2}\left \lbrack\frac{(\nabla\rho)^2}{\rho}-d\Delta\rho\right \rbrack,
\end{equation}
and the relation
\begin{equation}
\label{mad21c}
P_{ii}^{(1)}=P_{ii}^{(2)}+(d-1)\frac{\hbar^2}{4m^2}\Delta\rho.
\end{equation}
According to Eq. (\ref{mad19}) we note that
\begin{equation}
\label{zqf}
\langle {\bf F}_Q\rangle=\int \rho {\bf F}_Q\, d{\bf r}={\bf 0}
\end{equation}
so there is no resultant of the quantum force.

\subsection{Vortices}
\label{sec_vorticity}

In the Madelung transformation, the velocity field defined by ${\bf u}=\nabla
S/m$ is potential. This implies that the flow is irrotational:
\begin{equation}
\label{vorticity0}
\nabla \times {\bf u}={\bf 0}\qquad \forall {\bf r} \quad {\rm where}\quad
\rho({\bf r})\neq 0.
\end{equation}
This relation is valid only at the points where ${\bf u}=\nabla S/m$
is well defined, i.e., at the points where the wave function (or the density)
does
not vanish. When the wave function vanishes, its phase does not have any meaning
and neither $S$ nor $\nabla S$ is well defined (the velocity is singular). At
such points, known as nodal points, $\nabla \times {\bf u}$ does not vanish in
general, leading to the appearance of singular vortices. If we consider the
circulation of the velocity around a nodal point, we have
\begin{equation}
\label{vorticity1}
\Gamma=\oint {\bf u}\cdot d{\bf l}=\frac{1}{m}\oint \nabla S\cdot d{\bf
l}=\frac{1}{m}\oint dS=2\pi n\frac{\hbar}{m}\qquad n=\pm 1,\pm 2,...
\end{equation}
since the phase $S/\hbar$, when it exists, is defined up to a multiple of
$2\pi$. This relation shows that the circulation around a nodal point is
quantized in units of $h/m$. The integer $n$ is
the circulation number of the vortex.  Using the Stokes theorem, we have
\begin{equation}
\label{vorticity1b}
\Gamma=\int \nabla \times {\bf u}\, d{\bf S}=2\pi n\frac{\hbar}{m}.
\end{equation}
Therefore, the vorticity $\nabla \times {\bf u}$ vanishes everywhere except on
certain singular lines where it has singularities of the $\delta$-type. This
allows for the existence of point vortices with quantized circulation $n
h/m$. This result was
stated by Onsager in a
footnote \cite{onsager}, and by Feynman \cite{feynman,feynman2}, in the context
of
superfluidity. Actually, the quantization
of the circulation of singular vortices was discovered by Dirac
\cite{diracmm} in a more general context in which an electromagnetic field may
be
present.

\subsection{Time-independent GP equation}
\label{sec_tigp}

If we consider a wave function of the form
\begin{equation}
\label{tigp1}
\psi({\bf r},t)=\phi({\bf r})e^{-i E t/\hbar},
\end{equation}
where $\phi({\bf r})=\sqrt{\rho({\bf r})}$ is real, and substitute Eq. (\ref{tigp1}) into Eqs. (\ref{mfgp9}) and (\ref{mfgp14}), we obtain the  time-independent GPP equations
\begin{eqnarray}
\label{tigp2}
-\frac{\hbar^2}{2m}\Delta\phi+m(\Phi+h(\rho)+\Phi_{\rm ext})\phi=E\phi,
\end{eqnarray}
\begin{equation}
\label{tigp2b}
\Delta\Phi=S_d G \phi^2.
\end{equation}
Equations (\ref{tigp2}) and (\ref{tigp2b}) define an eigenvalue problem for the 
wave function $\phi({\bf r})$ where the eigenvalue $E$ is the energy
(eigenenergy). The fundamental eigenmode corresponds to the smallest value of
$E$. For this mode, the wave function $\phi(r)$ is spherically symmetric and 
has no node so that the density profile decreases monotonically with the
distance.
Dividing Eq. (\ref{tigp2}) by $\phi({\bf r})$ and using $\rho=\phi^2$,  we get
\begin{equation}
\label{tigp4}
m\Phi+mh(\rho)+m\Phi_{\rm ext}+Q=E.
\end{equation}
This relation can also be derived from the damped quantum 
Hamilton-Jacobi equation (\ref{mad7}) by setting  $S=-Et$. We note that
dissipative effects do not alter the time-independent solutions of the GPP
equations because $S=-Et$ is uniform so that $\xi(S-\langle S\rangle)=0$.
We also note that $E({\bf r},t)=-\partial S/\partial t=E$ when $S=-Et$, so that
Eq.
(\ref{tigp4}) corresponds to the static value (${\bf u}={\bf
0}$) of the energy defined by Eq. (\ref{enxi}).

\subsection{Hydrostatic equilibrium}
\label{sec_he}

The time-independent GP equation (\ref{tigp4})  can also be obtained from the
damped quantum barotropic Euler equation (\ref{mad10}) since it is equivalent to
the generalized GP equation. The equilibrium state of the damped quantum
barotropic Euler equation (\ref{mad10}), obtained by taking $\partial_t=0$ and
${\bf u}={\bf 0}$, satisfies
\begin{equation}
\label{he1}
\nabla P+\rho\nabla\Phi+\rho\nabla \Phi_{\rm ext}+\frac{\rho}{m}\nabla Q={\bf
0}.
\end{equation}
This equation generalizes the usual condition of hydrostatic equilibrium by
incorporating the contribution of the quantum potential. From the hydrodynamic
representation, we clearly understand why frictional effects do not influence
the equilibrium state since they vanish when ${\bf u}={\bf 0}$. Equation
(\ref{he1}) describes the balance between the gravitational attraction,  the
external potential, the quantum potential arising from the Heisenberg
uncertainty principle, and the pressure due to short-range interactions
(scattering). This equation is equivalent to Eq. (\ref{tigp4}). Indeed,
integrating Eq. (\ref{he1}) using Eq. (\ref{mad11}), we obtain Eq. (\ref{tigp4})
where the eigenenergy  $E$ appears as a constant of integration. On the other
hand, combining Eq. (\ref{he1}) with the Poisson equation (\ref{mad14}), we
obtain the fundamental differential equation of hydrostatic equilibrium
including the quantum potential
\begin{equation}
\label{he2}
-\nabla\cdot \left (\frac{\nabla P}{\rho}\right )+\frac{\hbar^2}{2m^2}\Delta
\left (\frac{\Delta\sqrt{\rho}}{\sqrt{\rho}}\right )=S_d G\rho+\Delta\Phi_{\rm
ext}.
\end{equation}
For the harmonic potential (\ref{mfgp8}), we get $\Delta\Phi_{\rm
ext}=d\omega_0^2$. Some interesting limits can be mentioned:

(i) In the absence of short-range interactions ($P=0$), the equations of
hydrostatic equilibrium (\ref{he1}) and (\ref{he2}) reduce to
\begin{equation}
\label{he3}
\rho\nabla\Phi+\rho\nabla \Phi_{\rm ext}+\frac{\rho}{m}\nabla Q={\bf 0},
\end{equation}
\begin{equation}
\label{he4}
\frac{\hbar^2}{2m^2}\Delta \left (\frac{\Delta\sqrt{\rho}}{\sqrt{\rho}}\right )=S_d G\rho+\Delta\Phi_{\rm ext}.
\end{equation}

(ii) In the TF limit where we can neglect the quantum potential
($Q\simeq 0$), Eqs. (\ref{he1}) and (\ref{he2}) reduce to the classical
equations of hydrostatic equilibrium
\begin{equation}
\label{he5}
\nabla P+\rho\nabla\Phi+\rho\nabla \Phi_{\rm ext}={\bf 0},
\end{equation}
\begin{equation}
\label{he6}
-\nabla\cdot \left (\frac{\nabla P}{\rho}\right )=4\pi G\rho+\Delta\Phi_{\rm
ext}.
\end{equation}

{\it Remark:} Using Eq. (\ref{mad11}), we can rewrite Eq. (\ref{he2}) under the
form
\begin{equation}
\label{he7}
\Delta \left (h+\frac{Q}{m}+\Phi_{\rm ext}\right )=-S_d G\rho.
\end{equation}
This equation can also be directly obtained by substituting Eq. (\ref{tigp4})
into Eq. (\ref{tigp2b}).

\section{Thermodynamics of self-gravitating BECs}
\label{sec_thermo}

In this section (and later in Sec. \ref{sec_eos}), we develop a
thermodynamical formalism associated with the generalized GPP equations
(\ref{mfgp9}) and (\ref{mfgp14}). We stress from the start that this
thermodynamical formalism is {\it effective} since we are basically considering
a boson gas at $T=0$. However, a strong analogy with thermodynamics arises from
the nonlinear term in the generalized GP equation (\ref{mfgp9}) giving rise to a
pressure force in the Euler equation (\ref{mad13}). We must keep in mind,
however, that this pressure has not a thermal origin.

\subsection{The free energy}
\label{sec_ef}

The free energy associated with the generalized GPP equations
(\ref{mfgp9}) and
(\ref{mfgp14}), or equivalently with the damped quantum barotropic EP equations
(\ref{mad12})-(\ref{mad14}),  
can be written as 
\begin{eqnarray}
\label{ef1}
F=\Theta_c+\Theta_Q+U+W+W_{\rm ext}.
\end{eqnarray}
The first two terms in Eq. (\ref{ef1}) correspond to the total
kinetic energy
\begin{eqnarray}
\label{ef2}
\Theta=\frac{1}{m}\left \langle \psi\left |-\frac{\hbar^2}{2m}\Delta\right
|\psi\right \rangle=
-\frac{\hbar^2}{2m^2}\int \psi^*\Delta\psi\, d{\bf r}=\frac{\hbar^2}{2m^2}\int
|\nabla\psi|^2 \, d{\bf r}.
\end{eqnarray}
Using the Madelung transformation, the kinetic energy can be decomposed
into  the  classical kinetic energy
\begin{eqnarray}
\label{ef3}
\Theta_c=\int\rho  \frac{{\bf u}^2}{2}\, d{\bf r}
\end{eqnarray}
and the quantum kinetic energy\footnote{This functional was introduced 
by von Weizs\"acker \cite{wei} and is related to the Fisher 
\cite{fisher} entropy
$S_F=(1/m)\int {(\nabla\rho)^2}/{\rho}\, d{\bf r}$ by the relation
$\Theta_Q=({\hbar^2}/{8m})S_F$. Actually, the functional (\ref{ef5}) was already
introduced by Madelung
\cite{madelung,madelungearly} under the equivalent form
$\Theta_Q=-(\hbar^2/2m^2)\int \sqrt{\rho}\Delta\sqrt{\rho}\, d{\bf r}$ [see
Eq. (\ref{ef5b})].}
\begin{eqnarray}
\label{ef5}
\Theta_Q=\frac{\hbar^2}{8m^2}\int \frac{(\nabla\rho)^2}{\rho}\, d{\bf r}.
\end{eqnarray}
Using Eq. (\ref{mad8}), integrating by parts, and assuming that the boundary
term can be neglected, we get
\begin{eqnarray}
\label{ef5b}
\frac{1}{m}\int \rho Q\, d{\bf r}=-\frac{\hbar^2}{2m^2}\int
\sqrt{\rho}\Delta\sqrt{\rho}\, d{\bf r}
=\frac{\hbar^2}{2m^2}\int (\nabla\sqrt{\rho})^2\, d{\bf
r}=\frac{\hbar^2}{8m^2}\int \frac{(\nabla\rho)^2}{\rho}\, d{\bf r}.
\end{eqnarray}
Therefore, the quantum
kinetic energy can be rewritten as
\begin{equation}
\label{ef4}
\Theta_Q=\frac{1}{m}\int \rho Q\, d{\bf r}.
\end{equation}
It can be interpreted  as a potential energy associated with the quantum
potential $Q$.\footnote{This is not obvious since $Q$ is a function of the
density (it is {\it not} an external potential).} The third term in Eq.
(\ref{ef1}) is the internal
energy
\begin{eqnarray}
\label{ney1}
U=\int\rho\int^{\rho}\frac{P(\rho')}{{\rho'}^2}\, d\rho'\, d{\bf r}.
\end{eqnarray}
The density of internal energy $\rho u$ satisfies the first law
of thermodynamic $du=-Pd(1/\rho)$. We note that the internal energy is defined
up to a term of the
form $AM+B$ where $M$ is the
total mass and $A$ and $B$ are constants. In the following, we shall use the
expression of the internal energy given by
\begin{eqnarray}
\label{ney1b}
U=\int \left\lbrack \rho h(\rho)-P(\rho)\right \rbrack\, d{\bf r}=\int V(\rho)\,
d{\bf r},
\end{eqnarray}
which can be obtained from Eq. (\ref{ney1}) by a part integration, using 
Eqs. (\ref{mad11}) and (\ref{mad11b}). For a given self-interaction
potential $V(\rho)$, the enthalpy
$h(\rho)$, the pressure $P(\rho)$, and the internal energy $U$   are completely
determined by Eqs.
(\ref{kggp}), (\ref{mad11b}) and (\ref{ney1b}). The
fourth term
in Eq. (\ref{ef1}) is
the gravitational potential energy
\begin{eqnarray}
\label{ney3}
W=\frac{1}{2}\int\rho\Phi\, d{\bf r}.
\end{eqnarray}
The fifth term in Eq. (\ref{ef1}) is the external potential energy
\begin{eqnarray}
\label{ef1b}
W_{\rm ext}=\int\rho\Phi_{\rm ext}\, d{\bf r}.
\end{eqnarray}
For the harmonic potential (\ref{mfgp8}), we have
\begin{eqnarray}
\label{ef1c}
W_{\rm ext}=\frac{1}{2}\omega_0^2 I,
\end{eqnarray}
where
\begin{eqnarray}
\label{ef1d}
I=\int \rho r^2\, d{\bf r}
\end{eqnarray}
is the moment of inertia. We note that $I=M\langle r^2\rangle$ where $\langle
r^2\rangle$ measures the dispersion of the particles (or the size of the
condensate). Regrouping all these results, the free energy can be explicitly
written as
\begin{eqnarray}
\label{ae9}
F=\int\rho \frac{{\bf u}^2}{2}\, d{\bf r}+\frac{1}{m}\int \rho Q\, d{\bf r}+\int
V(\rho)\, d{\bf r}
+\frac{1}{2}\int\rho\Phi\, d{\bf r}+\int\rho\Phi_{\rm ext}\, d{\bf r}.
\end{eqnarray}
On the other hand, the free energy associated with the quantum
barotropic SP
equations
(\ref{mad16}) and (\ref{mad17}) is given by 
\begin{eqnarray}
\label{ef12}
F=\Theta_Q+U+W+W_{\rm ext}=\frac{1}{m}\int
\rho Q\, d{\bf r}+\int
V(\rho)\, d{\bf r}
+\frac{1}{2}\int\rho\Phi\, d{\bf r}+\int\rho\Phi_{\rm ext}\, d{\bf r}
\end{eqnarray}
since the classical kinetic energy $\Theta_c$, which is of order $O(\xi^{-2})$,
can be 
neglected in the overdamped limit $\xi\rightarrow +\infty$. 

\subsection{Difference between the average energy and the free energy}
\label{sec_ae}

The average value $\langle E\rangle$ of the energy $E({\bf r},t)$ defined by Eq.
(\ref{enxi}) is given by
\begin{equation}
\label{ae7}
N\langle E\rangle =\int \frac{\rho}{m}E\, d{\bf r}=\int \rho\frac{{\bf
u}^2}{2}\,
d{\bf r}+\int\rho \Phi\, d{\bf r}+\int \rho h(\rho)\, d{\bf r}+\int\rho
\Phi_{\rm ext}\, d{\bf r}+\frac{1}{m}\int \rho Q\, d{\bf r},
\end{equation}
i.e.
\begin{equation}
\label{ae7b}
N\langle E\rangle =\Theta_c+\Theta_Q+\int \rho h(\rho)\, d{\bf r}+2W+W_{\rm
ext}.
\end{equation}
It coincides with the average value of the energy operator (see
Appendix \ref{sec_gh}). In a static state where $E({\bf
r},t)=E$, we have $\langle E\rangle =E$, where $E$ is the eigenenergy.
On the other hand, comparing Eqs. (\ref{ae9}) and (\ref{ae7}), we find that
\begin{eqnarray}
\label{ae11}
F=N\langle E\rangle-\frac{1}{2}\int\rho\Phi\, d{\bf r}+\int \lbrack V(\rho)-\rho
h(\rho)\rbrack\, d{\bf r}.
\end{eqnarray}
Using Eq. (\ref{mad11b}), we get
\begin{eqnarray}
\label{ae12}
F=N\langle E\rangle-\frac{1}{2}\int\rho\Phi\, d{\bf r}-\int P\, d{\bf
r}=N\langle E\rangle-W-\int P\, d{\bf
r}.
\end{eqnarray}
In general the free energy is different from the average energy:
\begin{equation}
\label{ae13}
F\neq N\langle E\rangle.
\end{equation}
It is only in the case of the linear Schr\"odinger equation ($\Phi=P=0$) that
$F=N\langle E\rangle$.  For nonlinear Schr\"odinger equations ($P\neq 0$), or
for
systems
with long-range interactions ($\Phi\neq 0$), $F$ differs from $N\langle
E\rangle$ by a nontrivial functional $-\frac{1}{2}\int\rho\Phi\, d{\bf r}-\int
P\, d{\bf
r}$. As a result, $F$ and $\langle
E\rangle$  have  different
properties in general.\footnote{One exception is when $P=\rho k_BT/m$ and
$\Phi=0$
because, in that case, $F$ and  $N\langle
E\rangle$ just differ by a constant $-Nk_B T$. This corresponds to the
logarithmic GP equation discussed in Sec. \ref{sec_bol}}

\subsection{The $H$-theorem}
\label{sec_eff}

It is shown in Appendices \ref{sec_var} and  \ref{sec_hth} that the time
derivative of the free energy (\ref{ef1}) satisfies the identity
\begin{eqnarray}
\label{ef11}
\dot F=-\xi\int \rho {\bf u}^2\, d{\bf r}=-2\xi\Theta_c.
\end{eqnarray}
The local free energy equation is given in
Appendix \ref{sec_lee}. We have to consider two situations:

(i) For dissipationless systems ($\xi=0$), Eq. (\ref{ef11}) 
shows that the GPP equations, or the quantum barotropic EP equations, conserve
the free energy ($\dot F=0$).\footnote{For dissipationless systems, $F$ is
called the total energy $E_{\rm tot}$ of the system, not the free energy.
However, for convenience, we shall always refer to $F$ as the free energy.}  In
that case, it can be shown from general arguments  \cite{holm} that a minimum of
free energy at fixed mass
determines a steady state of the GPP equations, or quantum barotropic EP
equations, that is formally nonlinearly dynamically stable.

(ii) For dissipative systems ($\xi>0$), Eq. (\ref{ef11}) 
shows that the generalized  GPP equations, or the damped quantum
barotropic EP
equations, decrease the free energy ($\dot F\le 0$). When $\dot F=0$,  Eq.
(\ref{ef11}) implies that ${\bf u}={\bf 0}$. From the Euler equation
(\ref{mad13}), we obtain the condition of hydrostatic equilibrium (\ref{he1}).
Therefore, Eq. (\ref{ef11}) forms an $H$-theorem for the generalized GPP 
equations or for the
damped quantum barotropic EP equations: $\dot F\le 0$ and $\dot F=0$
if, and only if, the system is at equilibrium. In that case, $F$
is called a Lyapunov functional. From Lyapunov's direct method,
one
can show that the system will relax, for $t\rightarrow +\infty$, towards an
equilibrium state that is a (local) minimum of free energy at fixed mass. 
Maxima or saddle points of free energy are unstable. If several local minima of
free energy exist, the selection depends on the initial condition and on a
notion of basin of attraction.

The free energy associated with the quantum barotropic SP
equations
(\ref{mad16}) and (\ref{mad17}) is given by Eq. (\ref{ef12}).
Its time derivative
 satisfies the identity
\begin{eqnarray}
\label{ef13}
\dot F=-\frac{1}{\xi}\int \frac{1}{\rho}\left (\nabla
P+\rho\nabla\Phi+\rho\nabla\Phi_{\rm ext}+\frac{\rho}{m}\nabla Q\right )^2\,
d{\bf r}.
\end{eqnarray}
This identity can be obtained from the quantum barotropic 
SP equations (see Appendix \ref{sec_hth}). It can also be  directly obtained
from Eq. (\ref{ef11}) by using Eq. (\ref{mad15}) which is valid in the strong
friction limit. When $\dot F=0$, Eq. (\ref{ef13}) implies that the term in
parenthesis vanishes, leading to the condition of hydrostatic equilibrium
(\ref{he1}).  Therefore, Eq. (\ref{ef13}) forms an $H$-theorem for the quantum
barotropic SP equations.

{\it Remark:} Since the dissipative ($\xi\neq 0$) GPP equations, the damped
quantum barotropic EP equations, and the quantum SP equations are
relaxation equations, they  can be used as {\it numerical algorithms} to compute
stable equilibrium states of the conservative ($\xi=0$) GPP equations, or
quantum barotropic EP equations. This can be very useful on a practical
point of view because it
is generally not easy to solve the time-independent equations directly and be
sure that
the solution is stable.\footnote{See Appendix E of \cite{nfp}, and \cite{vp},
for
numerical algorithms under the form of relaxation equations that can be used to
construct stable steady states of the
Vlasov-Poisson and 2D Euler-Poisson equations.}

 \subsection{The equilibrium state}
\label{sec_efff}

According to the previous discussion, the equilibrium state of the generalized
GPP equations, or quantum barotropic EP equations, is the solution of
the minimization problem
\begin{eqnarray}
\label{ef14}
F(M)=\min_{\rho,{\bf u}} \left\lbrace F[\rho,{\bf u}]\quad |\quad M\quad {\rm fixed}\right\rbrace.
\end{eqnarray}
A critical point of free energy at fixed mass is determined by the variational principle
\begin{eqnarray}
\label{ney4}
\delta F-\frac{\mu}{m}\delta M=0,
\end{eqnarray}
where $\mu$ is a Lagrange multiplier taking  into account the mass constraint.
Using the results of Appendix \ref{sec_var}, this variational problem gives
${\bf u}={\bf 0}$ (the equilibrium state is static) and the condition
\begin{eqnarray}
\label{ef14b}
m\Phi+m\Phi_{\rm ext}+m h(\rho)+Q=\mu.
\end{eqnarray}
Taking the gradient of Eq. (\ref{ef14b}), and using Eq. (\ref{mad11}), 
we recover the condition of hydrostatic equilibrium (\ref{he1}). Equation
(\ref{ef14b}) is also equivalent to the time-independent GP equation
(\ref{tigp4}) provided that we make the identification
\begin{eqnarray}
\mu=E.
\end{eqnarray}
This shows that the Lagrange multiplier 
(chemical potential) in the variational problem associated with Eq. (\ref{ef14})
can be identified with the eigenenergy $E$.  Inversely, the eigenenergy $E$ may
be
interpreted as a chemical potential. According to Eq. (\ref{ef14b}), the
equilibrium state is given by
\begin{eqnarray}
\label{ef14bz}
\rho=h^{-1}\left (\frac{\mu}{m}-\frac{Q}{m}-\Phi-\Phi_{\rm
ext}\right ).
\end{eqnarray}
When $Q=\Phi=0$, this equation determines the equilibrium distribution
$\rho({\bf r})$. More generally,  Eq. (\ref{ef14bz}) is a differential, or an
integrodifferential, equation. Considering the second
order variations of free energy, we find that the equilibrium is stable if,
and only if,
\begin{eqnarray}
\label{ef16}
\delta^2 F= \frac{1}{2}\int h'(\rho)(\delta\rho)^2\, d{\bf r}+\frac{1}{2}\int \delta\rho\delta\Phi\, d{\bf r}
+\frac{\hbar^2}{8m^2}\int\frac{1}{\rho}\left\lbrack \left (\frac{\Delta\rho}{\rho}-\frac{(\nabla\rho)^2}{\rho^2}\right ){(\delta\rho)^2}+{(\nabla\delta\rho)^2}\right\rbrack\, d{\bf r}>0,
\end{eqnarray}
for all perturbations that conserve mass: $\int \delta\rho\, d{\bf r}=0$. This inequality can also be written as
\begin{eqnarray}
\label{ef16b}
\delta^2 F= \frac{1}{2}\int h'(\rho)(\delta\rho)^2\, d{\bf r}+\frac{1}{2}\int \delta\rho\delta\Phi\, d{\bf r}
+\frac{\hbar^2}{8m^2}\int  \left \lbrack \nabla \left (\frac{\delta\rho}{\sqrt{\rho}}\right )\right\rbrack^2\, d{\bf r}+\frac{\hbar^2}{8m^2}\int \frac{\Delta\sqrt{\rho}}{\rho^{3/2}}(\delta\rho)^2\, d{\bf r}>0.
\end{eqnarray}

\subsection{The Poincar\'e theorem}
\label{sec_poi}

In the minimization problem of Sec. \ref{sec_efff}, the chemical potential (or
eigenenergy) $\mu/m=E/m=\partial F/\partial M$ is the quantity conjugate to the
mass
$M$ (constraint) with respect to the free  energy $F$ (thermodynamical
potential). Therefore, if we plot $\mu=E$  as a function of $M$, we can
determine the stability of the system by a direct application of
the Poincar\'e theory of linear series of equilibria
\cite{poincare}.\footnote{See Refs.
\cite{katzpoincare,ijmpb} for the application of the Poincar\'e theorem in
connection to the thermodynamical stability of self-gravitating systems and
\cite{prd1,prd2,bectcoll} for the application of the Poincar\'e theorem  in
connection to the dynamical stability of self-gravitating BECs.}
According to the Poincar\'e
theorem, a change of
stability can only occur at a turning point of mass,  or at a bifurcation point,
in the series of equilibria. Therefore, if we know a limit in which the
configuration is stable, we can use the Poincar\'e theorem to deduce the
stability of the whole series of equilibria. In general, the series of
equilibria becomes unstable at the first turning point of mass, corresponding to
the maximum mass $M_{\rm max}$. Furthermore, since $\delta F=0$ at a
turning point of mass where $\delta M=0$ [see Eq. (\ref{ney4})], the curve
$F(M)$ present cusps.

\subsection{Functional derivatives}
\label{sec_fd}

In this section, we show that the hydrodynamic equations associated with the
generalized GPP equations can be expressed in terms of functional
derivatives of the free energy. This reveals their Hamiltonian structure (in
the conservative case) and facilitates the derivation of the $H$-theorem (in
the dissipative case). The same is true for the generalized GPP equations as
shown in Appendix \ref{sec_gh}.

Using Eq. (\ref{mad5}), the free energy (\ref{ae9}) can be written as
\begin{eqnarray}
\label{ae9w}
F=\int\rho \frac{(\nabla S)^2}{2m^2}\, d{\bf r}+\frac{1}{m}\int \rho Q\, d{\bf
r}+\int
V(\rho)\, d{\bf r}
+\frac{1}{2}\int\rho\Phi\, d{\bf r}+\int\rho\Phi_{\rm ext}\, d{\bf r}.
\end{eqnarray}
Taking the functional derivatives of the free energy (\ref{ae9w}) with respect
to
$\rho$ and $S$, and using the relations of Appendix \ref{sec_var}, we
obtain
\begin{equation}
\label{fd1}
\frac{\delta F}{\delta\rho}=\frac{{\bf
u}^2}{2}+\Phi+h(\rho)+\Phi_{\rm
ext}+\frac{Q}{m}=\frac{E({\bf r},t)}{m}\qquad {\rm and}\qquad \frac{\delta
F}{\delta S}=-\frac{1}{m}\nabla\cdot (\rho {\bf u}).
\end{equation}
Therefore, the hydrodynamic equations (\ref{mad6}) and (\ref{mad7}) can
be rewritten as
\begin{equation}
\label{fd2}
\frac{\partial\rho}{\partial t}=m \frac{\delta
F}{\delta S},\qquad \frac{\partial S}{\partial
t}=-m \frac{\delta
F}{\delta \rho}-\xi (S-\langle S\rangle).
\end{equation}
For conservative systems ($\xi=0$), these equations can be interpreted as
Hamilton equations for the density $\rho$ and its canonical action $S$. This
shows that the free energy $F$ represents the true Hamiltonian of the system
(see also Appendices  \ref{sec_gh} and \ref{sec_lh}).  This formulation directly
implies the conservation of the  free energy $F$ since 
\begin{equation}
\label{fd3}
\dot F=\int \frac{\delta F}{\delta\rho}\frac{\partial\rho}{\partial t}\, d{\bf
r}+\int \frac{\delta F}{\delta S}\frac{\partial S}{\partial t}\, d{\bf
r}=0.
\end{equation}
For dissipative systems, one directly recovers the $H$-theorem (\ref{ef11})
from Eq. (\ref{fd2}) since
\begin{equation}
\label{fd4}
\dot F=-\xi\int \frac{\delta F}{\delta S}(S-\langle S\rangle)\, d{\bf
r}=\frac{\xi}{m}\int \nabla\cdot (\rho {\bf u})(S-\langle S\rangle)\, d{\bf
r}=-\frac{\xi}{m}\int \rho {\bf u}\cdot \nabla S\, d{\bf r}=-\xi\int
\rho {\bf u}^2\, d{\bf r}.
\end{equation}

On the other hand, taking the functional derivative of the free energy
(\ref{ae9}) with respect to
${\bf u}$, we
obtain
\begin{equation}
\label{fd5}
\frac{\delta
F}{\delta {\bf u}}=\rho {\bf u}.
\end{equation}
Therefore, the hydrodynamic equations (\ref{mad12}) and (\ref{mad13}) can
be rewritten as
\begin{equation}
\label{fd6}
\frac{\partial\rho}{\partial t}=-\nabla\cdot \left (\frac{\delta
F}{\delta {\bf u}}\right ),\qquad \frac{\partial {\bf u}}{\partial
t}=-\nabla\left (\frac{\delta
F}{\delta \rho}\right )-\xi{\bf u}.
\end{equation}
One directly recovers the $H$-theorem (\ref{ef11}) from Eq. (\ref{fd6}) since
\begin{equation}
\label{fd7}
\dot F=\int \frac{\delta F}{\delta\rho}\frac{\partial\rho}{\partial t}\, d{\bf
r}+\int \frac{\delta F}{\delta {\bf u}}\cdot \frac{\partial {\bf u}}{\partial
t}\, d{\bf
r}=-\int \frac{\delta F}{\delta\rho}\nabla\cdot \left (\frac{\delta
F}{\delta {\bf u}}\right )     \, d{\bf
r}-\int \frac{\delta F}{\delta {\bf u}}\cdot \nabla\left (\frac{\delta
F}{\delta \rho}\right )\, d{\bf
r}-\xi\int \frac{\delta F}{\delta {\bf u}}\cdot {\bf u}\,
d{\bf
r}=-\xi\int
\rho {\bf u}^2\, d{\bf r}.
\end{equation}
We note that the two first terms of the second equality cancel out after an
integration by parts. In the strong friction
limit $\xi\rightarrow +\infty$, Eqs. (\ref{fd5}) and (\ref{fd6}) can be
combined into a single equation
\begin{equation}
\label{fd8}
\xi\frac{\partial\rho}{\partial t}=\nabla\cdot\left (\rho\nabla \frac{\delta
F}{\delta {\bf \rho}}\right ),
\end{equation}
which is equivalent to the quantum
barotropic Smoluchowski equation (\ref{mad16}). Equation (\ref{fd8}) is called
a flow gradient equation in the mathematical literature. In this equation, the
free
energy $F$ is given by Eq. (\ref{ef12}). Again, one directly recovers the
$H$-theorem  (\ref{ef13}) from Eq. (\ref{fd8}) since
\begin{equation}
\label{fd9}
\dot F=\int \frac{\delta F}{\delta\rho}\frac{\partial\rho}{\partial t}\, d{\bf
r}=\frac{1}{\xi}\int \frac{\delta F}{\delta\rho} \nabla\cdot\left (\rho\nabla
\frac{\delta
F}{\delta {\bf \rho}}\right )  \, d{\bf
r}=-\frac{1}{\xi}\int \rho\left (\nabla
\frac{\delta
F}{\delta {\bf \rho}}\right )^2  \, d{\bf
r}.
\end{equation}

{\it Remark:} We note the identity
\begin{equation}
\label{rw1}
N\langle E\rangle=\int\rho\frac{\delta F}{\delta\rho}\, d{\bf r},
\end{equation}
obtained from Eq. (\ref{fd1}), which shows the intrinsic difference between
$N\langle E\rangle$ and $F$. It is the Hamiltonian structure
of the hydrodynamic equations (\ref{fd2}), (\ref{fd6}) and
(\ref{fd8}) [see also Eqs. (\ref{ab}) and (\ref{ae12w})] that justifies to
consider the free energy $F$ as a Hamiltonian, or as a Lyapunov functional,
instead of the average energy $N\langle E\rangle$. We have seen in Sec.
\ref{sec_efff} that the eigenenergy $E$ can be regarded as a chemical potential
$\mu=\delta F/\delta N$. Therefore, the average energy  $\langle E\rangle$,
which is
related to the free energy by Eq. (\ref{rw1}),  can
be regarded as a time-dependent chemical potential $\mu(t)$ that becomes
constant at equilibrium.  Similarly, the local energy $E({\bf r},t)$, which is
related to the free energy by Eq. (\ref{fd1}), can be regarded as an
out-of-equilibrium chemical potential $\mu({\bf r},t)$ that becomes uniform
and constant at
equilibrium.

\section{The virial theorem}
\label{sec_virial}

\subsection{General case}

From the damped quantum barotropic EP equations (\ref{mad12})-(\ref{mad14}), we can derive the time-dependent scalar virial theorem (see Appendix \ref{sec_virialannexe}):
\begin{equation}
\label{wt1}
\frac{1}{2}\ddot I+\frac{1}{2}\xi\dot I=2(\Theta_c+\Theta_Q)+d\int P\, d{\bf
r}+W_{ii}+W_{ii}^{\rm ext}.
\end{equation}
In the strong friction limit $\xi\rightarrow +\infty$, corresponding to the
quantum barotropic SP equations  (\ref{mad16}) and (\ref{mad17}), we get
\begin{equation}
\label{wt2}
\frac{1}{2}\xi\dot I=2\Theta_Q+d\int P\, d{\bf r}+W_{ii}+W_{ii}^{\rm ext}.
\end{equation}
At equilibrium ($\ddot I=\dot I=\Theta_c=0$), the virial theorem becomes
\begin{equation}
\label{wt3}
2\Theta_Q+d\int P\, d{\bf r}+W_{ii}+W_{ii}^{\rm ext}=0.
\end{equation}
On the other hand, the free energy (\ref{ef1}) reduces to
\begin{eqnarray}
\label{wt4}
F=\Theta_Q+U+W+W_{\rm ext}.
\end{eqnarray}
Multiplying Eq. (\ref{tigp4}) by $\rho$ and integrating over the whole
domain, we obtain
\begin{equation}
\label{wt5}
NE=\Theta_Q+\int \rho h\, d{\bf r}+2W+W_{\rm ext}.
\end{equation}
Comparing Eqs. (\ref{wt4}) and (\ref{wt5}), and using Eq. (\ref{mad11b}), we get
\begin{equation}
\label{wt5a}
F=NE-W-\int P\, d{\bf r}.
\end{equation}
Equations (\ref{wt5}) and (\ref{wt5a}) are the equilibrium forms
of Eqs. (\ref{ae7b}) and (\ref{ae12}). Eliminating $\int P\, d{\bf r}$ between
Eqs. (\ref{wt3}) and (\ref{wt5a}), we find that
\begin{equation}
\label{wt5b}
F=NE-W+\frac{2}{d}\Theta_Q+\frac{1}{d}W_{ii}+\frac{1}{d}W_{ii}^{\rm ext}.
\end{equation}

\subsection{Harmonic potential}

For the harmonic potential (\ref{mfgp8}), the time-dependent virial
theorem can be written as (see Appendix \ref{sec_virialannexe}):
\begin{equation}
\label{wt6}
\frac{1}{2}\ddot I+\frac{1}{2}\xi\dot I+\omega_0^2 I=2(\Theta_c+\Theta_Q)+d\int
P\, d{\bf r}+W_{ii}.
\end{equation}
In the strong friction limit $\xi\rightarrow +\infty$,  it reduces to
\begin{equation}
\label{wt7}
\frac{1}{2}\xi\dot I+\omega_0^2 I=2\Theta_Q+d\int P\, d{\bf r}+W_{ii}.
\end{equation}
At equilibrium ($\ddot I=\dot I=\Theta_c=0$), the virial theorem, the free
energy, and the energy take the form
\begin{equation}
\label{wt8}
2\Theta_Q+d\int P\, d{\bf r}+W_{ii}-\omega_0^2I=0,
\end{equation}
\begin{eqnarray}
\label{wt9}
F=\Theta_Q+U+W+\frac{1}{2}\omega_0^2 I,
\end{eqnarray}
\begin{equation}
\label{wt10}
NE=\Theta_Q+\int \rho h\, d{\bf r}+2W+\frac{1}{2}\omega_0^2 I.
\end{equation}

{\it Remark:} If we consider the nongravitational limit ($G=0$) and the case
$\xi=0$ where the free energy
$F$ is conserved, we can combine Eqs. (\ref{ef1}) and
(\ref{wt6}) to obtain the exact equation
\begin{equation}
\label{wt6b}
\frac{1}{2}\ddot I+2\omega_0^2 I=2F-2U+d\int
P\, d{\bf r}=2F-(d+2)\int V(\rho)\, d{\bf r}+d\int\rho V'(\rho)\, d{\bf r}.
\end{equation}
An application of this equation will be given in Sec. \ref{sec_es4}.

\section{Particular equations of state and generalized entropies}
\label{sec_eos}

The free energy (\ref{ef1}) associated with the generalized GPP equations
(\ref{mfgp9}) and (\ref{mfgp14}), or with the damped quantum barotropic EP
equations (\ref{mad12})-(\ref{mad14}),  can be written as
\begin{eqnarray}
\label{f1}
F=E_*+U,
\end{eqnarray}
where $U$ is the internal energy (\ref{ney1b}) and
\begin{eqnarray}
\label{f2}
E_*=\Theta_c+\Theta_Q+W+W_{\rm ext}
\end{eqnarray}
is the energy that includes the classical kinetic energy $\Theta_c$, the quantum kinetic energy $\Theta_Q$, the gravitational potential energy $W$, and the external potential energy $W_{\rm ext}$. It can be written explicitly as
\begin{eqnarray}
\label{f3}
E_*=\int\rho \frac{{\bf u}^2}{2}\, d{\bf r}+\frac{1}{m}\int \rho Q\, d{\bf r}
+\frac{1}{2}\int\rho\Phi\, d{\bf r}+\int\rho\Phi_{\rm ext}\, d{\bf r}.
\end{eqnarray}
The effective potential $h(\rho)$ which accounts for short-range interactions
(collisions) between the bosons in the GP equation (\ref{mfgp9}) determines a
barotropic equation of state $P(\rho)$ in the quantum Euler equation
(\ref{mad13}) through 
the relations (\ref{mad11}) and (\ref{mad11b}). Inversely, for a given equation
of state $P=P(\rho)$, we can obtain the corresponding effective potential 
$h(\rho)$. The effective potential and the equation of state determine the
internal energy $U$ through Eq. (\ref{ney1b}). As we
shall see, the internal energy can be interpreted as the opposite
of a generalized (effective) entropy $S_{\rm eff}$ multiplied by a generalized
(effective) temperature $T_{\rm eff}$. Consequently, the free energy can be put
in the standard form
\begin{eqnarray}
\label{f1b}
F=E_*-T_{\rm eff}S_{\rm eff}.
\end{eqnarray}
In this section, we construct generalized forms of GP equations associated 
with specific equations of state, and determine the corresponding entropies.
The case of composite models is discussed in Appendix \ref{sec_com}.

\subsection{Isothermal equation of state: Boltzmann entropy}
\label{sec_bol}

\subsubsection{Gross-Pitaevskii equation}
\label{sec_bol_gp}

The isothermal equation of state \cite{chandra}:
\begin{equation}
\label{f4}
P=\rho\frac{k_BT}{m},\qquad c_s^2=\frac{k_BT}{m},
\end{equation}
is associated with an effective  potential of the form
\begin{equation}
\label{f5}
h(\rho)=\frac{k_B T}{m}\ln\rho,\qquad V(\rho)=\frac{k_B T}{m}\rho(\ln\rho-1).
\end{equation}
The corresponding generalized GP equation is
\begin{eqnarray}
\label{f6}
i\hbar \frac{\partial\psi}{\partial t}=-\frac{\hbar^2}{2m}\Delta\psi+m\Phi\psi+\frac{1}{2}m\omega_0^2r^2\psi+2k_B T\ln|\psi|\psi
-i\frac{\hbar}{2}\xi\left\lbrack \ln\left (\frac{\psi}{\psi^*}\right )-\left\langle \ln\left (\frac{\psi}{\psi^*}\right )\right\rangle\right\rbrack\psi.
\end{eqnarray}
It has a logarithmic nonlinearity. The internal energy is given by
\begin{equation}
\label{f7}
U_B=\frac{k_B T}{m}\int \rho(\ln\rho-1)\, d{\bf r}.
\end{equation}
The free energy and the average energy
are
\begin{eqnarray}
\label{f9b}
F_B=\Theta_c+\Theta_Q+W+W_{\rm ext}+U_B,
\end{eqnarray}
\begin{eqnarray}
\label{f9c}
N\langle E\rangle=\Theta_c+\Theta_Q+2W+W_{\rm ext}+U_B+Nk_B T.
\end{eqnarray}
They satisfy the relation $F_B=N\langle E\rangle-W-Nk_B T$.
The free energy can be written as
\begin{eqnarray}
\label{f8}
F_B=E_*-TS_B,
\end{eqnarray}
where $T$ is the temperature and
\begin{eqnarray}
\label{f9}
S_B=-k_B \int \frac{\rho}{m}(\ln\rho-1)\, d{\bf r}
\end{eqnarray}
is the Boltzmann entropy.

{\it Remark:} In order to have a dimensionless quantity in the logarithm, we can
replace Eqs. (\ref{f5}), (\ref{f7}) and (\ref{f9}) by
\begin{equation}
\label{f10}
h(\rho)=\frac{k_B T}{m}\ln\left (\frac{\rho}{\rho_0}\right ),\quad U_B=\frac{k_B T}{m}\int \rho \left\lbrack \ln\left (\frac{\rho}{\rho_0}\right )-1\right\rbrack\, d{\bf r},\quad S_B=-k_B \int \frac{\rho}{m}\left\lbrack\ln\left (\frac{\rho}{\rho_0}\right )-1\right\rbrack\, d{\bf r},
\end{equation}
where $\rho_0$ is a reference density.

\subsubsection{Hydrodynamic representation}

The generalized GP equation (\ref{f6}) is equivalent to the damped quantum
isothermal Euler equations
\begin{equation}
\label{f11}
\frac{\partial\rho}{\partial t}+\nabla\cdot (\rho {\bf u})=0,
\end{equation}
\begin{eqnarray}
\label{f12}
\frac{\partial {\bf u}}{\partial t}+({\bf u}\cdot \nabla){\bf u}=-\frac{k_B T}{m}\nabla \ln\rho
-\nabla\Phi-\omega_0^2{\bf r}-\frac{1}{m}\nabla Q-\xi{\bf u}.
\end{eqnarray}
In the strong friction limit $\xi\rightarrow +\infty$, we obtain the quantum Smoluchowski equation
\begin{eqnarray}
\label{f13}
\xi\frac{\partial\rho}{\partial t}=\nabla\cdot\biggl (\frac{k_B T}{m}\nabla\rho
+\rho\nabla\Phi+\rho \omega_0^2{\bf r}+\frac{\rho}{m}\nabla Q\biggr ).
\end{eqnarray}
We note that it corresponds to a normal classical diffusion.

\subsubsection{Hydrostatic equilibrium}
\label{sec_he2}

The condition of hydrostatic equilibrium writes
\begin{eqnarray}
\label{f14}
\frac{k_B T}{m}\nabla\rho+\rho\nabla\Phi
+\rho\omega_0^2{\bf r}+\frac{\rho}{m}\nabla Q={\bf 0}.
\end{eqnarray}
Combined with the Poisson equation (\ref{mad14}), we obtain
\begin{equation}
\label{f15}
-\frac{k_B T}{m}\Delta\ln\rho+\frac{\hbar^2}{2m^2}\Delta \left (\frac{\Delta\sqrt{\rho}}{\sqrt{\rho}}\right )=S_d G\rho+d\omega_0^2.
\end{equation}
In the TF approximation ($Q=0$), the
foregoing equation reduces to
\begin{equation}
\label{f16}
-\frac{k_B T}{m}\Delta\ln\rho=S_d G\rho+d\omega_0^2.
\end{equation}

\subsubsection{The equilibrium state}
\label{sec_es}

The minimization of the Boltzmann free energy at fixed mass (see Sec.
\ref{sec_efff}) leads to the equation
\begin{eqnarray}
\label{f17}
Q+m\Phi+\frac{1}{2}m\omega_0^2r^2
+k_B T\ln\rho=\mu.
\end{eqnarray}
This equation is equivalent to the condition of hydrostatic equilibrium. It can be rewritten as
\begin{eqnarray}
\label{f19}
\rho=e^{-\beta(m\Phi+Q+\frac{1}{2}m\omega_0^2r^2-\mu)},
\end{eqnarray}
which can be interpreted as a generalized Boltzmann distribution including the contribution of the quantum potential. In the TF approximation ($Q=0$), we recover the Boltzmann distribution.

{\it Remark:} If we use the definitions of Eq. (\ref{f10}), we obtain
\begin{eqnarray}
\label{f20}
\rho=\rho_0e^{-\beta(m\Phi+Q+\frac{1}{2}m\omega_0^2r^2-\mu)}.
\end{eqnarray}

\subsubsection{The virial theorem}
\label{sec_es2}

The scalar virial theorem writes
\begin{equation}
\label{f21}
\frac{1}{2}\ddot I+\frac{1}{2}\xi\dot I+\omega_0^2 I=2(\Theta_c+\Theta_Q)+dN k_B T+W_{ii}.
\end{equation}
In the strong friction limit, we get
\begin{equation}
\label{f21b}
\frac{1}{2}\xi\dot I+\omega_0^2 I=2\Theta_Q+dN k_B T+W_{ii}.
\end{equation}
At equilibrium, the virial theorem, the free energy and the eigenenergy reduce
to
\begin{equation}
\label{f22}
2\Theta_Q+dN k_B T+W_{ii}-\omega_0^2 I=0,
\end{equation}
\begin{eqnarray}
\label{f23}
F_B=\Theta_Q+W+\frac{1}{2}\omega_0^2 I+U_B,
\end{eqnarray}
\begin{equation}
\label{f24}
NE=\Theta_Q+U_B+Nk_B T+2W+\frac{1}{2}\omega_0^2 I.
\end{equation}

{\it Remark:} In the strong friction limit $\xi\rightarrow +\infty$, in the TF
approximation  where we can neglect the quantum potential ($Q=0$),
and in $d=2$ dimensions where the virial of the gravitational force $W_{ii}$ is
given by Eq. (\ref{virial21}), the virial theorem (\ref{f21b}) takes the form
\begin{equation}
\label{ex1}
\frac{1}{2}\xi\dot I+\omega_0^2 I=2N k_B T-\frac{GM^2}{2}.
\end{equation}
If we introduce the critical temperature\footnote{The dimension $d=2$
is critical for isothermal self-gravitating systems.}
\begin{equation}
\label{ex2}
k_B T_c=\frac{GMm}{4},
\end{equation}
it can be rewritten as
\begin{equation}
\label{ex3}
\frac{1}{2}\xi\dot I+\omega_0^2 I=2N k_B (T-T_c).
\end{equation}
At equilibrium,
\begin{equation}
\label{ex7}
\omega_0^2 I=2N k_B (T-T_c).
\end{equation}
Remarkably, Eq. (\ref{ex3}) is a {\it closed} equation. Therefore, although we
cannot solve the SP equations  analytically, it turns out that we can obtain the
evolution of the moment of inertia $I(t)$, or equivalently the evolution of the
mean square displacement  $\langle r^2\rangle(t)=I(t)/M$, analytically. Indeed,
we obtain
\begin{equation}
\label{ex4}
\langle r^2\rangle (t)=\langle r^2\rangle_0 e^{-2\omega_0^2t/\xi}+\frac{2k_B}{m\omega_0^2}(T-T_c)\left (1-e^{-2\omega_0^2t/\xi}\right ).
\end{equation}
If $\omega_0=0$, we get
\begin{equation}
\label{ex5}
\langle r^2\rangle (t)=\frac{4k_B}{\xi m}(T-T_c)t+\langle r^2\rangle_0.
\end{equation}
The mean square displacement  behaves like in a  pure diffusion process, $\langle r^2\rangle=4D_{\rm eff}t+\langle r^2\rangle_0$, with an effective diffusion coefficient
\begin{equation}
\label{ex6}
D_{\rm eff}=\frac{k_B T}{\xi m}\left (1-\frac{T_c}{T}\right ).
\end{equation}
This exact result \cite{exact} generalizes the Einstein relation \cite{einstein}
to the case of
2D Brownian particles in gravitational interaction. The original Einstein
relation
$D=k_BT/\xi m$ is recovered for $T_c=0$, i.e., in the absence of gravitational
interaction ($G=0$). A detailed discussion of this model can be
found in \cite{exact,ijmpb12}.

\subsection{Polytropic equation of state: Tsallis entropy}
\label{sec_tsa}

\subsubsection{Gross-Pitaevskii equation}

The polytropic equation of state \cite{chandra}:
\begin{equation}
\label{f28}
P=K\rho^{\gamma},\qquad  \gamma=1+\frac{1}{n},\qquad
c_s^2=K\gamma\rho^{\gamma-1},
\end{equation}
is associated with an effective potential of the form
\begin{equation}
\label{f29}
h(\rho)=\frac{K\gamma}{\gamma-1}\rho^{\gamma-1},\qquad
V(\rho)=\frac{K}{\gamma-1} \rho^{\gamma}.
\end{equation}
We note that $P(\rho)=(\gamma-1)V(\rho)$. The corresponding
generalized
GP equation
is
\begin{eqnarray}
\label{f30}
i\hbar \frac{\partial\psi}{\partial t}=-\frac{\hbar^2}{2m}\Delta\psi+m\Phi\psi+\frac{1}{2}m\omega_0^2r^2\psi+\frac{K\gamma m}{\gamma-1} |\psi|^{2(\gamma-1)}\psi
-i\frac{\hbar}{2}\xi\left\lbrack \ln\left (\frac{\psi}{\psi^*}\right )-\left\langle \ln\left (\frac{\psi}{\psi^*}\right )\right\rangle\right\rbrack\psi.
\end{eqnarray}
It has a power-law nonlinearity. The internal energy is given by
\begin{equation}
\label{f31}
U_{\gamma}=\frac{K}{\gamma-1}\int \rho^{\gamma}\, d{\bf r}.
\end{equation}
The free energy and the average energy
are 
\begin{eqnarray}
\label{f31b}
F_\gamma=\Theta_c+\Theta_Q+W+W_{\rm ext}+U_\gamma,
\end{eqnarray}
\begin{eqnarray}
\label{f31c}
N\langle E\rangle=\Theta_c+\Theta_Q+2W+W_{\rm ext}+\gamma U_\gamma.
\end{eqnarray}
They satisfy the relation $F_\gamma=N\langle E\rangle-W-(\gamma-1)U_\gamma$.
The
free energy can
be written as
\begin{eqnarray}
\label{f32}
F_{\gamma}=E_*-K S_{\gamma},
\end{eqnarray}
where $K$ is the polytropic temperature and
\begin{eqnarray}
\label{f33}
S_{\gamma}=-\frac{1}{\gamma-1}\int \rho^{\gamma}\, d{\bf r}
\end{eqnarray}
is the Tsallis, or power-law, entropy of index $\gamma$.

{\it Remark:} In order to recover the results of Sec. \ref{sec_bol_gp} in the
limit $\gamma\rightarrow 1$, we can define
\begin{equation}
\label{imp1b}
h(\rho)=\frac{K\gamma}{\gamma-1}\left (\rho^{\gamma-1}-1\right ),\qquad
V(\rho)=\frac{K}{\gamma-1} \left ( \rho^{\gamma}-\gamma \rho\right ),
\end{equation}
\begin{equation}
\label{imp2b}
U_{\gamma}=\frac{K}{\gamma-1} \int\left (\rho^{\gamma}-\gamma \rho\right )\, d{\bf r},\qquad
S_{\gamma}=-\frac{1}{\gamma-1} \int\left ( \rho^{\gamma}-\gamma \rho\right )\, d{\bf r},
\end{equation}
or
\begin{equation}
\label{imp1c}
h(\rho)=\frac{K\gamma}{\gamma-1}\left\lbrack\left (\frac{\rho}{\rho_0}\right )^{\gamma-1}-1\right\rbrack,\qquad
V(\rho)=\frac{K\rho_0}{\gamma-1} \left\lbrack\left (\frac{\rho}{\rho_0}\right
)^{\gamma}-\gamma \frac{\rho}{\rho_0}\right\rbrack,
\end{equation}
\begin{equation}
\label{imp2c}
U_{\gamma}=\frac{K\rho_0}{\gamma-1} \int\left\lbrack\left (\frac{\rho}{\rho_0}\right )^{\gamma}-\gamma \frac{\rho}{\rho_0}\right\rbrack\, d{\bf r},\qquad
S_{\gamma}=-\frac{\rho_0}{\gamma-1} \int\left\lbrack\left (\frac{\rho}{\rho_0}\right )^{\gamma}-\gamma \frac{\rho}{\rho_0}\right\rbrack\, d{\bf r}.
\end{equation}
We note that the entropy 
defined by Eqs. (\ref{imp2b}) and (\ref{imp2c})  slightly differs from the usual
Tsallis entropy because of the factor $\gamma$ in front of $\rho$. On the other
hand, in
order to recover the correct dimensions of temperature and entropy when
$\gamma\rightarrow 1$, we must define the polytropic temperature $T_\gamma$
such that $K=k_B T_\gamma/{m}$ and the Tsallis entropy $S_{\gamma}$ such
that the Tsallis free energy writes $F_\gamma=E_*-T_\gamma S_\gamma$. This
amounts  to multiplying
the entropies (\ref{imp2b}) and (\ref{imp2c}) by $k_B/m$. When
$\gamma\rightarrow 1$, we recover Eqs.
(\ref{f9}) and (\ref{f10}).

\subsubsection{Hydrodynamic representation}

The generalized GP equation (\ref{f30}) is equivalent to the damped
quantum polytropic Euler equations
\begin{equation}
\label{f36}
\frac{\partial\rho}{\partial t}+\nabla\cdot (\rho {\bf u})=0,
\end{equation}
\begin{eqnarray}
\label{f37}
\frac{\partial {\bf u}}{\partial t}+({\bf u}\cdot \nabla){\bf u}=-\frac{K\gamma}{\gamma-1}\nabla\rho^{\gamma-1}
-\nabla\Phi-\omega_0^2{\bf r}-\frac{1}{m}\nabla Q-\xi{\bf u}.
\end{eqnarray}
In the strong friction limit $\xi\rightarrow +\infty$, we obtain the
quantum polytropic Smoluchowski equation
\begin{eqnarray}
\label{f38}
\xi\frac{\partial\rho}{\partial t}=\nabla\cdot\biggl (K\nabla\rho^{\gamma}
+\rho\nabla\Phi+\rho \omega_0^2{\bf r}+\frac{\rho}{m}\nabla Q\biggr ).
\end{eqnarray}

\subsubsection{Hydrostatic equilibrium}
\label{sec_he3}

The condition of hydrostatic equilibrium writes
\begin{eqnarray}
\label{f39}
K\nabla\rho^{\gamma}+\rho\nabla\Phi
+\rho\omega_0^2{\bf r}+\frac{\rho}{m}\nabla Q={\bf 0}.
\end{eqnarray}
Combined with the Poisson equation (\ref{mad14}), we obtain
\begin{equation}
\label{f40}
-\frac{K\gamma}{\gamma-1}\Delta\rho^{\gamma-1}+\frac{\hbar^2}{2m^2}\Delta \left (\frac{\Delta\sqrt{\rho}}{\sqrt{\rho}}\right )=S_d G\rho+d\omega_0^2.
\end{equation}
In the TF approximation ($Q=0$), the foregoing equation reduces to
\begin{equation}
\label{f41}
-\frac{K\gamma}{\gamma-1}\Delta\rho^{\gamma-1}=S_d G\rho+d\omega_0^2.
\end{equation}

\subsubsection{The equilibrium state}
\label{sec_es3}

The minimization of the Tsallis free energy at fixed mass (see Sec.
\ref{sec_efff}) leads to the equation
\begin{eqnarray}
\label{f42}
Q+m\Phi+\frac{1}{2}m\omega_0^2r^2+
\frac{K\gamma m}{\gamma-1}\rho^{\gamma-1}=\mu.
\end{eqnarray}
This equation is equivalent to the condition of hydrostatic equilibrium. It can be rewritten as
\begin{eqnarray}
\label{f44}
\rho=\left\lbrack -\frac{\gamma-1}{K\gamma m}\left (m\Phi+Q+\frac{1}{2}m\omega_0^2r^2-\mu\right )\right\rbrack^{1/(\gamma-1)},
\end{eqnarray}
which can be interpreted as a generalized Tsallis distribution including the contribution of the quantum potential. In the TF approximation ($Q=0$), we recover the Tsallis distribution.

{\it Remark:} If we use the definition of Eq. (\ref{imp1c}), we get
\begin{eqnarray}
\label{f45}
\rho=\rho_0\left\lbrack 1-\frac{\gamma-1}{K\gamma m \rho_0^{\gamma-1}}\left (m\Phi+Q+\frac{1}{2}m\omega_0^2r^2-\mu\right )\right\rbrack^{1/(\gamma-1)}
\end{eqnarray}
which returns the Boltzmann distribution (\ref{f20}) when $\gamma\rightarrow 1$.

\subsubsection{The virial theorem}
\label{sec_es4}

The scalar virial theorem writes
\begin{equation}
\label{f47}
\frac{1}{2}\ddot I+\frac{1}{2}\xi\dot I+\omega_0^2 I=2(\Theta_c+\Theta_Q)+d(\gamma-1)U_{\gamma}+W_{ii},
\end{equation}
where we have used the identity $\int P\, d{\bf r}=(\gamma-1)U_{\gamma}$.
In the strong friction limit, we get
\begin{equation}
\label{f47b}
\frac{1}{2}\xi\dot I+\omega_0^2 I=2\Theta_Q+d(\gamma-1)U_{\gamma}+W_{ii}.
\end{equation}
At equilibrium, the virial theorem, the free energy and the eigenenergy reduce
to
\begin{equation}
\label{f48}
2\Theta_Q+d(\gamma-1)U_{\gamma}+W_{ii}-\omega_0^2I=0,
\end{equation}
\begin{eqnarray}
\label{f49}
F_{\gamma}=\Theta_Q+U_{\gamma}+W+\frac{1}{2}\omega_0^2 I,
\end{eqnarray}
\begin{equation}
\label{f51}
NE=\Theta_Q+\gamma U_{\gamma}+2W+\frac{1}{2}\omega_0^2 I,
\end{equation}
where we have used the identity $\int \rho h\, d{\bf r}=\gamma U_{\gamma}$
valid for a polytropic equation of state.

{\it Remark:} If we consider the nongravitational limit ($G=0$) and the case of
conservative systems ($\xi=0$) for which the free energy $F_{\gamma}$ is
conserved, we can combine Eqs.
(\ref{f31b}) and (\ref{f47}) to obtain
\begin{equation}
\label{f51c}
\frac{1}{2}\ddot I+2\omega_0^2 I=2F_{\gamma}+\lbrack d(\gamma-1)-2\rbrack
U_{\gamma}.
\end{equation}
This is a particular case of Eq. (\ref{wt6b}). For the critical index
$\gamma_c=1+2/d$ (corresponding to $n_c=d/2$) \cite{sulem}, we obtain
a {\it closed} equation for the moment of inertia
\begin{equation}
\label{f51b}
\ddot I+4\omega_0^2 I=4F_{\gamma_c}.
\end{equation}
It has the solution $I(t)=A\cos(2\omega_0 t+\phi)+F_{\gamma_c}/\omega_0^2$. This
result is valid for repulsive ($K>0$) and attractive ($K<0$)  interactions. For
$d=2$, we find
$\gamma_c=2$ which corresponds to the standard BEC. 

{\it Remark:} If we consider classical polytropes ($\hbar=0$) without external
potential ($\omega_0=0$) we can combine Eqs. (\ref{f48}) and (\ref{f49}) to
obtain, at equilibrium, and for $d\neq 2$:
\begin{equation}
\label{f51d}
F_\gamma=\frac{\gamma-\gamma_{4/3}}{\gamma-1}W=\left (1-\frac{n}{n_3}\right )W,
\end{equation}
where $\gamma_{4/3}=2-2/d$ (corresponding to $n_3=d/(d-2)$). In $d=3$, we
get $F_{\gamma}=(\gamma-4/3)W/(\gamma-1)=(1-n/3)W$.  According
to the Poincar\'e argument \cite{chandra}, a necessary (but not sufficient)
condition of nonlinear dynamical stability is that $F<0$ (we assume $n>0$).
Since $W<0$ [see Eq. (\ref{virial7w})], we conclude that the system is unstable
for $\gamma<{4/3}$ (i.e. $n>3$).

\subsection{Standard BEC: quadratic entropy}
\label{sec_st}

\subsubsection{Gross-Pitaevskii equation}

For a standard BEC  in $d=3$ \cite{revuebec}, the generalized GP equation
writes
\begin{eqnarray}
\label{f58}
i\hbar \frac{\partial\psi}{\partial t}=-\frac{\hbar^2}{2m}\Delta\psi+m\Phi\psi+\frac{1}{2}m\omega_0^2r^2\psi+\frac{4\pi a_s\hbar^2}{m^2} |\psi|^{2}\psi
-i\frac{\hbar}{2}\xi\left\lbrack \ln\left (\frac{\psi}{\psi^*}\right )-\left\langle \ln\left (\frac{\psi}{\psi^*}\right )\right\rangle\right\rbrack\psi.
\end{eqnarray}
It has a cubic nonlinearity. The effective potential is given by
\begin{equation}
\label{f59}
h(\rho)=\frac{4\pi a_s\hbar^2}{m^3}\rho,\qquad V(\rho)=\frac{2\pi
a_s\hbar^2}{m^3}\rho^2,
\end{equation}
where $a_s$ is the s-scattering length of the bosons. This potential 
arises from close contact interactions (see Sec. \ref{sec_mfgp}). The
corresponding equation of state is\footnote{This equation of state was first
derived by Bogoliubov \cite{bogoliubov} from a hard spheres model.}
\begin{equation}
\label{f60}
P=\frac{2\pi a_s\hbar^2}{m^3}\rho^{2},\qquad c_s^2=\frac{4\pi
a_s\hbar^2}{m^3}\rho.
\end{equation}
This is a polytropic equation of state of the form of Eq. (\ref{f28}) with $n=1$, $\gamma=2$ and  $K=g/2={2\pi a_s\hbar^2}/{m^3}$. It is quadratic. The internal energy is given by
\begin{equation}
\label{f61}
U_2=\frac{2\pi a_s\hbar^2}{m^3}\int \rho^2\, d{\bf r}.
\end{equation}
The free energy and the average energy
are
\begin{eqnarray}
\label{f61b}
F_2=\Theta_c+\Theta_Q+W+W_{\rm ext}+U_2,
\end{eqnarray}
\begin{eqnarray}
\label{f61c}
N\langle E\rangle=\Theta_c+\Theta_Q+2W+W_{\rm ext}+2 U_2.
\end{eqnarray}
They satisfy the relation  $F_2=N\langle E\rangle-W-U_2$. The free energy
can be
written as
\begin{eqnarray}
\label{f62}
F_{2}=E_*-K S_{2},
\end{eqnarray}
where $K$ is a generalized temperature and
\begin{eqnarray}
\label{f63}
S_{2}=-\int \rho^{2}\, d{\bf r}
\end{eqnarray}
is the Tsallis entropy of index $\gamma=2$. We shall call it the quadratic
entropy. This quadratic functional is similar to the enstrophy in
two-dimensional
turbulence \cite{frish}.

\subsubsection{Hydrodynamic representation}

The generalized GP equation (\ref{f58}) is equivalent to the damped
quantum  Euler equations
\begin{equation}
\label{f64}
\frac{\partial\rho}{\partial t}+\nabla\cdot (\rho {\bf u})=0,
\end{equation}
\begin{eqnarray}
\label{f65}
\frac{\partial {\bf u}}{\partial t}+({\bf u}\cdot \nabla){\bf u}=-\frac{4\pi a_s\hbar^2}{m^3}\nabla\rho
-\nabla\Phi-\omega_0^2{\bf r}-\frac{1}{m}\nabla Q-\xi{\bf u}.
\end{eqnarray}
In the strong friction limit $\xi\rightarrow +\infty$, we obtain the
quantum Smoluchowski equation
\begin{eqnarray}
\label{f66}
\xi\frac{\partial\rho}{\partial t}=\nabla\cdot\biggl (\frac{2\pi a_s\hbar^2}{m^3}\nabla\rho^2
+\rho\nabla\Phi+\rho \omega_0^2{\bf r}+\frac{\rho}{m}\nabla Q\biggr ).
\end{eqnarray}
We note that it corresponds to an anomalous classical diffusion.

\subsubsection{Hydrostatic equilibrium}
\label{sec_he4}

The condition of hydrostatic equilibrium writes
\begin{eqnarray}
\label{f67}
\frac{2\pi a_s\hbar^2}{m^3}\nabla\rho^2+\rho\nabla\Phi
+\rho\omega_0^2{\bf r}+\frac{\rho}{m}\nabla Q={\bf 0}.
\end{eqnarray}
Combined with the Poisson equation (\ref{mad14}), we obtain
\begin{equation}
\label{f68}
-\frac{4\pi a_s\hbar^2}{m^3}\Delta\rho+\frac{\hbar^2}{2m^2}\Delta \left (\frac{\Delta\sqrt{\rho}}{\sqrt{\rho}}\right )=4\pi G\rho+3\omega_0^2.
\end{equation}
In the TF approximation ($Q=0$), the foregoing equation reduces to
\begin{equation}
\label{f69}
-\frac{4\pi a_s\hbar^2}{m^3}\Delta\rho=4\pi G\rho+3\omega_0^2.
\end{equation}

\subsubsection{The equilibrium state}
\label{sec_es5}

The minimization of the quadratic free energy at fixed mass (see Sec.
\ref{sec_efff}) leads to the equation
\begin{eqnarray}
\label{f73}
Q+m\Phi+\frac{1}{2}m\omega_0^2r^2+
\frac{4\pi a_s\hbar^2}{m^2}\rho=\mu.
\end{eqnarray}
This equation is equivalent to the condition of hydrostatic equilibrium. It can be rewritten as
\begin{eqnarray}
\label{f75}
\rho=\frac{m^2}{4\pi a_s\hbar^2}\left (\mu-m\Phi-Q-\frac{1}{2}m\omega_0^2r^2\right ),
\end{eqnarray}
which can be interpreted as a generalized Tsallis distribution of index
$\gamma=2$  
including the contribution of the quantum potential. In the TF approximation
($Q=0$), we recover the Tsallis distribution of index $\gamma=2$ which is just
an affine relationship.

\subsubsection{The virial theorem}
\label{sec_es6}

The scalar virial theorem writes
\begin{equation}
\label{f76}
\frac{1}{2}\ddot I+\frac{1}{2}\xi\dot I+\omega_0^2 I=2(\Theta_c+\Theta_Q)+3U_2+W.
\end{equation}
In the strong friction limit, we get
\begin{equation}
\label{f77}
\frac{1}{2}\xi\dot I+\omega_0^2 I=2\Theta_Q+3U_2+W.
\end{equation}
At equilibrium, the virial theorem, the free energy and the eigenenergy reduce
to
\begin{equation}
\label{f78}
2\Theta_Q+3U_2+W-\omega_0^2 I=0,
\end{equation}
\begin{eqnarray}
\label{f79}
F_2=\Theta_Q+W+\frac{1}{2}\omega_0^2 I+U_2,
\end{eqnarray}
\begin{equation}
\label{f80}
NE=\Theta_Q+2U_2+2W+\frac{1}{2}\omega_0^2 I.
\end{equation}

\subsection{Logotropic equation of state: logarithmic entropy}
\label{sec_tsl}

\subsubsection{Gross-Pitaevskii equation}

The logotropic equation of state\footnote{The logotropic equation of state was
introduced  in \cite{pud} in astrophysics. It was further
discussed in \cite{logo} in the context of generalized
thermodynamics and NFP equations. More recently, it was used in
\cite{delog,delogplb} in cosmology. It can be
viewed as a polytropic equation of
state of the form $P=K(\rho^{\gamma}-1)$ with
$\gamma\rightarrow 0$ and $K\rightarrow \infty$ in such a way
that $A=\gamma K$ is finite \cite{logo}. }
\begin{equation}
\label{f85}
P=A\ln\rho,\qquad c_s^2=\frac{A}{\rho},
\end{equation}
is associated with an effective potential of the form
\begin{equation}
\label{f86}
h(\rho)=-\frac{A}{\rho},\qquad V(\rho)=-A\ln\rho-A.
\end{equation}
The corresponding generalized GP equation is
\begin{eqnarray}
\label{f87}
i\hbar \frac{\partial\psi}{\partial t}=-\frac{\hbar^2}{2m}\Delta\psi+m\Phi\psi+\frac{1}{2}m\omega_0^2r^2\psi-A m \frac{1}{|\psi|^{2}}\psi
-i\frac{\hbar}{2}\xi\left\lbrack \ln\left (\frac{\psi}{\psi^*}\right )-\left\langle \ln\left (\frac{\psi}{\psi^*}\right )\right\rangle\right\rbrack\psi.
\end{eqnarray}
It has a hyperbolic nonlinearity. The
internal energy is given by
\begin{equation}
\label{f88}
U_{L}=-A\int \ln\rho\, d{\bf r}.
\end{equation}
We have omitted a constant term
proportional to the volume. The free energy and
the average energy
are
\begin{eqnarray}
\label{f88b}
F_L=\Theta_c+\Theta_Q+W+W_{\rm ext}+U_L,
\end{eqnarray}
\begin{eqnarray}
\label{f88c}
N\langle E\rangle=\Theta_c+\Theta_Q+2W+W_{\rm ext}.
\end{eqnarray}
They satisfy the relation  $F_L=N\langle E\rangle-W+U_L$. The free energy
can be written as
\begin{eqnarray}
\label{f89}
F_{L}=E_*-A S_{L},
\end{eqnarray}
where $A$ is the logotropic temperature and
\begin{eqnarray}
\label{f90}
S_{L}=\int \ln\rho\, d{\bf r}
\end{eqnarray}
is the logarithmic entropy \cite{logo,delog,delogplb}.

{\it Remark:} In order to have a dimensionless quantity in the logarithm, we can
replace Eqs. (\ref{f85}), (\ref{f88}) and (\ref{f90}) by
\begin{equation}
\label{f91}
P=A\ln\left (\frac{\rho}{\rho_0}\right ),\qquad U_{L}=-A\int \ln\left (\frac{\rho}{\rho_0}\right )\, d{\bf r},\qquad S_{L}=\int \ln\left (\frac{\rho}{\rho_0}\right )\, d{\bf r},
\end{equation}
where $\rho_0$ is a reference density.

\subsubsection{Hydrodynamic representation}

The generalized GP equation (\ref{f87}) is equivalent to the damped
quantum logotropic Euler equations
\begin{equation}
\label{f92}
\frac{\partial\rho}{\partial t}+\nabla\cdot (\rho {\bf u})=0,
\end{equation}
\begin{eqnarray}
\label{f93}
\frac{\partial {\bf u}}{\partial t}+({\bf u}\cdot \nabla){\bf u}=A\nabla\left (\frac{1}{\rho}\right )
-\nabla\Phi-\omega_0^2{\bf r}-\frac{1}{m}\nabla Q-\xi{\bf u}.
\end{eqnarray}
In the strong friction limit $\xi\rightarrow +\infty$, we obtain the
quantum logotropic Smoluchowski equation
\begin{eqnarray}
\label{f94}
\xi\frac{\partial\rho}{\partial t}=\nabla\cdot\biggl (A\nabla\ln\rho
+\rho\nabla\Phi+\rho \omega_0^2{\bf r}+\frac{\rho}{m}\nabla Q\biggr ).
\end{eqnarray}

\subsubsection{Hydrostatic equilibrium}
\label{sec_he5}

The condition of hydrostatic equilibrium writes
\begin{eqnarray}
\label{f95}
A\nabla\ln\rho+\rho\nabla\Phi
+\rho\omega_0^2{\bf r}+\frac{\rho}{m}\nabla Q={\bf 0}.
\end{eqnarray}
Combined with the Poisson equation (\ref{mad14}), we obtain
\begin{equation}
\label{f96}
{A}\Delta\left (\frac{1}{\rho}\right )+\frac{\hbar^2}{2m^2}\Delta \left (\frac{\Delta\sqrt{\rho}}{\sqrt{\rho}}\right )=S_d G\rho+d\omega_0^2.
\end{equation}
In the TF approximation ($Q=0$), the foregoing equation reduces to
\begin{equation}
\label{f97}
{A}\Delta\left (\frac{1}{\rho}\right )=S_d G\rho+d\omega_0^2.
\end{equation}

\subsubsection{The equilibrium state}
\label{sec_es7}

The minimization of the logotropic free energy at fixed mass (see Sec.
\ref{sec_efff}) leads to the equation
\begin{eqnarray}
\label{f98}
Q+m\Phi+\frac{1}{2}m\omega_0^2r^2-
\frac{Am}{\rho}=\mu.
\end{eqnarray}
This equation is equivalent to the condition of hydrostatic equilibrium.  It can be rewritten as
\begin{eqnarray}
\label{f99}
\rho=\frac{Am}{m\Phi+Q+\frac{1}{2}m\omega_0^2r^2-\mu},
\end{eqnarray}
which can be interpreted as a generalized logotropic distribution including 
the contribution of the quantum potential. In the TF approximation ($Q=0$), we
recover the logotropic distribution \cite{pud,logo,delog,delogplb}. For
$\Phi=0$, it
reduces to
the  Lorentzian distribution.

\subsubsection{The virial theorem}
\label{sec_es8}

The scalar virial theorem writes
\begin{equation}
\label{f100}
\frac{1}{2}\ddot I+\frac{1}{2}\xi\dot I+\omega_0^2 I=2(\Theta_c+\Theta_Q)-d U_{L}+W_{ii},
\end{equation}
were we have used the identity $\int P\, d{\bf r}=-U_{L}$
valid for a logotropic equation of state.
In the strong friction limit, we get
\begin{equation}
\label{f102}
\frac{1}{2}\xi\dot I+\omega_0^2 I=2\Theta_Q-d U_{L}+W_{ii}.
\end{equation}
At equilibrium, the virial theorem, the free energy and the eigenenergy reduce
to
\begin{equation}
\label{f103}
2\Theta_Q-d U_{L}+W_{ii}-\omega_0^2I=0,
\end{equation}
\begin{eqnarray}
\label{f104}
F_{L}=\Theta_Q+U_{L}+W+\frac{1}{2}\omega_0^2 I,
\end{eqnarray}
\begin{equation}
\label{f106}
NE=\Theta_Q+2W+\frac{1}{2}\omega_0^2 I.
\end{equation}

\subsection{An improved form of Tsallis entropy}

For a given equation of state $P(\rho)$, the effective potential $h(\rho)$ is
defined up to an additive
constant $C_1$ and the potential $V(\rho)$ is defined up to a function of the
form
$C_1\rho+C_2$. In the previous sections, we have determined the constants $C_1$
and $C_2$ in order to obtain the simplest expressions of $h$ and $V$. However,
the expressions of $h$ and $V$ that we have given in Sec. \ref{sec_tsa}
(polytropes) do not exactly return the results of Sec. \ref{sec_bol} (isothermal
distributions) in the limit $\gamma\rightarrow 1$ nor the results of Sec.
\ref{sec_tsl} (logotropes) in the limit $\gamma\rightarrow 0$, $K\rightarrow
+\infty$ and $A=K\gamma$ finite because the constants $C_1$ and $C_2$ have not
been calibrated for that purpose. In order to obtain unified results, we can
define\footnote{We have introduced a reference density $\rho_0$ in the equations
in order to have dimensionless quantities in the powers and in the logarithms,
but we can take $\rho_0=1$ to make the connection with the results obtained in
the
previous sections.}
\begin{equation}
\label{imp1}
h(\rho)=\frac{K\gamma}{\gamma-1}\left\lbrack\left (\frac{\rho}{\rho_0}\right )^{\gamma-1}-1\right\rbrack,\qquad
V(\rho)=\frac{K\rho_0}{\gamma-1} \left\lbrack\left (\frac{\rho}{\rho_0}\right
)^{\gamma}-\gamma \frac{\rho}{\rho_0}+\gamma-1\right\rbrack,
\end{equation}
\begin{equation}
\label{imp2}
U_{\gamma}=\frac{K\rho_0}{\gamma-1} \int\left\lbrack\left (\frac{\rho}{\rho_0}\right )^{\gamma}-\gamma \frac{\rho}{\rho_0}+\gamma-1\right\rbrack\, d{\bf r},\qquad
S_{\gamma}=-\frac{\rho_0}{\gamma-1} \int\left\lbrack\left (\frac{\rho}{\rho_0}\right )^{\gamma}-\gamma \frac{\rho}{\rho_0}+\gamma-1\right\rbrack\, d{\bf r}.
\end{equation}
For $\gamma\rightarrow 1$ and $K=k_B T/m$, they reduce to
\begin{equation}
\label{imp3}
h(\rho)=\frac{k_B T}{m}\ln\left (\frac{\rho}{\rho_0}\right ),\qquad
V(\rho)=\frac{k_B T}{m}\rho\left\lbrack \ln\left (\frac{\rho}{\rho_0}\right
)-1\right \rbrack+\frac{k_B T}{m}\rho_0,
\end{equation}
\begin{equation}
\label{imp4}
U=\frac{k_B T}{m}\int \left\lbrace \rho\left\lbrack \ln\left (\frac{\rho}{\rho_0}\right )-1\right \rbrack+\rho_0\right\rbrace\, d{\bf r},\qquad
S=-\int \left\lbrace \rho\left\lbrack \ln\left (\frac{\rho}{\rho_0}\right )-1\right \rbrack+\rho_0\right\rbrace\, d{\bf r}.
\end{equation}
For $\gamma\rightarrow 0$ and $K\rightarrow +\infty$ with $A=K\gamma$ finite,
they
reduce to
\begin{equation}
\label{imp5}
h(\rho)=-A\left (\frac{\rho_0}{\rho}-1\right ),\qquad
V(\rho)=-A\rho_0\left\lbrack\ln\left (\frac{\rho}{\rho_0}\right
)-\frac{\rho}{\rho_0}+1\right \rbrack,
\end{equation}
\begin{equation}
\label{imp6}
U=-A\rho_0\int \left\lbrack\ln\left (\frac{\rho}{\rho_0}\right )-\frac{\rho}{\rho_0}+1\right \rbrack\, d{\bf r},\qquad
S\sim \gamma\rho_0\int \left\lbrack\ln\left (\frac{\rho}{\rho_0}\right )-\frac{\rho}{\rho_0}+1\right \rbrack\, d{\bf r}.
\end{equation}
We can interpret the entropy $S_{\gamma}$ given by Eq. (\ref{imp2}) as an
improved form of Tsallis entropy that unifies the entropies associated
with isothermal, polytropic and logotropic equations of
state.

\section{A generalized BEC model of dark matter halos}
\label{sec_standmod}

\subsection{A generalized Gross-Pitaevskii equation}

We propose to give a special emphasis to the following GPP equations
\begin{eqnarray}
\label{mi1}
i\hbar \frac{\partial\psi}{\partial t}=-\frac{\hbar^2}{2m}\Delta\psi+m\Phi\psi+\frac{1}{2}m\omega_0^2r^2\psi
+\frac{4\pi a_s\hbar^2}{m^2}|\psi|^{2}\psi+2k_B T\ln|\psi|\psi
-i\frac{\hbar}{2}\xi\left\lbrack \ln\left (\frac{\psi}{\psi^*}\right )-\left\langle \ln\left (\frac{\psi}{\psi^*}\right )\right\rangle\right\rbrack\psi,
\end{eqnarray}
\begin{equation}
\label{mi2}
\Delta\Phi=4\pi G |\psi|^2,
\end{equation}
that could provide a relevant BEC model of dark matter halos. This model
takes into account the Heisenberg uncertainty principle (quantum potential),
the self-gravity of the system, an external (harmonic) potential,
the self-interaction of the bosons (scattering), an effective  temperature, and
a source of dissipation. The effective potential entering into the GP equation
(\ref{mi1}) is
given by
\begin{equation}
\label{mi3}
h(\rho)=\frac{k_B T}{m}\ln\rho+\frac{4\pi a_s\hbar^2}{m^3}\rho.
\end{equation}
It leads to an equation of state
\begin{equation}
\label{mi4}
P=\rho \frac{k_B T}{m}+\frac{2\pi a_s\hbar^2}{m^3}\rho^{2}
\end{equation}
which has a linear part and a quadratic part. The linear part corresponds to
the effective temperature and the quadratic part corresponds to the
self-interaction of the bosons. This is a composite model with a core-halo
structure  (see Appendix \ref{sec_com}). The quadratic equation of state
dominates in the core where the density is high and the isothermal equation of
state dominates in the halo where the density is low. This leads to dark matter
halos with a solitonic/BEC core surrounded by an isothermal envelope (see
Sec. \ref{sec_gcc}). The internal energy is given by
\begin{equation}
\label{mi5}
U=U_B+U_2,
\end{equation}
where $U_B$ is given by Eq. (\ref{f7}) and $U_2$ is given by Eq.
(\ref{f61}). The
free energy and the average energy are
\begin{equation}
\label{mi5b}
F=\Theta_c+\Theta_Q+W+W_{\rm ext}+U_B+U_2,
\end{equation}
\begin{equation}
\label{mi5c}
N\langle E\rangle=\Theta_c+\Theta_Q+2W+W_{\rm ext}+U_B+Nk_B T+2 U_2.
\end{equation}
They satisfy the relation  $F=N\langle E\rangle-W-Nk_B T-U_2$.

\subsection{Mixed entropy: Boltzmann and Tsallis}
\label{sec_mix}

The free energy can be written as
\begin{equation}
\label{mi6}
F=E_*-TS_B-KS_2,
\end{equation}
where $S_B$ is the Boltzmann entropy (\ref{f9}) and $S_2$ in the Tsallis entropy
(\ref{f63}) of index $\gamma=2$ (quadratic entropy). We can also write
\begin{eqnarray}
\label{mi7}
F=E_*-K S_{\rm mix}
\end{eqnarray}
where
\begin{eqnarray}
\label{mi8}
S_{\rm mix}=-\int \rho^{2}\, d{\bf r}-\lambda\int \rho(\ln\rho-1)\, d{\bf r}
\end{eqnarray}
is a mixed entropy combining the Tsallis and Boltzmann entropies,
and
\begin{eqnarray}
\label{mi8b}
\lambda=\frac{k_B T}{Km}
\end{eqnarray}
is the ratio between the ordinary temperature $T$ and the polytropic temperature
$K=2\pi a_s\hbar^2/m^3$. Alternatively, we can choose to include the internal
energy $U_2$ in an ``augmented'' energy 
\begin{eqnarray}
\label{mi10}
E_0\equiv E_*+U_2=\Theta_c+\Theta_Q+W+W_{\rm ext}+U_2
\end{eqnarray}
and write the free energy as
\begin{eqnarray}
\label{mi9}
F=E_0-T S_B,
\end{eqnarray}
so that it only involves the Boltzmann entropy and the ordinary
temperature.

{\it Remark:} Instead of considering a polytrope of index $\gamma=2$, we could
consider a polytrope of arbitrary index $\gamma$. In that case, the mixed
entropy takes the form
\begin{eqnarray}
\label{mi10b}
S_{\rm mix}=-\frac{1}{\gamma-1}\int \rho^{\gamma}\, d{\bf r}-\lambda\int \rho(\ln\rho-1)\, d{\bf r}.
\end{eqnarray}

\subsection{Hydrodynamic representation}

The generalized GPP equations (\ref{mi1}) and (\ref{mi2}) are equivalent
to the
damped quantum isothermal-polytropic EP equations
\begin{equation}
\label{mi11}
\frac{\partial\rho}{\partial t}+\nabla\cdot (\rho {\bf u})=0,
\end{equation}
\begin{eqnarray}
\label{mi12}
\frac{\partial {\bf u}}{\partial t}+({\bf u}\cdot \nabla){\bf u}=-\frac{k_B T}{m}\nabla \ln\rho
-\frac{4\pi a_s\hbar^2}{m^3}\nabla\rho
-\nabla\Phi-\omega_0^2{\bf r}-\frac{1}{m}\nabla Q-\xi{\bf u},
\end{eqnarray}
\begin{equation}
\label{mi13}
\Delta\Phi=4\pi G\rho.
\end{equation}
In the strong friction limit $\xi\rightarrow +\infty$, we obtain the quantum
isothermal-polytropic SP
equations
\begin{eqnarray}
\label{mi14}
\xi\frac{\partial\rho}{\partial t}=\nabla\cdot\biggl (\frac{k_B T}{m}\nabla\rho
+\frac{2\pi a_s\hbar^2}{m^3}\nabla\rho^2
+\rho\nabla\Phi+\rho \omega_0^2{\bf r}+\frac{\rho}{m}\nabla Q\biggr ),
\end{eqnarray}
\begin{equation}
\label{mi15}
\Delta\Phi=4\pi G \rho.
\end{equation}

\subsection{Hydrostatic equilibrium}
\label{sec_he6}

The condition of hydrostatic equilibrium writes
\begin{eqnarray}
\label{mi16}
\frac{k_B T}{m}\nabla\rho+\frac{2\pi a_s\hbar^2}{m^3}\nabla\rho^2+\rho\nabla\Phi
+\rho\omega_0^2{\bf r}+\frac{\rho}{m}\nabla Q={\bf 0}.
\end{eqnarray}
Combined with the Poisson equation (\ref{mad14}), we obtain
\begin{equation}
\label{mi17}
-\frac{k_B T}{m}\Delta\ln\rho-\frac{4\pi a_s\hbar^2}{m^3}\Delta\rho+\frac{\hbar^2}{2m^2}\Delta \left (\frac{\Delta\sqrt{\rho}}{\sqrt{\rho}}\right )=4\pi G\rho+3\omega_0^2.
\end{equation}
In the TF approximation ($Q=0$), we get
\begin{equation}
\label{mi19}
-\frac{k_B T}{m}\Delta\ln\rho-\frac{4\pi a_s\hbar^2}{m^3}\Delta\rho=4\pi G\rho+3\omega_0^2.
\end{equation}

\subsection{The equilibrium state}
\label{sec_es9}

The minimization of the mixed free energy at fixed mass (see Sec.
\ref{sec_efff}) leads to the equation
\begin{eqnarray}
\label{mi20}
Q+m\Phi+\frac{1}{2}m\omega_0^2r^2+k_B T\ln\rho+\frac{4\pi a_s\hbar^2}{m^2}\rho=\mu.
\end{eqnarray}
This equation is equivalent to the  condition of hydrostatic equilibrium. It can be rewritten as
\begin{eqnarray}
\label{mi21}
\rho=\frac{m^2 k_B T}{4\pi |a_s|\hbar^2}W\left\lbrack \frac{4\pi |a_s|\hbar^2}{m^2 k_B T} e^{-\beta(m\Phi+Q+\frac{1}{2}m\omega_0^2r^2-\mu)}\right\rbrack.
\end{eqnarray}
In the case of repulsive interactions ($a_s>0$), $W(z)$ is the Lambert function defined implicitly by the equation $We^W=z$. Therefore, Eq. (\ref{mi21}) can be interpreted as a generalized Lambert distribution including the contribution of the quantum potential. In the TF approximation ($Q=0$) we recover the Lambert distribution. In the case of attractive interactions ($a_s<0$), $W(z)$ is a new function defined implicitly by the equation $We^{-W}=z$. In that case, the density (\ref{mi21}) is defined only for
\begin{eqnarray}
\label{mi21b}
\frac{4\pi |a_s|\hbar^2}{m^2 k_B T} e^{-\beta(m\Phi+Q+\frac{1}{2}m\omega_0^2r^2-\mu)}\le \frac{1}{e},
\end{eqnarray}
and Eq. (\ref{mi20}) exhibits two different solutions. This interesting feature
will be studied in a specific paper.\footnote{The same discussion 
applies to Sec. 7.2. of \cite{entropy}.}

{\it Remark:} It is actually possible to generalize these results to the case of
a polytrope of arbitrary index $\gamma$ instead of $\gamma=2$. In that case, we
obtain
\begin{eqnarray}
\label{mi21c}
\rho=\left\lbrace \frac{k_B T}{|K|\gamma m}W\left\lbrack \frac{|K|\gamma m}{k_B T} e^{-\beta(\gamma-1)(m\Phi+Q+\frac{1}{2}m\omega_0^2r^2-\mu)}\right\rbrack\right\rbrace^{\frac{1}{\gamma-1}}.
\end{eqnarray}
When $K>0$, there is only one solution. When $K<0$, there
are
two solutions when the term in bracket is less than $1/e$ and no solution
otherwise.

\subsection{The virial theorem}
\label{sec_es10}

The scalar virial theorem writes
\begin{equation}
\label{mi22}
\frac{1}{2}\ddot I+\frac{1}{2}\xi\dot I+\omega_0^2 I=2(\Theta_c+\Theta_Q)+3N k_B T+3U_2+W.
\end{equation}
In the strong friction limit, we get
\begin{equation}
\label{mi23}
\frac{1}{2}\xi\dot I+\omega_0^2 I=2\Theta_Q+3N k_B T+3U_2+W.
\end{equation}
At equilibrium, the virial theorem, the free energy and the eigenenergy reduce
to
\begin{equation}
\label{mi24}
2\Theta_Q+3N k_B T+3U_2+W-\omega_0^2 I=0,
\end{equation}
\begin{eqnarray}
\label{mi25}
F=\Theta_Q+W+\frac{1}{2}\omega_0^2 I+U_B+U_2,
\end{eqnarray}
\begin{equation}
\label{mi26}
NE=\Theta_Q+U_B+Nk_B T+2U_2+2W+\frac{1}{2}\omega_0^2 I.
\end{equation}

\section{Discussion}
\label{sec_pi}

\subsection{Comparison with other works}

The generalized BEC model (\ref{mi1}) and (\ref{mi2}) includes two new terms
with respect to the standard BEC model (\ref{mfgp13}) and (\ref{mfgp14a}): a
temperature term $T$ and a friction term $\xi$. The standard BEC model is
recovered when $T=\xi=0$. In this section, we connect the generalized GP
equation (\ref{mi1}) to nonlinear Schr\"odinger equations that have been
introduced in the past from different arguments. 

A wave equation with a logarithmic nonlinearity $-b\ln|\psi|$ similar to the
one present in Eq. (\ref{mi1}), namely
\begin{equation}
\label{bial}
i\hbar\frac{\partial\psi}{\partial t}=-\frac{\hbar^2}{2m}\Delta\psi+m\Phi\psi-2b\ln|\psi|\, \psi,
\end{equation}
has been introduced by Bialynicki-Birula \& Mycielski \cite{bm} 
as a possible generalization of the Schr\"odinger equation in quantum mechanics.
 In that case, the logarithmic nonlinearity has a fundamental origin and the
coefficient $b$ is interpreted as a fundamental constant of physics. In the
interpretation of Bialynicki-Birula \& Mycielski \cite{bm}, the Schr\"odinger
equation is an approximation of this nonlinear wave equation. The coefficient
$b$ can be positive or negative. When it is positive, it plays the role of an
attractive interaction that  can balance the repulsion due to the quantum
potential and lead to a stationary solution of the wave
equation with a  Gaussian profile and a finite
width called a gausson. The radius of the gausson is $R=\hbar/(2mb)^{1/2}$. The
logarithmic nonlinearity may be a way to prevent the spreading of the wave
packet in the Schr\"odinger equation. Of course, the constant $b$ must be
sufficiently small in order to satisfy the constraints set by laboratory
experiments. Bialynicki-Birula and Mycielski \cite{bm} obtained $b<4\times
10^{-10}{\rm eV}$ which implies a bound to the electron soliton spatial width of
$10\, \mu{\rm m}$.  Shull {\it et al.} \cite{shull} obtained  $b<3.4\times
10^{-13} {\rm eV}$. Finally, an upper limit $b<3.3\times 10^{-15}\, {\rm eV}$
was obtained by G\"ahler {\it et al.} \cite{gkz} from precise measurements of
Fresnel diffraction with slow neutrons. This implies a bound to the electron
soliton spatial width of $3\, {\rm mm}$. For $m\rightarrow +\infty$, the density
probability becomes a delta-function and the particle is localized. This means
that a particle with a sufficiently big mass has a classical motion.
In our approach, the logarithmic potential  $k_B T \ln|\psi|$ in Eq. (\ref{mi1})
is obtained by looking for the generalized Schr\"odinger equation that leads,
through the Madelung transformation, to a quantum Euler equation including
a pressure term with an
isothermal equation of state. Our approach gives another interpretation to the
logarithmic Schr\"odinger equation (\ref{bial}) from a hydrodynamic
representation of the wave equation (see Sec. \ref{sec_mad})
associated with an
effective thermodynamic formalism (see Sec. \ref{sec_thermo}).\footnote{The
motivation of
Bialynicki-Birula  \& Mycielski \cite{bm} to introduce a logarithmic
nonlinearity in the Schr\"odinger equation is that this term still satisfies the
additivity property for noninteracting subsystems while solving the spreading of
the wave packet problem. As we have seen in our effective thermodynamic
formalism, this logarithmic nonlinearity is associated with the Boltzmann
entropy.
We note that a power-law nonlinearity  $-|\psi|^{2(\gamma-1)}$, which
corresponds to a polytropic equation of state, also solves the spreading of the
wave packet problem but does not satisfy the additivity property for
noninteracting subsystems. It is associated with the Tsallis entropy that has
been introduced precisely in order to deal with non-extensive and non-additive
systems \cite{tsallisbook}. Therefore, the Tsallis entropy may find applications
in
relation to nonlinear Schr\"odinger equations with a power-law nonlinearity.}
Comparing Eqs.
(\ref{mi1}) and (\ref{bial}) we find that $b=-k_B
T$ so that a positive coefficient $b$ corresponds to a negative effective
temperature.

A Schr\"odinger equation with a damping term similar to Eq. (\ref{mi1}), namely
\begin{equation}
\label{koss}
i\hbar\frac{\partial\psi}{\partial t}=-\frac{\hbar^2}{2m}\Delta\psi+m\Phi\psi-i\frac{\hbar}{2}\xi\left\lbrack \ln\left (\frac{\psi}{\psi^*}\right )-\left\langle \ln\left (\frac{\psi}{\psi^*}\right )\right\rangle\right\rbrack \psi,
\end{equation}
has been introduced by Kostin \cite{kostin}. He derived it from the
Heisenberg-Langevin equation describing a quantum Brownian particle interacting
with a thermal bath environment. In our approach, the damping term is
obtained by looking for the Schr\"odinger equation that leads, through the
Madelung transformation, to a quantum Euler equation with a linear friction
force proportional and opposite to the velocity. This gives another
interpretation to the generalized  Schr\"odinger equation (\ref{koss})
from a
hydrodynamic representation of the wave
equation  (see Sec. \ref{sec_mad}) associated with  an effective thermodynamic
formalism (see Sec. \ref{sec_thermo}).

In Ref. \cite{papernottale}, we have derived a generalized
Schr\"odinger equation similar to Eq. (\ref{mi1}) that unifies the
logarithmic Schr\"odinger equation (\ref{bial}) and 
the damped Schr\"odinger equation (\ref{koss}). We have shown that 
the temperature and friction terms in the generalized
Schr\"odinger equation
(\ref{mi1}) have a common origin and that they can be obtained from a unified
description based on Nottale's theory of scale relativity \cite{nottale}. They
satisfy a sort
of fluctuation-dissipation theorem. In this approach, one can show
that $\xi\sim 1$ while $b=-k_B T\sim \hbar$. Therefore $b=-k_B T\sim \hbar$ has
its origin in quantum mechanics (it vanishes in the classical limit
$\hbar\rightarrow 0$) while $\xi\sim 1$ survives in the classical limit. The
fact that $b=-k_B T$ is proportional to the Planck constant may explain its
small value and why it is not detectable in earth experiments.  We also note,
parenthetically, that the scaling  $T\sim\hbar$ is similar to the one arising in
the expression of the Hawking temperature $k_B T=\hbar c/4\pi R=\hbar c^3/8\pi
GM$ of the black holes. This may suggest
a relation to quantum gravity.

\subsection{Physical interpretation: coarse-grained parametrization of
gravitational
cooling}
\label{sec_gcc}

The fluid equations (\ref{mi11})-(\ref{mi13}) associated with the dissipative
GPP equations (\ref{mi1}) and (\ref{mi2}) generalize the
hydrodynamic
equations of the CDM model by accounting for a quantum force due to the
Heisenberg uncertainty principle, a pressure force due to the self-interaction
of the bosons, an external harmonic potential, a temperature, and a friction.
The hydrodynamic equations of the CDM model are recovered for
$\hbar=a_s=\omega_0=T=\xi=0$.

Basically, a self-gravitating BEC is described by
the conservative GPP equations (\ref{mfgp13})
and (\ref{mfgp14a}) at $T=0$ without friction. These equations are
extremely complex because they experience a process of gravitational
cooling in the same manner that the Vlasov-Poisson equations describing
collisionless self-gravitating systems experience a process of violent
relaxation \cite{lb}. This relaxation process (virialization) is
accompanied by damped oscillations of the system and by an ejection of
mass \cite{seidel94,gul0,gul}. We propose that the
generalized GPP equations (\ref{mi1})
and
(\ref{mi2}), or
the corresponding 
damped quantum  EP equations  (\ref{mi11})-(\ref{mi13}), provide an
effective coarse-grained model of dark matter halos experiencing gravitational
cooling in the same manner that the generalized Fokker-Planck
equations of \cite{csr,mnras,dubrovnik} provide a coarse-grained parametrization
of the Vlasov-Poisson equations.\footnote{We should write $\overline{\psi}$
instead of $\psi$ to distinguish the coarse-grained  wave function that is the
solution of Eqs. (\ref{mi1}) and (\ref{mi2}) from the fine-grained wave function
that is the solution of Eqs.  (\ref{mfgp13}) and (\ref{mfgp14a}) but, for
commodity, we shall denote them with the same symbol.} As we
have seen, the generalized GPP equations satisfy an $H$-theorem and relax
towards an equilibrium
state described by an equation of state $P=\rho k_B T/m+2\pi
a_s\hbar^2\rho^{2}/m^3$.
Therefore, the damping term heuristically explains {\it how} a system of
self-gravitating bosons rapidly reaches an equilibrium state by
dissipating some free energy. This
equilibrium state is characterized by an effective temperature $T$ [the one
that appears in Eq. (\ref{mi1})]
even if $T=0$ fundamentally [see Eqs. (\ref{mfgp13}) and  (\ref{mfgp14a})]. The
equilibrium structure of the system, determined
by Eq. (\ref{mi17}), results from the balance between the repulsive quantum
force, the repulsive pressure due to the scattering of the bosons (when
$a_s>0$), the repulsive pressure due to the effective temperature (when 
$T>0$), and the gravitational attraction.  For the sake of generality, we
have included in Eq. (\ref{mi1}) a harmonic potential that can mimic a confining
force due to tidal interactions ($\omega_0^2>0$) or a centrifugal force due to a
solid rotation ($\omega_0^2<0$). We note that the scattering length of the
bosons $a_s$ and the effective temperature $T$ may be positive or negative
leading to a rich diversity of behaviors.\footnote{In some cases, there
is no equilibrium state (see, e.g., \cite{prd1,prd2,bectcoll}). A complete study
of
these
different behaviors will be
given in a forthcoming paper (in preparation).} When $a_s>0$, $T>0$ and
$\omega_0^2>0$, the equilibrium
state of the system has a core-halo structure. It is made of a compact core
(BEC/soliton)
with an equation of state $P=2\pi a_s\hbar^2\rho^{2}/m^3$, which is a stable
stationary solution of the GPP equations (\ref{mfgp13}) and (\ref{mfgp14a}) at
$T=0$
(ground state), surrounded by
an isothermal atmosphere with an equation of state $P=\rho k_B T/m$
mimicking a halo of scalar radiation at
temperature $T$. The polytropic equation of state $P=2\pi
a_s\hbar^2\rho^{2}/m^3$ dominates in the core where the density is high and the
isothermal equation of state $P=\rho k_B T/m$ dominates in the halo where the
density is low. The density of a self-gravitating isothermal
halo decreases for $r\rightarrow +\infty$ as $\rho(r)\sim k_B T/2\pi G m r^{2}$
(corresponding to an accumulated mass $M(r)\sim 2k_B Tr/Gm$) and leads to flat
rotation curves $v_c^2(r)=GM(r)/r\rightarrow 2k_B T/m$ in agreement with the
observations. However, such a profile has an infinite mass \cite{bt}.
Furthermore, it
differs from the numerical NFW \cite{nfw} density profile and from the
observational Burkert \cite{burkert} density profile that both decrease as
$r^{-3}$ for
$r\rightarrow +\infty$. The difference of behavior could be due to incomplete
relaxation, tidal effects, or stochastic perturbations that can steepen the
logarithmic slope from $-2$
to $-3$ as shown in \cite{kingclassique,kingfermionic} in the context of the
King model.\footnote{In this respect, we note that a decay exponent in the
density profile $\rho\sim r^{-\alpha}$ equal to $\alpha=-3$ (NFW/Burkert) is the
closest exponent to $\alpha=-2$ (isothermal) that yields a halo with a
(marginally) {\it finite} mass.} In the
present
model, this confinement can be taken into account by the harmonic potential.
Another possibility would be to change the nonlinearity in the generalized
GPP equations (\ref{mi1}) and (\ref{mi2}), or equivalently the
equation of state in the damped quantum  EP equations (\ref{mi11})-(\ref{mi13}),
so as to yield a density profile decreasing as $r^{-3}$ at large
distances similarly to the NFW and Burkert profiles. 
However, an isothermal halo can account heuristically for 
an atmosphere of scalar radiation. Therefore, Eqs.
(\ref{mi1}) and (\ref{mi2}) can
provide a relevant parametrization of dark matter halos experiencing
gravitational
cooling. As mentioned in the Introduction, it is the 
atmosphere of scalar radiation that fixes the size of the halos. The presence
of the atmosphere solves the apparent paradox (see Appendix
F of \cite{kingfermionic}) that BEC halos
at $T=0$
should all have the same radius (in the self-interacting case) or that their
radius should decrease with their mass as $R\propto M^{-1}$  (in the
non-interacting case), in
contradiction with the observations that reveal that their radius increases
with their mass as $R\propto M^{1/2}$ corresponding to a constant surface
density (see
\cite{vega3,kingclassique,kingfermionic}). In
Appendix
\ref{sec_ce}, we propose an improved model in which the temperature $T(t)$
changes with time in order to exactly conserve the energy $E_0$ of the
(fine-grained) GPP equations (\ref{mfgp13}) and (\ref{mfgp14a}).

\subsection{Speculations: fundamental wave equations}
\label{sec_spec}

We would like to close this discussion by making speculations. We have
previously 
justified the generalized BEC model (\ref{mi1}) and (\ref{mi2})  as an effective
coarse-grained description of a dissipationless BEC at $T=0$ described by Eqs.
(\ref{mfgp13})
and (\ref{mfgp14a}) undergoing a
process of gravitational cooling. In that case, $\xi$ is an effective friction
which accounts for the relaxation process. On the other hand, $T$ is an
effective temperature which represents the temperature of the halo made of
scalar radiation. Alternatively, following Bialynicki-Birula \& Mycielski
\cite{bm}, we may argue that the
generalized GP equation (\ref{mi1}) is a fundamental equation of
physics from which the standard GP equation (\ref{mfgp13}) is an
approximation. The new terms in Eq. (\ref{mi1}) may arise from the
interaction of the system with an external medium, a sort of aether. In that
case, $\xi$ represents the friction with the aether and $T$ represents the
temperature of the aether. The constant  $b=-k_B T$ in Eq. (\ref{mi1}) may
naturally account for the nonzero temperature of the halos and for the flat
rotation curves of the galaxies.   In this interpretation, the flat rotation
curves of the galaxies would have a fundamental origin related to the nonzero
value of $b=-k_B T$ in the generalized GP equation (\ref{mi1}). This
interpretation can be correct only if $b$ is very small so that it is not
detectable in earth laboratory experiments and manifests itself only on
astrophysical or cosmological scales. This actually sets a constraint on
the mass $m$ of the bosons that may compose dark matter halos as we now show. 

An
isothermal halo leads to a flat rotation curve with an asymptotic  circular
velocity $v_c^2(r)=GM(r)/r$ given by
\begin{equation}
\label{pi2}
v_c^2\rightarrow \frac{2k_B T}{m}.
\end{equation}
For a typical galaxy such as the Medium Spiral, the 
circular velocity is of the order of $v_c=150\, {\rm km/s}$. In
Ref. \cite{kingfermionic}, we have estimated the mass of the bosons that may
compose dark matter
halos by assuming that the smallest observed halos such as Willman 1 are
completely condensed (ground state). For noninteracting bosons, we found
$m=2.57\times 10^{-20}\, {\rm eV}/c^2$. From Eq. (\ref{pi2}), we obtain
$T=3.73\times 10^{-23}\, {\rm K}$ and $k_B T=-b=3.22\times 10^{-27}\, {\rm eV}$.
This value of $b$ is below
the experimental bounds found in \cite{bm,shull,gkz} so that an ultra small
boson mass is acceptable.\footnote{We note that the results of
\cite{bm,shull,gkz} assume $b>0$ while, in the present case, $b<0$. Therefore,
our comparisons are only indicative.} For self-interacting bosons, we found a
maximum boson mass
$m=1.69\times 10^{-2}\, {\rm eV}/c^2$. From Eq. (\ref{pi2}), we obtain
$T=2.45\times 10^{-5}\, {\rm K}$ and $k_B T=-b=2.115\times 10^{-9}\, {\rm eV}$.
This value of $b$ is above the experimental bounds found in \cite{bm,shull,gkz}
so that a too large  boson mass must be rejected. Using the constraint $k_B
T=-b<3.3\times 10^{-15}\, {\rm eV}$ \cite{gkz}, we get $T<3.83\times 10^{-11}\,
{\rm K}$ leading to a boson mass $m<2.64\times 10^{-8}\, {\rm eV}/c^2$.

It would be fascinating to find that the flat rotation 
curves of dark matter halos arise from a fundamental nonlinear term in the
Schr\"odinger equation that cannot be detected in earth experiments but that
would have an effect at large (cosmological) scales. This is not completely
impossible if the
mass of the bosons is ultra small (as shown above) so that $T$ and $b$ are also
very small,
beyond experimental reach on earth. There is nevertheless is difficulty with
this scenario. Indeed, if the temperature of the halos is related to a
fundamental constant $b$ in the Schr\"odinger equation, i.e., if $T$ is a
fundamental constant representing the temperature of the aether, the halos
should all have the same temperature, or the same value of $v_c$, which is not
the case. Indeed, it appears that the temperature of the halos increases
linearly with their size, namely $T\propto R$ (see
\cite{vega3,kingclassique,kingfermionic}).\footnote{It is an
observational evidence that
the dark
matter halos have the {\it same} surface density $\Sigma_0\equiv
\rho_0r_h=141\, M_{\odot}/{\rm pc}^2$ \cite{donato}. This implies that their
mass and temperature scale with the radius as $M_h\sim \rho_0 r_h^3\propto
r_h^2$ and $k_B T\sim GM_hm/r_h\propto r_h$.} We do not know at that stage
how to escape this difficulty. This suggests that our interpretation of the
generalized BEC model (\ref{mi1}) and (\ref{mi2}) as an effective equation
describing gravitational cooling is more relevant than its interpretation
 in terms of a fundamental wave equation. In the first 
interpretation, $T$ and $\xi$ are effective coefficients determined by the
efficiency of gravitational cooling allowing them to change from halo to halo.
However, even in this case, we have to explain why the (effective) temperature
increases linearly with the size of the halos.

{\it Remark:} In Ref. \cite{delog,delogplb}, we have shown 
that many
properties of dark matter halos (in particular the universal value of their
surface density $\Sigma_0$) can be remarkably well explained by assuming that
their equation of state is logotropic (see Sec. \ref{sec_tsl}). As a result, we
are led to postulating that the fundamental nonlinear wave equation (from which
the Schr\"odinger equation would be an approximation) is the hyperbolic GP
equation
\begin{eqnarray}
\label{f87graal}
i\hbar \frac{\partial\psi}{\partial
t}=-\frac{\hbar^2}{2m}\Delta\psi+m\Phi\psi-A m
\frac{1}{|\psi|^{2}}\psi
\end{eqnarray}
instead of the logarithmic GP equation (\ref{bial}). In this equation, $A$
should be regarded as a new fundamental constant of
physics (superseeding the cosmological constant). In \cite{delog,delogplb}, we
have
shown that its value is $A=B\rho_{\Lambda}c^2=2.13\times 10^{-9}\, {\rm g}
\,{\rm m}^{-1}\,{\rm s}^{-2}$, where $\rho_{\Lambda}=6.72\times 10^{-24}\, {\rm
g} \,{\rm m}^{-3}$ is the cosmological density
and $B=3.53\times 10^{-3}$ is a dimensionless constant approximately equal to
$B\simeq 1/[123\ln(10)]$ where $123$ is the famous number appearing in the
ratio between the Planck density and the cosmological density:
$\rho_P/\rho_\Lambda\sim 10^{123}$.
We note that the energy scale associated with the last term in Eq.
(\ref{f87graal}) is of the order of $Am|\psi|^2\sim (\rho_\Lambda/\rho)mc^2$.
In view of the smallness of $\rho_\Lambda$, this term is completely negligible
for the typical values of the density $\rho\gg \rho_\Lambda$ considered in
laboratory experiments.
However, this term becomes important on cosmological scales where
$\rho\sim \rho_\Lambda$ and could explain the structure of the core of dark
matter halos. When applied to dark matter halos, the logotropic model
\cite{delog,delogplb}
yields a universal rotation
curve that coincides, up to the halo radius $r_h$, with the empirical
Burkert \cite{burkert}
profile that fits a lot of observational rotation curves.
However, the density of a self-gravitating logotropic halo
decreases for $r\rightarrow +\infty$ as $\rho(r)\sim (A/8\pi G)^{1/2}r^{-1}$
(corresponding to an accumulated mass $M(r)\sim (A\pi/2G)^{1/2}r^2$). Therefore,
this profile has an infinite mass \cite{logo}. As
explained previously,  the confinement of dark matter halos may be due to
complicated physical processes such as incomplete
relaxation, evaporation, stochastic forcing from the external environment etc.
As a result, the density profiles of the halos decrease at large distances as
$r^{-3}$ like the NFW \cite{nfw} and Burkert \cite{burkert} profiles
instead of $r^{-1}$ as predicted by the logotropic model \cite{delog,delogplb}.
This extra
confinement
could be taken into account in our model by introducing a  harmonic potential
in Eq. (\ref{f87graal}) or
by using a more complicated nonlinearity in the generalized GP equation, or a
more complicated equation of state in the corresponding EP equations,  that
yields a logotropic profile  in the core and  a profile in the halo decreasing
as $r^{-3}$ at large distances, 
like for the NFW  \cite{nfw} and Burkert profiles \cite{burkert}.

\section{The Gaussian ansatz}
\label{sec_ansatz}

\subsection{The generalized Gross-Pitaevskii-Poisson equations}
\label{sec_efg4}

In order to obtain exact results, the generalized GPP equations
(\ref{mfgp9}) and
(\ref{mfgp14}) must be solved numerically. However, we can obtain approximate
analytical results by making a Gaussian ansatz for the wave function.  In order
to be sufficiently general, we consider the  GPP equations
\begin{equation}
\label{a1}
i\hbar \frac{\partial\psi}{\partial t}=-\frac{\hbar^2}{2m}\Delta\psi+m\Phi\psi+\frac{1}{2}m\omega_0^2r^2\psi
+\frac{K\gamma m}{\gamma-1}|\psi|^{2(\gamma-1)}\psi+2k_B T\ln|\psi|\psi
-i\frac{\hbar}{2}\xi\left\lbrack \ln\left (\frac{\psi}{\psi^*}\right )-\left\langle \ln\left (\frac{\psi}{\psi^*}\right )\right\rangle\right\rbrack\psi,
\end{equation}
\begin{equation}
\label{a2}
\Delta\Phi=S_d G |\psi|^2,
\end{equation}
in a space of dimension $d$. They take into account the Heisenberg uncertainty
principle (quantum potential), the self-gravity of the system, an external
harmonic potential, a power-law self-interaction, an effective  temperature, and
a source of dissipation. All the relevant
equations associated with this model can be obtained from the results of Secs.
\ref{sec_bol} and
\ref{sec_tsa}  by adding the contribution of the isothermal and polytropic
equations of state as we did in Sec. \ref{sec_standmod} for the particular case
$\gamma=2$ (see the remark at the end of Sec. \ref{sec_mix} and Appendix
\ref{sec_com}). For example, the
equation of state writes
\begin{equation}
\label{a2w}
P=K\rho^{\gamma}+\rho\frac{k_B T}{m}.
\end{equation}
The free energy and the average energy are given by
\begin{equation}
\label{a2b}
F=\Theta_c+\Theta_Q+W+W_{\rm ext}+U_B+U,
\end{equation}
\begin{equation}
\label{a2cb}
N\langle E\rangle=\Theta_c+\Theta_Q+2W+W_{\rm ext}+U_B+Nk_B T+\gamma U.
\end{equation}
They satisfy $F=N\langle E\rangle-W-Nk_B T-(\gamma-1)U$. The
scalar
virial theorem takes the form
\begin{equation}
\label{a2c}
\frac{1}{2}\ddot I+\frac{1}{2}\xi\dot I+\omega_0^2 I=2(\Theta_c+\Theta_Q)+dN k_B
T+d(\gamma-1)U+W_{ii}.
\end{equation}
In the  strong friction limit, we get
\begin{equation}
\label{a2d}
\frac{1}{2}\xi\dot I+\omega_0^2 I=2\Theta_Q+dN k_B
T+d(\gamma-1)U+W_{ii}.
\end{equation}
At equilibrium, the virial theorem, the free energy and the
eigenenergy reduce to
\begin{equation}
\label{a2e}
2\Theta_Q+dN k_B T+d(\gamma-1)U+W_{ii}-\omega_0^2 I=0,
\end{equation}
\begin{equation}
\label{a2f}
F=\Theta_Q+W+W_{\rm ext}+U_B+U,
\end{equation}
\begin{equation}
\label{a2g}
N E=\Theta_Q+2W+W_{\rm ext}+U_B+Nk_B T+\gamma U.
\end{equation}

\subsection{The free energy}
\label{sec_efg2}

We shall calculate the free energy functional (\ref{a2b}) by making a Gaussian
ansatz for the wave function
\begin{eqnarray}
\label{a3}
\psi({\bf r},t)=\left \lbrack \frac{2M}{S_d \Gamma(d/2)R(t)^d}\right
\rbrack^{1/2} e^{-\frac{r^2}{2R(t)^2}}e^{imH(t)r^2/2\hbar}e^{iS_0(t)/\hbar},
\end{eqnarray}
where $R(t)$ measures the size of the system (wave packet). It corresponds
to the typical radius of the BEC. The wave function is normalized such that
$\int |\psi|^2\, d{\bf r}=M$. $\Gamma(x)$ is the Gamma function:
$\Gamma(3/2)=\sqrt{\pi}/2$ in $d=3$, $\Gamma(1)=1$ in $d=2$, and
$\Gamma(1/2)=\sqrt{\pi}$ in  $d=1$. Comparing Eq. (\ref{a3}) with Eq.
(\ref{mad1}), we find that the density and the action are given by
\begin{eqnarray}
\label{a4}
\rho({\bf r},t)=\frac{2M}{S_d
\Gamma(d/2)R(t)^d}e^{-\frac{r^2}{R(t)^2}}\qquad {\rm and}\qquad S({\bf
r},t)=\frac{1}{2} m H(t) r^2+S_0(t).
\end{eqnarray}
The velocity defined by Eq. (\ref{mad5}) is then given by
\begin{eqnarray}
\label{a3x}
{\bf u}({\bf r},t)=H(t){\bf r}.
\end{eqnarray}
It is proportional to the radial distance ${\bf r}$ with a proportionality
constant $H(t)$ depending only on time. It is shown in Appendix \ref{sec_velf}
that Eqs. (\ref{a4}) and (\ref{a3x}) constitute an exact solution of the 
continuity equation (\ref{mad12}) provided that 
\begin{eqnarray}
\label{hubb}
H=\frac{\dot R}{R}.
\end{eqnarray}
Equation (\ref{hubb}) is similar to the Hubble parameter in cosmology (see
Sec. \ref{sec_cosmo}). Using the Gaussian ansatz, one can show that (see
Appendix \ref{sec_det}):
\begin{eqnarray}
\label{a5}
I=\alpha MR^2,\qquad \Theta_c=\frac{1}{2}\alpha M\left (\frac{dR}{dt}\right )^2,\qquad \Theta_Q=\sigma \frac{\hbar^2M}{m^2R^2},\qquad U=\frac{K\zeta}{\gamma-1}\left (\frac{M}{R^d}\right )^{\gamma} R^d,
\end{eqnarray}
\begin{eqnarray}
\label{a6}
U_B=-d\frac{k_B T}{m}M\ln R+C,\qquad W_{\rm ext}=\frac{1}{2}\omega_0^2 \alpha M R^2,
\end{eqnarray}
\begin{eqnarray}
\label{a7}
W=-\frac{\nu}{d-2} \frac{GM^2}{R^{d-2}}\quad (d\neq 2),\qquad W=\frac{1}{2}GM^2\ln R+W_0\quad (d=2),
\end{eqnarray}
\begin{eqnarray}
\label{a7b}
W_{ii}=-\nu \frac{GM^2}{R^{d-2}},
\end{eqnarray}
with the coefficients
\begin{eqnarray}
\label{a8}
\alpha=\frac{d}{2}, \qquad \sigma=\frac{d}{4},\qquad  \zeta=\left\lbrack \frac{2}{S_d \Gamma(d/2)}\right\rbrack^{\gamma-1}\frac{1}{\gamma^{d/2}},\qquad \nu=\frac{1}{\Gamma(d/2)2^{d/2}},
\end{eqnarray}
\begin{eqnarray}
\label{a9}
C=\frac{k_B T}{m}M\left\lbrack \ln\left (\frac{2M}{S_d\Gamma(d/2)}\right )-1-\alpha\right\rbrack,\qquad
W_0=\frac{1}{4}(\ln 2-\gamma_E)GM^2=0.0289828... \, GM^2.
\end{eqnarray}
We have assumed that $\gamma>0$ otherwise the internal energy $U$ is infinite
within the Gaussian ansatz.  For $d=2$, we note that $\nu=1/2$.

Using these results, the free energy functional (\ref{a2b}) can be written
as a function of $R$ and $\dot R$  (for a fixed mass $M$) as
\begin{eqnarray}
\label{a10}
F=\frac{1}{2}\alpha M\left (\frac{dR}{dt}\right )^2+V(R),
\end{eqnarray}
with
\begin{eqnarray}
\label{a11}
V(R)=\sigma \frac{\hbar^2M}{m^2R^2}-\frac{\nu}{d-2}
\frac{GM^2}{R^{d-2}}+\frac{1}{2}\omega_0^2\alpha MR^2
+\frac{\zeta}{\gamma-1}\frac{KM^{\gamma}}{R^{d(\gamma-1)}}-d\frac{M k_B T}{m}\ln R+C\qquad (d\neq 2),
\end{eqnarray}
\begin{eqnarray}
\label{a12}
V(R)=\sigma \frac{\hbar^2M}{m^2R^2}+\frac{1}{2}
GM^2\ln R+\frac{1}{2}\omega_0^2\alpha MR^2
+\frac{\zeta}{\gamma-1}\frac{KM^{\gamma}}{R^{2(\gamma-1)}}-2\frac{M k_B T}{m}\ln R+W_0\qquad (d=2).
\end{eqnarray}
Equation (\ref{a10}) can be interpreted as the total energy of a
fictive   particle with effective mass $\alpha M$ and position $R$  moving in a
potential
$V(R)$. The first term is the classical kinetic energy $\Theta_c$ and the second
term is
the potential energy $V$ including the quantum kinetic energy $\Theta_Q$, the
gravitational potential energy $W$, the external potential energy $W_{\rm ext}$,
and the internal energies $U$ and $U_B$. We shall come back to this mechanical
analogy in Sec.
\ref{sec_vg}.

{\it Remark:} The average energy $\langle E\rangle$ can be
obtained similarly from
Eq. (\ref{a2cb}). At equilibrium, it reduces to the eigenenergy (\ref{a2g})
which, within the Gaussian ansatz, takes the form
\begin{eqnarray}
\label{a11a}
NE=\sigma \frac{\hbar^2M}{m^2R^2}-\frac{2\nu}{d-2}
\frac{GM^2}{R^{d-2}}+\frac{1}{2}\omega_0^2\alpha MR^2
+\frac{\zeta\gamma}{\gamma-1}\frac{KM^{\gamma}}{R^{d(\gamma-1)}}-d\frac{M k_B
T}{m}\ln R+Nk_B T+C\qquad (d\neq 2),
\end{eqnarray}
\begin{eqnarray}
\label{a12a}
NE=\sigma \frac{\hbar^2M}{m^2R^2}+
GM^2\ln R+\frac{1}{2}\omega_0^2\alpha MR^2
+\frac{\zeta\gamma}{\gamma-1}\frac{KM^{\gamma}}{R^{2(\gamma-1)}}-2\frac{M k_B
T}{m}\ln R+Nk_B T+W_0\qquad (d=2).
\end{eqnarray}

\subsection{The mass-radius relation}
\label{sec_mr}

We have seen in Sec. \ref{sec_efff} that a stable equilibrium state of
the generalized  GPP equations is
a minimum of free energy $F[\rho,{\bf u}]$, given by Eq.
(\ref{ae9}),  at fixed mass $M$. Within the Gaussian ansatz, we are led to
determining the minimum of the function $F(R,\dot R)$, given by Eq.
(\ref{a10}), at fixed mass $M$. Clearly, we must have $\dot R=0$, implying that
a minimum of energy at fixed mass is a static state. Then, we must determine the
minimum of the potential energy $V(R)$.  Computing the first derivative of
${V}(R)$ giving
\begin{eqnarray}
\label{a13}
V'(R)=-2\sigma \frac{\hbar^2M}{m^2R^3}+\nu \frac{GM^2}{R^{d-1}}+\omega_0^2\alpha
MR-d \zeta \frac{KM^{\gamma}}{R^{d(\gamma-1)+1}}-d \frac{M k_B T}{m R},
\end{eqnarray}
and writing $V'(R)=0$, we obtain the mass-radius relation
\begin{eqnarray}
\label{a13b}
-2\sigma \frac{\hbar^2M}{m^2R^3}+\nu \frac{GM^2}{R^{d-1}}+\omega_0^2\alpha
MR-d \zeta \frac{KM^{\gamma}}{R^{d(\gamma-1)+1}}-d \frac{M k_B T}{m R}=0
\end{eqnarray}
or, equivalently,
\begin{eqnarray}
\label{a14}
\alpha \omega_0^2 R^{d(\gamma-1)+2}+\nu G M R^{d(\gamma-1)+2-d}
-2\sigma\frac{\hbar^2}{m^2}R^{d(\gamma-1)-2}-dK\zeta M^{\gamma-1}-d\frac{k_B T}{m}R^{d(\gamma-1)}=0.
\end{eqnarray}
These relations may also be obtained from the equilibrium virial theorem
(\ref{a2e}) by making the Gaussian ansatz (see Sec. \ref{sec_vg}). A
critical point of ${V}(R)$, satisfying $V'(R)=0$, is a free energy minimum if, and
only if, ${V}''(R)>0$. Computing the second derivative of ${V}(R)$ and using the
mass-radius relation (\ref{a13b}), we get
\begin{eqnarray}
\label{a15}
{V}''(R)=6\sigma \frac{\hbar^2 M}{m^2R^4}-(d-1)\nu \frac{GM^2}{R^d}+\alpha\omega_0^2
M+[d(\gamma-1)+1]d\zeta \frac{K M^{\gamma}}{R^{d(\gamma-1)+2}}+d\frac{k_B T M}{mR^2}.
\end{eqnarray}

\subsection{The virial theorem}
\label{sec_vg}

Using the Gaussian ansatz, the time-dependent virial theorem (\ref{a2c})
can be
written as
\begin{eqnarray}
\label{a16}
\frac{1}{2}\alpha M\frac{d^2R^2}{dt^2}+\frac{1}{2}\xi\alpha
M\frac{dR^2}{dt}+\alpha\omega_0^2 MR^2=\alpha M\left (\frac{dR}{dt}\right
)^2
+2\sigma\frac{\hbar^2 M}{m^2 R^2}+d\frac{M k_B T}{m}
+d\zeta \frac{K M^{\gamma}}{R^{d(\gamma-1)}}-\nu\frac{GM^2}{R^{d-2}}.
\end{eqnarray}
Since
\begin{eqnarray}
\label{a17}
\frac{d^2R^2}{dt^2}=2R\frac{d^2R}{dt^2}+2\left (\frac{dR}{dt}\right )^2,
\end{eqnarray}
we note the nice cancelation of terms in Eq. (\ref{a16}) leading to the final
equation
\begin{eqnarray}
\label{a18}
\alpha M\frac{d^2R}{dt^2}+\xi\alpha M\frac{d R}{dt}+\alpha\omega_0^2
MR=2\sigma\frac{\hbar^2 M}{m^2 R^3}+d\frac{M k_B T}{m R}+d\zeta \frac{K M^{\gamma}}{R^{d(\gamma-1)+1}}-\nu
\frac{G M^2}{R^{d-1}}.
\end{eqnarray}
The equilibrium virial theorem ($\ddot R=\dot R=0$) returns the mass-radius
relation (\ref{a13b}) obtained from the condition $d{V}/dR=0$. In fact, the
time-dependent virial theorem (\ref{a18}) can be written as
\begin{eqnarray}
\label{a19}
\alpha M\frac{d^2R}{dt^2}+\xi\alpha M\frac{dR}{dt}=-\frac{d{V}}{dR}.
\end{eqnarray}
This equation describes the damped motion of a fictive particle with
effective mass
$\alpha M$ and position $R$ in a potential $V(R)$. From Eq. (\ref{a19}), we find
that the rate of change of the free energy $F=\Theta_c+V$ defined by Eq.
(\ref{a10})  is given by
\begin{eqnarray}
\label{a20}
\frac{dF}{dt}=-\xi\alpha M\left (\frac{dR}{dt}\right )^2\le 0.
\end{eqnarray}
This equation can be obtained directly from the $H$-theorem (\ref{ef11}) by
making the Gaussian ansatz.  For $\xi>0$ the free energy is dissipated and for
$\xi=0$ it is conserved. In this mechanical analogy, a stable equilibrium state
corresponds to a {\it minimum} of  ${V}(R)$ as we have previously indicated.
In the dissipative case ($\xi>0$), the
system relaxes
towards an equilibrium state for $t\rightarrow +\infty$ by
exhibiting damped oscillations due to the friction. As explained in Sec.
\ref{sec_pi}, this may account for the process of gravitational
cooling. In the
dissipationless case ($\xi=0$), the system oscillates permanently, without
reaching equilibrium, except if it is prepared in a very particular initial
state.\footnote{We note that the Gaussian ansatz is not relevant to describe
the dynamics of the conservative GPP equations  (\ref{mfgp13}) and
(\ref{mfgp14a}) because it would lead to undamped everlasting
oscillations
and would not account for the process of gravitational cooling. By contrast, the
Gaussian ansatz is able to describe the dynamics of the dissipative GPP
equations (\ref{a1}) and (\ref{a2}) that already incorporate the process of
gravitational cooling.}

In the strong friction limit $\xi\rightarrow +\infty$, the equation of motion (\ref{a19}) reduces to
\begin{eqnarray}
\label{a19d}
\xi\alpha M\frac{dR}{dt}=-\frac{d{V}}{dR}.
\end{eqnarray}
Furthermore, the free energy is $F=V$. Therefore, we obtain
\begin{eqnarray}
\label{a20d}
\frac{dF}{dt}=\frac{dV}{dt}=V'(R)\frac{dR}{dt}=-\frac{1}{\xi\alpha M}\left (\frac{dV}{dR}\right )^2\le 0.
\end{eqnarray}
The free energy decreases and the system relaxes towards an equilibrium state
which corresponds to the minimum of $V(R)$. In that case, the
relaxation is overdamped, showing no oscillation.

{\it Remark:} We can get the equation of dynamics (\ref{a19})
in a
different manner, without using the virial theorem. In the dissipationless case,
the free energy $F$ is conserved. Canceling the time derivative of Eq.
(\ref{a10}), we
obtain Eq. (\ref{a19}) with $\xi=0$. In the dissipative case, taking the time
derivative of Eq. (\ref{a10})
and using the $H$-theorem of Eq. (\ref{a20}) which can be obtained from Eq.
(\ref{ef11}), we obtain Eq. (\ref{a19}) with
$\xi\neq 0$.

\subsection{The general solution of the problem}
\label{sec_coll}

For dissipationless systems ($\xi=0$), the equation of
motion (\ref{a19}) reduces to
\begin{eqnarray}
\label{a22}
\alpha M\frac{d^2R}{dt^2}=-\frac{d{V}}{dR}.
\end{eqnarray}
The free energy $F$ is conserved ($\dot F=0$), leading to the first integral of
motion (\ref{a10}), where $F$ is a constant. The general solution of this
equation is
\begin{eqnarray}
\label{a24}
\int_{R_0}^{R(t)}\frac{d R}{\sqrt{F-V(R)}}=\pm\left (\frac{2}{\alpha
M}\right )^{1/2}t,
\end{eqnarray}
with $+$ for solutions describing an expansion (evaporation) and $-$ for
solutions describing a contraction (collapse).

For overdamped systems ($\xi\rightarrow +\infty$), the
equation of motion (\ref{a19}) reduces to Eq. (\ref{a19d}) whose general
solution is
\begin{eqnarray}
\label{a26}
\int_{R_0}^{R(t)}\frac{dR}{-V'(R)}=\frac{t}{\xi\alpha M}.
\end{eqnarray}

\subsection{The pulsation equation}
\label{sec_pulse}

To study the linear dynamical stability of an equilibrium state of Eq. (\ref{a19}), we make a small perturbation about that state
and write $R(t)=R+\epsilon(t)$ where $R$ is the equilibrium radius and
$\epsilon(t)\ll R$ is the perturbation. Using $V'(R)=0$ and keeping only terms
that are linear in $\epsilon$, we obtain the equation
\begin{eqnarray}
\label{a27}
\frac{d^2\epsilon}{dt^2}+\xi \frac{d\epsilon}{dt}+\omega^2\epsilon=0,
\end{eqnarray}
where
\begin{eqnarray}
\label{a28}
\omega^2=\frac{1}{\alpha M}{V}''(R).
\end{eqnarray}

Looking for solutions of Eq. (\ref{a27}) under the form $\epsilon\sim e^{\lambda
t}$, we obtain $\lambda^2+\xi\lambda+\omega^2=0$ yielding
$\lambda_{\pm}=(-\xi\pm\sqrt{\xi^2-4\omega^2})/2$. If $\omega^2>\xi^2/4$, the two
modes ($\pm$) are damped at a rate $-\xi/2<0$ and oscillate with a pulsation
$\pm\sqrt{4\omega^2-\xi^2}/2$. If $0<\omega^2<\xi^2/4$, the two modes are damped
at a rate $(-\xi\pm\sqrt{\xi^2-4\omega^2})/2<0$. If $\omega^2<0$, one mode is
damped at a rate $(-\xi-\sqrt{\xi^2-4\omega^2})/2<0$ and the other one grows
at a rate $(-\xi+\sqrt{\xi^2-4\omega^2})/2>0$. For $\omega^2=\xi^2/4$, the
perturbation decays as $\epsilon=(At+B) e^{-\xi t/2}$.  For dissipationless systems
($\xi=0$), the perturbation oscillates with a pulsation $\pm\omega$ if
$\omega^2>0$ and grows at a rate  $\sqrt{-\omega^2}>0$ if $\omega^2<0$ (the
other mode is damped at a rate $-\sqrt{-\omega^2}<0$). For overdamped systems
($\xi\rightarrow +\infty$), the perturbation is damped at a rate
$-\omega^2/\xi<0$ if $\omega^2>0$ and grows at a rate  $-\omega^2/\xi>0$ if
$\omega^2<0$. In conclusion, the equilibrium state of Eq. (\ref{a19}) is linearly stable if, and only
if, $\omega^2>0$ that is to say if, and only if, it is a (local) minimum of
the potential
energy $V(R)$.

For a self-gravitating BEC satisfying $\omega^2>\xi^2/4$, the radius of the condensate evolves as
\begin{eqnarray}
\label{osc}
R(t)=R+(R_0-R)e^{-\xi t/2}\cos\left
(\frac{\sqrt{4\omega^2-\xi^2}}{2}t\right ),
\end{eqnarray}
where $R_0$ is the initial radius and $R$ the equilibrium radius. The BEC
undergoes damped oscillations and relaxes towards an
equilibrium state for $t\rightarrow +\infty$.
These damped oscillations may account for the process of gravitational cooling.
They can explain how a self-gravitating BEC achieves an equilibrium state by
dissipating free energy.

Using Eqs. (\ref{a15}) and (\ref{a28}), we find that the complex pulsation is
given by
\begin{eqnarray}
\label{a29}
\omega^2=\omega_0^2+\frac{6\sigma}{\alpha}\frac{\hbar^2}{m^2 R^4}+\lbrack d(\gamma-1)+1\rbrack \frac{d\zeta}{\alpha} \frac{K M^{\gamma-1}}{R^{d(\gamma-1)+2}}-\frac{(d-1)\nu}{\alpha}\frac{GM}{R^d}+\frac{d}{\alpha}\frac{k_B T}{m R^2}.
\end{eqnarray}
Using Eqs. (\ref{a5}) and
(\ref{a7b}), it can be expressed under the form
\begin{eqnarray}
\label{a30}
\omega^2=\frac{6\Theta_Q+\lbrack d(\gamma-1)+1\rbrack d(\gamma-1) U+(d-1) W_{ii}+\omega_0^2 I+dNk_B T}{I}.
\end{eqnarray}
Alternative expressions of the pulsation can be obtained by using the
equilibrium
virial theorem (\ref{a2e}) or the equilibrium free energy (\ref{a2f}). This may
be useful in the dissipationless case ($\xi=0$) where $F$ is conserved. Let us
consider particular
cases.

For classical polytropes ($\Theta_Q=T=0$), the virial theorem reduces to
$d(\gamma-1)U+W_{ii}-\omega_0^2I=0$ and the complex pulsation can be written
as
\begin{eqnarray}
\label{a34}
\omega^2=(2d-2-d\gamma) \frac{W_{ii}}{I}+(d\gamma-d+2)\omega_0^2.
\end{eqnarray}
For $d=3$ and $\omega_0=0$, using $W_{ii}=W$, we recover the usual Ledoux
formula $\omega^2=(4-3\gamma){W}/{I}$ \cite{ledoux}.

For classical isothermal spheres ($\Theta_Q=U=0$),  the virial theorem reduces
to $W_{ii}-\omega_0^2 I+dNk_B T=0$ and the complex pulsation can be written
as
\begin{eqnarray}
\label{a35}
\omega^2=(d-2) \frac{W_{ii}}{I}+2\omega_0^2\qquad {\rm or}\qquad
\omega^2=\frac{(d-2)dNk_BT}{I}+d\omega_0^2.
\end{eqnarray}
For $d=2$, we obtain $\omega^2=2\omega_0^2$ and
the virial theorem leads to the identity (\ref{ex7}).

In the noninteracting case ($U=0$), the virial theorem reduces
to $2\Theta_Q+W_{ii}-\omega_0^2 I+dNk_B T=0$ and the complex pulsation can be
written
as
\begin{eqnarray}
\label{a36}
\omega^2=\frac{(d-4)W_{ii}+4\omega_0^2I-2dNk_B T}{I}.
\end{eqnarray}

For nongravitational ($G=0$) polytropes ($T=0$), the virial theorem reduces
to $2\Theta_Q+d(\gamma-1)U-\omega_0^2 I=0$ and the complex
pulsation can be
written
as
\begin{eqnarray}
\label{a37}
\omega^2=\frac{d(\gamma-1)\lbrack
d(\gamma-1)-2\rbrack U+4\omega_0^2I}{I}.
\end{eqnarray}
For the critical index $\gamma_c=1+2/d$ \cite{sulem}, we obtain
$\omega^2=4\omega_0^2$ in
agreement with Eq. (\ref{f51b}). In the TF approximation ($\Theta_Q=0$), the
virial theorem reduces to $d(\gamma-1)U-\omega_0^2 I=0$ and the
complex pulsation becomes $\omega^2=\lbrack d(\gamma-1)+2\rbrack\omega_0^2$.

\subsection{The Poincar\'e theorem}
\label{sec_turning}

If we differentiate the mass-radius relation (\ref{a13b}) with respect to $R$
and substitute the result into Eq. (\ref{a15}), we obtain
\begin{eqnarray}
\label{turning1}
V''(R)-\left (\frac{2\sigma\hbar^2}{m^2R^3}-\frac{2\nu GM}{R^{d-1}}-\omega_0^2\alpha R+\frac{d K\zeta\gamma M^{\gamma-1}}{R^{d(\gamma-1)+1}}+\frac{dk_B T}{m R}\right )\frac{d M}{d R}=0.
\end{eqnarray}
Using Eqs. (\ref{a13b}) and (\ref{a28}), the foregoing equation can be rewritten
as\footnote{We
note that the third term in the parenthesis vanishes for the
index $\gamma=2$ corresponding to the standard BEC model.}
\begin{eqnarray}
\label{turning2}
\omega^2(R)=-\frac{1}{\alpha M}\left (\frac{2\sigma\hbar^2}{m^2R^3}-\omega_0^2\alpha R+\frac{d K\zeta(2-\gamma) M^{\gamma-1}}{R^{d(\gamma-1)+1}}+\frac{dk_B T}{m R}\right )\frac{d M}{d R}.
\end{eqnarray}
This equation relates the square of the complex pulsation $\omega^2$ determining
the stability of the system to the slope of the mass-radius relation $M(R)$. We
see that the change of stability along the series of equilibria
($\omega^2=V''(R)=0$) coincides with the turning point of mass ($M'(R)=0$) in
agreement with the Poincar\'e theorem.

{\it Remark:} More generally, the mass-radius relation $M(R)$ is given in
implicit form by
\begin{eqnarray}
\label{turning3}
\frac{\partial V}{\partial R}(R,M(R))=0.
\end{eqnarray}
Differentiating this relation with respect to $R$, we get
\begin{eqnarray}
\label{turning4}
\frac{\partial^2 V}{\partial R^2}(R,M)+\frac{\partial^2 V}{\partial R\partial M}(R,M)\frac{dM}{dR}=0,
\end{eqnarray}
which is the generalization of Eq. (\ref{turning1}). We first
note that the turning point of mass ($M'(R)=0$) corresponds to
$\partial^2V/\partial R^2=0$. On the other hand, the turning point of radius
($R'(M)=0$) corresponds to
$\partial^2V/\partial R\partial M=0$. A change of stability takes place
when $\partial^2V/\partial R^2$ changes sign, i.e. when $M'(R)$ changes sign
while $\partial^2V/\partial R\partial M$ does not change sign. This happens
at a turning point of mass, not at a turning point of radius. As a result, the
stability of the system is not directly related to the sign of the slope of the
mass-radius relation since there is no change of stability after a turning
point of radius although the slope changes.

\subsection{Analogy with cosmology}
\label{sec_cosmo}

In $d=3$, the free energy (\ref{a10}) is given by
\begin{eqnarray}
\label{cosmo0}
F=\frac{1}{2}\alpha M{\dot R}^2+\sigma \frac{\hbar^2M}{m^2R^2}-\nu
\frac{GM^2}{R}+\frac{1}{2}\omega_0^2\alpha MR^2
+\frac{\zeta}{\gamma-1}\frac{KM^{\gamma}}{R^{3(\gamma-1)}}-3\frac{Mk_B T}{m}\ln R+C.
\end{eqnarray}
For dissipationless systems ($\xi=0$), the free energy is conserved so that Eq.
(\ref{cosmo0}) can be seen as the first integral of the equation of motion
(\ref{a22}). It can be rewritten under the form
\begin{eqnarray}
\label{cosmo1}
\left (\frac{\dot R}{R}\right )^2=\frac{2(F-C)}{\alpha M
R^2}-\frac{2\sigma\hbar^2}{\alpha m^2R^4}+\frac{2\nu GM}{\alpha
R^3}-\omega_0^2-\frac{2K\zeta M^{\gamma-1}}{(\gamma-1)\alpha R^{3\gamma-1}}+\frac{6k_B T\ln R}{\alpha m R^2}.
\end{eqnarray}
For $\hbar=a_s=T=0$, it reduces to
\begin{eqnarray}
\label{cosmo2}
\left (\frac{\dot R}{R}\right )^2=\frac{2F}{\alpha M R^2}+\frac{2\nu
GM}{\alpha R^3}-\omega_0^2.
\end{eqnarray}
Equation (\ref{cosmo2}) is similar to the Friedmann equation in
cosmology\footnote{A short
account of the early development of modern cosmology is given in the
Introduction of
\cite{cosmopoly1}.}
\begin{eqnarray}
\label{cosmo3}
H^2=\left (\frac{\dot R}{R}\right )^2=-\frac{kc^2}{R^2}+\frac{8\pi
G}{3c^2}\epsilon+\frac{\Lambda}{3}
\end{eqnarray}
for a pressureless ($P=0$) universe whose energy density decreases as
$\epsilon\propto R^{-3}$. In
this analogy, $R$ plays the role of the scale factor, $H=\dot R/R$ plays the
role
of the Hubble parameter, $-2F/\alpha M$ plays the role of the curvature
constant $kc^2$, $2\nu M/\alpha R^3$ plays the role of the mass density $8\pi
 \epsilon/3c^2$, and $-3\omega_0^2$ plays the
role of the cosmological constant
$\Lambda$. When  $\omega_0^2<0$ we obtain the $\Lambda$CDM model ($\Lambda>0$)
and when $\omega_0^2>0$ we obtain the anti-$\Lambda$CDM model ($\Lambda<0$). We
can therefore
draw certain analogies between
the evolution of a self-gravitating BEC and the evolution of a 
Friedmann-Lema\^itre-Robertson-Walker (FLRW) universe. This
suggests the possibility of reproducing cosmological behaviors in BEC laboratory
experiments.

{\it Remark:}  In a universe filled with a fluid with an equation of state
$P=\alpha\epsilon$, the energy density is related to the scale factor by
$\epsilon\propto R^{-3(1+\alpha)}$. If we take into account all the terms in
Eq. (\ref{cosmo1}), we find that the quantum potential term is analogous to an
energy density $\epsilon\propto -1/R^4$ in cosmology. This corresponds to
$\alpha=1/3$ like for the standard radiation. However, the energy density is
negative. 
On the other hand, the temperature term in Eq. (\ref{cosmo1}) is analogous  to
an energy density $\epsilon\propto \ln R/R^{2}$ in cosmology. This corresponds
to $\alpha=-1/3$ (up to a logarithmic correction) like for a gas of cosmic
strings. Finally, the self-interaction term is analogous  to an energy density
$\epsilon\propto \mp/R^{3\gamma-1}$ in cosmology. This  corresponds to
$\alpha=\gamma-4/3$. The standard BEC 
($\gamma=2$, $n=1$, $K={2\pi a_s\hbar^2}/{m^3}$) is analogous  to an
energy density
$\epsilon\propto \mp/R^{5}$ in cosmology corresponding to
$\alpha=2/3$ (to our knowledge, this coefficient has no particular
interpretation
in
cosmology). The energy density is negative when $a_s>0$ and a positive 
when $a_s<0$. A gas of domain walls in cosmology  ($\alpha=-2/3$)
corresponds to
a BEC with $\gamma=2/3$ ($n=-3$).

\section{Conclusion}

In this paper, we have developed a general formalism applying to Newtonian
self-gravitating BECs. We have introduced and studied the generalized GPP
equations (\ref{mfgp9}) and (\ref{mfgp14}). We
have given their main properties, derived a hydrodynamic representation of these
equations, established the virial theorem, and showed that these equations are
consistent with a generalized thermodynamic formalism. In particular, they
satisfy an $H$-theorem for a free energy functional associated with a
generalized entropy.
We have shown how the
generalized free energy and the equation of state are related to the
nonlinearity in the generalized GP
equation and we have given several illustrative examples. Finally, by using a
Gaussian ansatz for the wave function, we have shown how the generalized GPP
equations
(\ref{mfgp9}) and (\ref{mfgp14}) could be reduced to a simple dynamical equation
giving the evolution of the size $R(t)$ of the BEC. We have established very
general equations that can describe many situations of physical and
astrophysical interest. In our following papers (in
preparation), these equations will be studied in detail.

The generalized GPP equations (\ref{mfgp9}) and (\ref{mfgp14}) may provide a
model of dark matter halos. The generalized GP equation (\ref{mfgp9}) includes a
source of
dissipation and an arbitrary nonlinearity $h(|\psi|^2)$. Concerning the
nonlinearity, we have given a special emphasis to the logarithmic nonlinearity
$2k_B T\ln|\psi|$ where $T$ plays the role of an effective temperature. 
This leads to a particularly interesting BEC dark matter model described by
the generalized GPP equations (\ref{mi1}) and (\ref{mi2}). In this model, the
system has a core-halo structure with a soliton/BEC core and an
isothermal halo. Furthermore, the dissipation term ensures that the system
relaxes towards an equilibrium state.

We have
given three possible justifications of the generalized GPP equations:

(i) The generalized GPP equations (\ref{mfgp9}) and (\ref{mfgp14}) may 
be justified by physical processes. The dissipation could be due to non ideal
effects or to the  interaction of the system with an external medium, and the
nonlinear potential $h(|\psi|^2)$ could account for the self-interaction of the
bosons. For bosons interacting via short-range interactions, the effective
potential is quadratic, given by $h(|\psi|^2)=(4\pi a_s\hbar^2/m^3)|\psi|^2$,
but in more general
situations other forms of potentials can emerge such as the
logarithmic potential $h(|\psi|^2)=(2k_B
T/m)\ln|\psi|$. In that interpretation, $T$ would be a formal measure of the
collisionless interactions inside the BEC at zero (thermal) temperature.

(ii) The generalized GPP equations (\ref{mfgp9}) and (\ref{mfgp14}) may 
provide an effective model of gravitational cooling.\footnote{In
particular, they are able to
account for the
damped oscillations of a system experiencing gravitational cooling
\cite{seidel94,gul0,gul}. This damping is apparent on the simplified equation of
motion
(\ref{a18}) derived within the Gaussian ansatz.} In that interpretation,
they
can be viewed as a coarse-grained description of the (fine-grained) GPP
equations
(\ref{mfgp13}) and (\ref{mfgp14a}). The friction
term accounts for the relaxation process (dissipation of free energy) and the
nonlinear term accounts for the
formation of a halo of scalar radiation. This interpretation may be particularly
relevant in the case of a logarithmic nonlinearity leading to an isothermal
halo, with a density profile decreasing as $r^{-2}$, that is relatively close to
the NFW/Burkert profile at large distances. This
model could be improved to match exactly the $r^{-3}$ (NFW/Burkert) profile at
large
distances by introducing a more complicated nonlinearity in the generalized GPP
equations (\ref{mfgp9}) and (\ref{mfgp14}), or by introducing an external
potential to confine the system. 

(iii) In Ref. \cite{papernottale}, we have derived 
the generalized GP equation (\ref{mfgp9}) with a logarithmic nonlinearity from
Nottale's theory of scale relativity relying on a fractal spacetime
\cite{nottale}. In that
interpretation, the friction and the temperature are obtained from a unified
formalism and they correspond to the real and imaginary parts of a complex
friction
coefficient arising in a scale-covariant equation of dynamics. These terms can
be the properties of a fractal spacetime or an aether. The
dissipation is due to the friction with the aether and the temperature
represents the temperature of the aether. In this interpretation,
the
generalized GP equation (\ref{mi1})  could be a fundamental
equation of physics
from which the standard GP and Schr\"odinger equations would be
approximations. In that case, $\xi$ and $T$ would be fundamental constants. We
have, however, pointed out a difficulty with this
interpretation in relation to the non constancy of the temperature of dark
matter halos.

\appendix

\section{The momentum tensor}
\label{sec_stress}

The equation of continuity (\ref{mad12}) can be written as
\begin{eqnarray}
\label{stress1}
\frac{\partial\rho}{\partial t}+\nabla\cdot {\bf j}=0,
\end{eqnarray}
where ${\bf j}=\rho{\bf u}$ is the density current. Using Eqs.
(\ref{mad1})-(\ref{mad5}), the density current can be expressed in terms of the
wave function as
\begin{eqnarray}
\label{stress2}
{\bf j}=\frac{\hbar}{2im}\left (\psi^*\nabla\psi-\psi\nabla\psi^*\right ).
\end{eqnarray}
As a result, Eq. (\ref{stress1}) takes the form
\begin{eqnarray}
\label{stress1b}
\frac{\partial |\psi|^2}{\partial t}+\frac{\hbar}{2im}\nabla\cdot \left
(\psi^*\nabla\psi-\psi\nabla\psi^*\right )=0.
\end{eqnarray}
On the other hand, the quantum Euler equation  (\ref{mad14b}) can be written as
\begin{eqnarray}
\label{stress3}
\frac{\partial {\bf j}}{\partial t}=-\nabla(\rho {\bf u}\otimes {\bf u})
-\nabla P-\rho\nabla\Phi-\rho\nabla\Phi_{\rm ext}-\frac{\rho}{m}\nabla Q-\xi{\bf
j}.
\end{eqnarray}
Introducing the quantum pressure tensor (\ref{mad20}), we
find that the equation for the density current is given by
\begin{eqnarray}
\label{stress4}
\frac{\partial {j_i}}{\partial t}=-\partial_j
T_{ij}-\rho\partial_i\Phi-\rho\partial_i\Phi_{\rm ext}-\xi{j}_i,
\end{eqnarray}
where
\begin{eqnarray}
\label{stress5}
T_{ij}=\rho u_i u_j+P\delta_{ij}+P_{ij}
\end{eqnarray}
is the momentum tensor. Using Eq. (\ref{mad20}), we get
\begin{eqnarray}
\label{stress6}
T_{ij}=\rho u_i u_j+P\delta_{ij}
-\frac{\hbar^2}{4m^2}\rho\partial_i\partial_j\ln\rho
\end{eqnarray}
or, alternatively,
\begin{eqnarray}
\label{stress7}
T_{ij}=\rho u_i u_j+\left (P-\frac{\hbar^2}{4m^2}\Delta\rho\right )\delta_{ij}
+\frac{\hbar^2}{4m^2}\frac{1}{\rho}\partial_i\rho\partial_j\rho.
\end{eqnarray}
Using Eqs. (\ref{mad1}) and (\ref{mad5}), we find after straightforward algebra
that
\begin{eqnarray}
\label{stress8}
\frac{\hbar^2}{4m^2}\frac{1}{\rho}\partial_i\rho\partial_j\rho=
\frac{\hbar^2}{4m^2}\frac{1}{|\psi|^2}
(\psi^*\partial_i\psi+\psi\partial_i\psi^*)
(\psi^*\partial_j\psi+\psi\partial_j\psi^*)
\end{eqnarray}
and
\begin{eqnarray}
\label{stress9}
\rho u_i
u_j=-\frac{\hbar^2}{4m^2}\frac{1}{|\psi|^2}
(\psi^*\partial_i\psi-\psi\partial_i\psi^*)
(\psi^*\partial_j\psi-\psi\partial_j\psi^*).
\end{eqnarray}
Therefore
\begin{eqnarray}
\label{stress10}
\rho u_i
u_j+\frac{\hbar^2}{4m^2}\frac{1}{\rho}\partial_i\rho\partial_j\rho=\frac{\hbar^2
}{m^2}{\rm Re}\left (\frac{\partial\psi}{\partial
x_i}\frac{\partial\psi^*}{\partial x_j}\right ).
\end{eqnarray}
Regrouping these results, the momentum tensor can be expressed in terms of the
wave function as
\begin{eqnarray}
\label{stress12}
T_{ij}=\frac{\hbar^2}{m^2}{\rm Re}\left (\frac{\partial\psi}{\partial
x_i}\frac{\partial\psi^*}{\partial x_j}\right )
+\left (P-\frac{\hbar^2}{4m^2}\Delta|\psi|^2\right )\delta_{ij}.
\end{eqnarray}

\section{The energy operator and the Hamiltonian}
\label{sec_gh}

The generalized GP equation (\ref{mfgp9}) can be written as
\begin{eqnarray}
\label{ae3}
i\hbar \frac{\partial\psi}{\partial
t}={\hat E}\psi-i\frac{\hbar}{2}\xi\left\lbrack \ln\left
(\frac{\psi}{\psi^*}\right )-\left\langle \ln\left (\frac{\psi}{\psi^*}\right
)\right\rangle\right\rbrack\psi
\end{eqnarray}
with the energy operator given by 
\begin{eqnarray}
\label{ae5}
{\hat E}=-\frac{\hbar^2}{2m}\Delta
+m\lbrack\Phi
+h(|\psi|^2)+\Phi_{\rm ext}\rbrack.
\end{eqnarray}
Its average value is
\begin{eqnarray}
\label{ae6}
N \langle E\rangle=\frac{1}{m}\langle
\psi|{\hat E}|\psi\rangle=\frac{\hbar^2}{2m^2}\int
|\nabla\psi|^2\, d{\bf r}+\int |\psi|^2 \Phi_{\rm ext}\, d{\bf r}+\int |\psi|^2
\Phi\, d{\bf r}+\int |\psi|^2 h(|\psi|^2)\, d{\bf r}.
\end{eqnarray}
Using the results of Sec. \ref{sec_ef}, we see that the average value of
the energy operator coincides with the average value of the energy given by Eq.
(\ref{ae7}). It differs from the free energy (\ref{ae9}) which can be written
as 
\begin{eqnarray}
\label{ae10}
F=\frac{\hbar^2}{2m^2}\int |\nabla\psi|^2\, d{\bf r}+\int |\psi|^2 \Phi_{\rm
ext}\, d{\bf r}+\frac{1}{2}\int |\psi|^2 \Phi\, d{\bf r}+\int V(|\psi|^2)\,
d{\bf r}.
\end{eqnarray}
We have 
\begin{eqnarray}
\label{ae10b}
F=N\langle E\rangle-\frac{1}{2}\int |\psi|^2 \Phi\, d{\bf r}+\int
\lbrack V(|\psi|^2)-|\psi|^2 h(|\psi|^2)\rbrack\,
d{\bf r},
\end{eqnarray}
where the term in brackets is the opposite of the pressure [see Eq.
(\ref{mad11b})]. Equation (\ref{ae10b}) is equivalent to Eqs. (\ref{ae11})
and (\ref{ae12}).

Taking the first variation of the free energy (\ref{ae10}), we get
\begin{eqnarray}
\label{ae11za}
\delta F=\frac{1}{m}\int\left\lbrack -\frac{\hbar^2}{2m}\Delta\psi^*+m\Phi_{\rm
ext}\psi^*+m\Phi\psi^*+mh(|\psi|^2)\psi^*\right\rbrack\delta\psi\, d{\bf
r}+{\rm c.c.}
\end{eqnarray}
The term in brackets coincides with the energy
operator (\ref{ae5}) applied on $\psi^*$. Indeed,
\begin{eqnarray}
\label{ae11zb}
m\frac{\delta F}{\delta\psi^*}=\hat{E}\psi.
\end{eqnarray}
This relation can be compared with Eq. (\ref{fd1}). We also note that
\begin{eqnarray}
\label{ae11zc}
N\langle E\rangle=\int \frac{\delta F}{\delta\psi^*}\psi^*\, d{\bf r}
\end{eqnarray}
which can be compared with Eq. (\ref{rw1}).

Let us first consider conservative systems
($\xi=0$). The generalized GP equation
(\ref{ae3}) can be rewritten as
\begin{eqnarray}
\label{ab}
i\hbar \frac{\partial\psi}{\partial
t}=m\frac{\delta F}{\delta\psi^*}.
\end{eqnarray}
This expression shows that $F$ represents the true
Hamiltonian of the system. Indeed, in terms of
the wavefunction $\psi({\bf
r},t)$ and its canonical momentum $\pi({\bf r},t)=i\hbar\psi^*({\bf r},t)$, the
generalized GP equation is exactly reproduced by the Hamilton equations
\begin{eqnarray}
\label{lh5w}
\frac{\partial\psi}{\partial t}=m\frac{\delta F}{\delta\pi},\qquad
\frac{\partial\pi}{\partial t}=-m\frac{\delta F}{\delta\psi}.
\end{eqnarray}
This formulation directly implies the
conservation of the free energy
$F$ since
\begin{eqnarray}
\label{lh5wb}
{\dot F}=\int \frac{\delta F}{\delta\psi}\frac{\partial\psi}{\partial t}\,
d{\bf r}+\int \frac{\delta F}{\delta\pi}\frac{\partial\pi}{\partial t}\, d{\bf
r}=0.
\end{eqnarray}

For dissipative systems, the generalized GP
equation
(\ref{ae3}) can be rewritten as
\begin{eqnarray}
\label{ae12w}
i\hbar \frac{\partial\psi}{\partial
t}=m\frac{\delta F}{\delta\psi^*}-i\frac{\hbar}{2}\xi\left\lbrack
\ln\left
(\frac{\psi}{\psi^*}\right )-\left\langle \ln\left (\frac{\psi}{\psi^*}\right
)\right\rangle\right\rbrack\psi.
\end{eqnarray}
From Eqs. (\ref{ae11zb}) and (\ref{ae12w}), we easily obtain the
identity
\begin{eqnarray}
\label{ae13hg}
\dot F=-\xi\int \frac{\hbar^2}{4m^2}|\psi|^2\left |\nabla\ln\left
(\frac{\psi}{\psi^*}\right )\right |^2\, d{\bf r},
\end{eqnarray}
which coincides with the $H$-theorem (\ref{ef11}).

Finally,  considering the minimization of the
free energy $F[\psi]$ given by
Eq. (\ref{ae10}) at fixed mass $M$ given by Eq. (\ref{mfgp3b}), and writing the
variational principle as 
\begin{eqnarray}
\label{ef14z}
\delta F-\frac{\mu}{m}\delta \int |\psi|^2\, d{\bf r}=0,
\end{eqnarray}
where $\mu$ is a Lagrange multiplier (chemical potential), we obtain the
time-independent GP equation (\ref{tigp2}) with $\mu=E$ (see Sec.
\ref{sec_efff} for a detailed discussion). This
variational principle was  introduced by Schr\"odinger
\cite{schrodinger1} in his first
paper on wave mechanics. Actually, this is how he originally obtained the
fundamental 
eigenvalue equation (\ref{tigp2}). A short account of the early development of
quantum
mechanics is given in the Introduction of
\cite{chavmatos}.

{\it Remark:} For nonlinear Schr\"odinger
equations, the true Hamiltonian of the
system is $F$, not $N\langle E\rangle$. As a result, it is the free
energy $F$ that is
conserved ($\dot F=0$) for dissipationless systems ($\xi=0$) and that satisfies
an $H$-theorem ($\dot F\le 0$) for dissipative systems  ($\xi\neq 0$), {\it
not} the average energy $N\langle E\rangle$. For the standard
(linear) Schr\"odinger
equation with $h=\Phi=0$, we have $F=N\langle E\rangle$. For the logarithmic
Schr\"odinger equation (\ref{f6}) with $\Phi=0$, since $F$ and $N\langle
E\rangle$ only differ by
a constant $Nk_B T$ [see Sec. \ref{sec_bol_gp}], we can also interpret $N\langle
E\rangle$ as the Hamiltonian of the
system. However, this identification is not true anymore for other nonlinear
Schr\"odinger equations.

\section{Variation of the energies}
\label{sec_var}

In this Appendix, we detail the first and second order variations of the
different functionals 
that compose the free energy (\ref{ef1}). 

The first and second order variations
of the classical kinetic energy (\ref{ef3}) are
\begin{equation}
\label{var1}
\delta\Theta_c=\int \frac{{\bf u}^2}{2}\delta\rho \, d{\bf r}+\int \rho {\bf
u}\cdot\delta {\bf u} \, d{\bf r},\qquad \delta^2\Theta_c=\frac{1}{2}\int \rho
(\delta{\bf u})^2 \, d{\bf r}+\int \delta\rho {\bf u}\cdot\delta {\bf u} \,
d{\bf r}.
\end{equation}
The first and second order variations of the quantum kinetic energy  (\ref{ef5})
are
\begin{equation}
\label{var2}
\delta\Theta_Q=\frac{1}{m}\int Q\delta\rho \, d{\bf r},\qquad
\delta^2\Theta_Q=\frac{\hbar^2}{8m^2}\int\frac{1}{\rho}\left\lbrack \left
(\frac{\Delta\rho}{\rho}-\frac{(\nabla\rho)^2}{\rho^2}\right
){(\delta\rho)^2}+{(\nabla\delta\rho)^2}\right\rbrack\, d{\bf r}.
\end{equation}
The first and second order variations of the internal energy (\ref{ney1b}) are
\begin{eqnarray}
\label{var3}
\delta U=\int h(\rho)\delta \rho \, d{\bf r},\qquad \delta^2 U=\frac{1}{2}\int
h'(\rho)(\delta \rho)^2 \, d{\bf r}.
\end{eqnarray}
The first and second order variations of the gravitational potential energy 
(\ref{ney3}) are
\begin{equation}
\label{var4}
\delta W=\int \Phi\delta\rho \, d{\bf r},\qquad \delta^2 W=\frac{1}{2}\int
\delta\rho\delta\Phi \, d{\bf r}.
\end{equation}
The first and second order variations of the external potential energy 
(\ref{ef1b}) are
\begin{equation}
\label{var5}
\delta W_{\rm ext}=\int \Phi_{\rm ext}\delta\rho \, d{\bf r},\qquad \delta^2
W_{\rm ext}=0.
\end{equation}

The calculations leading to these relations are straightforward
except, maybe, the ones leading to the first relation of Eq. (\ref{var2}).  We
give below two different
derivations of this relation:

(i) From Eqs. (\ref{mad8}), (\ref{ef5b}) and
(\ref{ef4}) we directly obtain 
\begin{equation}
\label{ide4}
\delta\Theta_Q=\frac{\hbar^2}{m^2}\int \nabla\sqrt{\rho}\cdot
\nabla\delta\sqrt{\rho}\, d{\bf r}=-\frac{\hbar^2}{m^2}\int
\delta\sqrt{\rho}\Delta\sqrt{\rho}\, d{\bf r}=-\frac{\hbar^2}{2m^2}\int
\delta\rho\frac{\Delta\sqrt{\rho}}{\sqrt{\rho}}\, d{\bf r}=\frac{1}{m}\int
Q\delta\rho\,
d{\bf r}.
\end{equation}

(ii) From the first equality of Eq. (\ref{mad8}), we
find that 
\begin{equation}
\label{var6}
\delta Q=-\frac{\hbar^2}{2m\rho}\nabla\cdot 
(\sqrt{\rho}\nabla\delta\sqrt{\rho}-\delta\sqrt{\rho}\nabla\sqrt{\rho}
)=-\frac{\hbar^2}{4m\rho}\nabla\cdot
(\rho\delta\nabla\ln\rho). 
\end{equation}
This implies the identity
\begin{equation}
\label{var7}
\int \rho \delta Q\, d{\bf r}=0,
\end{equation}
from which we get $\delta\Theta_Q=\frac{1}{m}\int
\rho\delta Q \, d{\bf r}+\frac{1}{m}\int Q\delta\rho \, d{\bf
r}=\frac{1}{m}\int Q\delta\rho \, d{\bf r}$ leading to the first relation of Eq.
(\ref{var2}). We note
that Eq. (\ref{var6}) is the equivalent of the tensorial equation (\ref{mad19})
with Eq. (\ref{mad20}) except that it applies to a perturbation $\delta$ instead
of a space derivative $\partial_i$.

\section{The $H$-theorem}
\label{sec_hth}

In this Appendix, we establish the $H$-theorem
associated with the damped quantum barotropic EP equations
(\ref{mad12})-(\ref{mad14}) that are equivalent to the generalized GPP
equations
(\ref{mfgp9}) and (\ref{mfgp14}).

Taking the time derivative of the free energy  (\ref{ef1}), and using the
results of Appendix \ref{sec_var}, we get
\begin{eqnarray}
\label{cons5}
\dot F=\int \left ( \frac{{\bf u}^2}{2}+\frac{Q}{m}+h(\rho)+\Phi+\Phi_{\rm
ext}\right ) \frac{\partial\rho}{\partial t}\, d{\bf r}+\int \rho {\bf u}\cdot
\frac{\partial {\bf u}}{\partial t}\, d{\bf r}.
\end{eqnarray}
Substituting the continuity equation  (\ref{mad12}) into Eq. (\ref{cons5}),
integrating by parts, 
and using the Euler equation (\ref{mad13}) together with  Eq.
(\ref{mad11}) to simplify some terms, we obtain
\begin{eqnarray}
\label{cons6}
\dot F=\int \rho {\bf u}\cdot \left \lbrack\nabla\left (\frac{{\bf
u}^2}{2}\right )-\xi {\bf u}- ({\bf u}\cdot \nabla){\bf u} \right \rbrack\,
d{\bf r}.
\end{eqnarray}
Using the identity $({\bf u}\cdot \nabla){\bf u}=\nabla\left ({{\bf
u}^2}/{2}\right )-{\bf u}\times (\nabla\times {\bf u})$, the foregoing equation
can be rewritten as
\begin{eqnarray}
\label{cons7}
\dot F=\int \rho {\bf u}\cdot \left \lbrack {\bf u}\times (\nabla\times {\bf u})
-\xi {\bf u} \right \rbrack\, d{\bf r}.
\end{eqnarray}
Since ${\bf u}$ is a potential flow, it is irrotational ($\nabla\times
{\bf u}={\bf 0}$), so that Eq. (\ref{cons7}) reduces to Eq. (\ref{ef11}).
Actually, we note that this result remains valid even if ${\bf u}$ is not a
potential flow since ${\bf u}\cdot [{\bf u}\times (\nabla\times {\bf u})]=0$ in
any case.\footnote{We could arrive directly at Eq.  (\ref{ef11}) from Eq.
(\ref{cons5}) by using Eq. (\ref{enxidf}) but we wanted to be more general so
that our derivation also applies to nonpotential flows (this may be useful in
other circumstances).}

In the strong friction limit, the free energy is given by Eq. (\ref{ef12}).
Taking its time derivative and using the results of Appendix \ref{sec_var}, we
get
\begin{eqnarray}
\label{cons8}
\dot F=\int \left (\frac{Q}{m}+h(\rho)+\Phi+\Phi_{\rm ext}\right )
\frac{\partial\rho}{\partial t}\, d{\bf r}.
\end{eqnarray}
Substituting the quantum barotropic Smoluchowski equation (\ref{mad16}) into Eq.
(\ref{cons8}), integrating by parts, and using Eq. (\ref{mad11}), we obtain Eq.
(\ref{ef13}).

\section{Local free energy equation}
\label{sec_lee}

The free energy (\ref{ae9}) can be written as
\begin{eqnarray}
\label{stress14z}
F=\int\rho e\, d{\bf r},
\end{eqnarray}
where $e({\bf r},t)$ is the free energy density given by
\begin{eqnarray}
\label{stress14}
e=\frac{{\bf u}^2}{2}+\frac{Q}{m}+\frac{V(\rho)}{\rho}+\frac{\Phi}{2}+\Phi_{\rm
ext}.
\end{eqnarray}
Using the equation of continuity (\ref{mad12}) and the damped quantum Euler
equation (\ref{mad13}), we obtain the local free energy equation
\begin{eqnarray}
\label{stress15q}
\frac{\partial}{\partial t}(\rho e)+\nabla\cdot (\rho e {\bf u})=-\nabla\cdot (P
{\bf u})-\frac{1}{2}\rho {\bf u}\cdot\nabla\Phi+\frac{\rho}{m}\frac{\partial
Q}{\partial t}
+\frac{1}{2}\rho\frac{\partial\Phi}{\partial t}-\xi\rho {\bf u}^2.
\end{eqnarray}
According to Eq. (\ref{var6}), we have
\begin{eqnarray}
\label{stress19b}
\frac{\rho}{m}\frac{\partial Q}{\partial t}=-\nabla\cdot {\bf J}_Q,
\end{eqnarray}
where
\begin{eqnarray}
\label{stress19bq}
{\bf J}_Q=\frac{\hbar^2}{2m^2}
\left (\sqrt{\rho}\frac{\partial\nabla\sqrt{\rho}}{\partial
t}-\frac{\partial\sqrt{\rho}}{\partial t}\nabla\sqrt{\rho}\right
)=\frac{\hbar^2}{4m^2}\rho\frac{\partial\nabla\ln\rho}{\partial t}
\end{eqnarray}
is the quantum current. Therefore, Eq. (\ref{stress15q}) can be rewritten as
\begin{eqnarray}
\label{stress15b}
\frac{\partial}{\partial t}(\rho e)+\nabla\cdot (\rho e {\bf
u})+\nabla\cdot {\bf J}_Q+\nabla\cdot (P{\bf u})=-\frac{1}{2}\rho {\bf
u}\cdot\nabla\Phi+\frac{1}{2}\rho\frac{\partial\Phi}{\partial t}-\xi\rho {\bf
u}^2.
\end{eqnarray}
The $H$-theorem directly results from this equation. Indeed, taking the time
derivative of the free energy (\ref{stress14z}), and using Eq.
(\ref{stress15b}),
we get
\begin{eqnarray}
\label{stress16}
\dot F=-\frac{1}{2}\int \rho {\bf u}\cdot\nabla\Phi\, d{\bf r}
+\frac{1}{2}\int \rho\frac{\partial\Phi}{\partial t}\, d{\bf r}-\xi\int\rho {\bf
u}^2\, d{\bf r}.
\end{eqnarray}
Using the Poisson equation (\ref{mad14}) and the equation of continuity
(\ref{mad12}), and integrating by parts, we find that
\begin{eqnarray}
\label{stress17}
\frac{1}{2}\int \rho\frac{\partial\Phi}{\partial t}\, d{\bf r}=\frac{1}{2 S_d
G}\int \Delta\Phi\frac{\partial\Phi}{\partial t}\, d{\bf r}=\frac{1}{2}\int
\Phi\frac{\partial\rho}{\partial t}\, d{\bf r}\nonumber\\
=-\frac{1}{2}\int \Phi \nabla\cdot (\rho {\bf u})\, d{\bf r}=\frac{1}{2}\int 
\rho {\bf u}\cdot \nabla\Phi\, d{\bf r}.
\end{eqnarray}
Substituting this identity into Eq. (\ref{stress16}), we obtain the $H$-theorem
(\ref{ef11}).

We can also consider the energy $E({\bf r},t)$ defined by Eq. (\ref{enxi}). We
note that
\begin{eqnarray}
\label{stress19}
\frac{E}{m}-e=\frac{1}{2}\Phi+h(\rho)-\frac{V(\rho)}{\rho}.
\end{eqnarray}
Proceeding as before, we obtain the local energy equation
\begin{eqnarray}
\label{stress20}
\frac{\partial}{\partial t}\left (\rho \frac{E}{m}\right )+\nabla\cdot \left
(\rho \frac{E}{m} {\bf u}\right )+\nabla\cdot {\bf J}_Q=\frac{\partial
P}{\partial t}+\rho\frac{\partial\Phi}{\partial t}-\xi\rho {\bf u}^2,
\end{eqnarray}
where we used Eq. (\ref{mad11}) in the course of the calculations. Taking the
time derivative of Eq. (\ref{ae7}) giving the average value of $E$, and using
Eq. (\ref{stress20}), we obtain
\begin{eqnarray}
\label{stress21}
N\frac{d\langle E\rangle}{dt}=-\xi\int \rho {\bf u}^2\, d{\bf r}+\int
\rho\frac{\partial\Phi}{\partial t}\, d{\bf r}+\frac{d}{dt}\int P\, d{\bf r}.
\end{eqnarray}
This relation can also be obtained by taking the time derivative of Eq.
(\ref{ae12}) and using Eq. (\ref{ef11}). It confirms that when $P\neq \rho
k_BT/m$ and
$\Phi\neq 0$ the average energy $\langle E\rangle$ is not conserved, even when
$\xi=0$, contrary to the free energy $F$.

The previous results remain valid in the strong friction limit
$\xi\rightarrow +\infty$ with ${\bf u}$ given by Eq. (\ref{mad15}).

{\it Remark:} In Ref. \cite{papernottale}, we have established the identity
\begin{equation}
\label{cqq7c}
\rho\frac{\partial Q}{\partial t}=-\frac{\hbar}{2}\nabla\cdot \left (\rho\,
\frac{\partial {\bf u}_Q}{\partial t}\right ),
\end{equation}
where ${\bf u}_Q=(\hbar/2m)\nabla\ln\rho$ is the quantum (or osmotic)
velocity. Therefore, the quantum current can be written as
\begin{equation}
\label{cqq7cd}
{\bf J}_Q=\frac{\hbar}{2m}\rho\frac{\partial {\bf u}_Q}{\partial t}.
\end{equation}
From Eqs. (\ref{stress19b}) and (\ref{cqq7c}), we obtain the identity [see also
Eq. (\ref{var7})]:
\begin{eqnarray}
\label{stress18}
\int \rho\frac{\partial Q}{\partial t}\, d{\bf r}=0.
\end{eqnarray}

\section{Lagrangian of a  self-gravitating BEC}
\label{sec_lh}

In this Appendix, we discuss the Lagrangian structure of the generalized GPP
equations  (\ref{mfgp9}) and (\ref{mfgp14}) and of the corresponding
hydrodynamic  EP equations
(\ref{mad12})-(\ref{mad14}). We take $\xi=0$ for simplicity. The Lagrangian of
the
generalized GPP equations is
\begin{eqnarray}
L=\int \biggl\lbrace\frac{i\hbar}{2m} \left (\psi^*\frac{\partial\psi}{\partial
t}-\psi\frac{\partial\psi^*}{\partial t}\right
)-\frac{\hbar^2}{2m^2}|\nabla\psi|^2-\frac{1}{2}\Phi|\psi|^2-\Phi_{\rm
ext}|\psi|^2-V(|\psi|^2)\biggr\rbrace\, d{\bf
r}.
\label{lh1}
\end{eqnarray}
We can view the Lagrangian (\ref{lh1}) as a functional of $\psi$, $\dot\psi$
and $\nabla\psi$. The action is $S=\int L\, dt$. The least action principle
$\delta S=0$, which is
equivalent to the Lagrange equation
\begin{eqnarray}
\label{lh2}
\frac{\partial}{\partial t}\left (\frac{\delta
L}{\delta\dot\psi}\right)+\nabla\cdot \left (\frac{\delta
L}{\delta\nabla\psi}\right)-\frac{\delta L}{\delta\psi}=0,
\end{eqnarray}
returns the GP equation (\ref{mfgp9}).  The free energy is
obtained from the
transformation
\begin{eqnarray}
F=\int \frac{i\hbar}{2m} \left (\psi^*\frac{\partial\psi}{\partial
t}-\psi\frac{\partial\psi^*}{\partial t}\right )\, d{\bf r}-L
\label{lh3}
\end{eqnarray}
leading to
\begin{eqnarray}
F=\frac{\hbar^2}{2m^2}\int|\nabla\psi|^2\, d{\bf
r}+\frac{1}{2}\int\Phi|\psi|^2\, d{\bf r}+\int \Phi_{\rm ext}|\psi|^2\, d{\bf r}
+\int V(|\psi|^2)\, d{\bf r}.
\label{lh4}
\end{eqnarray}
The first term is the kinetic energy, the second term
is the gravitational energy, the third term is the external potential energy and
the fourth term is the self-interaction energy.
Using the Lagrange equations, one can show that the free
energy is conserved (see also Appendix \ref{sec_gh}).

Using the Madelung transformation, we can rewrite the
Lagrangian in terms of hydrodynamic variables. According to
Eqs. (\ref{mad1}) and (\ref{mad2}) we have
\begin{eqnarray}
\label{lh6}
\frac{\partial S}{\partial t}=-i\frac{\hbar}{2}\frac{1}{|\psi|^2}\left
(\psi^*\frac{\partial\psi}{\partial t}-\psi\frac{\partial\psi^*}{\partial
t}\right )
\end{eqnarray}
and
\begin{eqnarray}
\label{lh7}
|\nabla\psi|^2=\frac{1}{4\rho}(\nabla\rho)^2+\frac{\rho}{\hbar^2}
(\nabla S)^2.
\end{eqnarray}
Substituting these identities into Eq. (\ref{lh1}), we get
\begin{eqnarray}
\label{lh8}
L=-\int \biggl \lbrace \frac{\rho}{m}\frac{\partial S}{\partial
t}+\frac{\rho}{2m^2}(\nabla S)^2
+\frac{\hbar^2}{8m^2}\frac{(\nabla\rho)^2}{\rho}
+\frac{1}{2}\rho\Phi+\rho\Phi_{\rm ext}+V(\rho)\biggr\rbrace\, d{\bf r}.
\end{eqnarray}
We can view the Lagrangian (\ref{lh8}) as a functional of $S$, $\dot S$,
$\nabla S$, $\rho$, $\dot\rho$, and $\nabla\rho$. The Lagrange equation for the
action
\begin{eqnarray}
\label{lh9}
\frac{\partial}{\partial t}\left (\frac{\delta L}{\delta\dot
S}\right)+\nabla\cdot \left (\frac{\delta L}{\delta\nabla S}\right)-\frac{\delta
L}{\delta S}=0
\end{eqnarray}
returns the equation of continuity (\ref{mad6}). The Lagrange equation for the
density
\begin{eqnarray}
\label{lh10}
\frac{\partial}{\partial t}\left (\frac{\delta L}{\delta\dot
\rho}\right)+\nabla\cdot \left (\frac{\delta L}{\delta\nabla
\rho}\right)-\frac{\delta L}{\delta \rho}=0
\end{eqnarray}
returns the quantum Hamilton-Jacobi (or Bernoulli) equation (\ref{mad7})
leading to the quantum Euler equation (\ref{mad10}). The free energy is obtained
from
the transformation
\begin{eqnarray}
\label{lh11}
F=-\int \frac{\rho}{m}\frac{\partial S}{\partial t}\, d{\bf r}-L
\end{eqnarray}
leading to
\begin{eqnarray}
\label{lh12}
F=\int  \frac{1}{2}\rho {\bf
u}^2\,
d{\bf r}+\int \frac{\hbar^2}{8m^2}\frac{(\nabla\rho)^2}{\rho}\,
d{\bf r}
+\frac{1}{2}\int \rho\Phi\,
d{\bf r}+\int \rho\Phi_{\rm ext}\,
d{\bf r}
+\int V(\rho)\,
d{\bf r}.
\end{eqnarray}
The first term is the classical kinetic energy, the
second term
is the quantum kinetic energy, the third term is the gravitational energy, the
fourth term is the external potential energy and
the fifth term is the self-interaction energy.
Using the Lagrange equations, one can show that the free energy is conserved
(see also Sec. \ref{sec_eff}).

We now consider the model of Sec. \ref{sec_ansatz}. With the Gaussian ansatz of
Eq. (\ref{a3}), we obtain
\begin{eqnarray}
\int \frac{i\hbar}{2m} \left (\psi^*\frac{\partial\psi}{\partial
t}-\psi\frac{\partial\psi^*}{\partial t}\right )\, d{\bf
r}
=-\int \frac{\rho}{m}\frac{\partial S}{\partial t}\, d{\bf r}
=-\frac{1}{2}\alpha MR^2\dot H
\label{lh13}
\end{eqnarray}
and
\begin{eqnarray}
F=\frac{1}{2}\alpha
M R^2 H^2+V(R),
\label{lh14}
\end{eqnarray}
where $V(R)$ is given by Eqs. (\ref{a11}) and (\ref{a12}). Substituting these
expressions into Eq. (\ref{lh3}) or into Eq. (\ref{lh11}), we
obtain the effective Lagrangian
\begin{eqnarray}
L(H,\dot H,R)=-\frac{1}{2}\alpha
MR^2(\dot H+H^2)-V(R).
\label{lh15}
\end{eqnarray}
We can view the Lagrangian (\ref{lh15}) as a function of $H$, $\dot H$ and $R$.
The Lagrange
equation for $H$
\begin{eqnarray}
\label{lh16}
\frac{\partial}{\partial t}\left (\frac{\delta L}{\delta\dot
H}\right)-\frac{\delta
L}{\delta H}=0
\end{eqnarray}
returns Eq. (\ref{hubb}). The Lagrange equation for $R$
\begin{eqnarray}
\label{lh17}
\frac{\delta L}{\delta R}=0,
\end{eqnarray}
together with Eq. (\ref{hubb}), returns the equation of motion (\ref{a22}). We
also note that, using Eq. (\ref{hubb}), the free energy (\ref{lh14}) can be
written in the form of Eqs. (\ref{a10}).

\section{Derivation of the virial theorem}
\label{sec_virialannexe}

\subsection{General case}

In this Appendix, we establish the  virial theorem
associated with the damped quantum barotropic EP equations
(\ref{mad12})-(\ref{mad14}) that are equivalent to the generalized GPP
equations
(\ref{mfgp9}) and (\ref{mfgp14}).  This generalizes the results of
\cite{prd1,bookspringer}.

Taking the time derivative
of the moment of inertia tensor
\begin{eqnarray}
\label{virial1}
I_{ij}=\int \rho x_i x_j \, d{\bf r}
\end{eqnarray}
and using the continuity equation (\ref{mad12}), we obtain after an
integration by parts
\begin{eqnarray}
\label{virial2}
\dot I_{ij}=\int \rho (x_i u_j+x_j u_i)\, d{\bf r}.
\end{eqnarray}
Taking the time derivative of Eq. (\ref{virial2}), we get
\begin{eqnarray}
\label{virial3}
\ddot I_{ij}=\int  x_i \frac{\partial}{\partial t}(\rho u_j) \, d{\bf r} + (i\leftrightarrow j),
\end{eqnarray}
where $\partial_t(\rho u_j)$ is given by Eq. (\ref{mad14b}). This equation
can be rewritten as
\begin{eqnarray}
\label{virial3b}
\frac{\partial}{\partial t}(\rho u_j)=-\frac{\partial}{\partial x_k}(\rho u_k
u_j)-\frac{\partial P}{\partial x_j}-\rho\frac{\partial \Phi}{\partial
x_j}-\rho\frac{\partial \Phi_{\rm ext}}{\partial
x_j}-\frac{\rho}{m}\frac{\partial Q}{\partial x_j}-\xi\rho u_j.
\end{eqnarray}
To obtain $\ddot I_{ij}$, we need to evaluate six terms. The first term is the kinetic energy tensor
\begin{eqnarray}
\label{virial5}
-\int  x_i \frac{\partial}{\partial x_k}(\rho u_k u_j) \, d{\bf r} =\int \rho u_i u_j\, d{\bf r}.
\end{eqnarray}
The second term is the pressure tensor
\begin{eqnarray}
\label{virial6}
-\int  x_i \frac{\partial P}{\partial x_j} \, d{\bf r} =\delta_{ij} \int P \,
d{\bf r}.
\end{eqnarray}
The third term is the gravitational potential energy tensor
\begin{eqnarray}
\label{virial7}
W_{ij}=-\int  \rho x_i \frac{\partial \Phi}{\partial x_j} \, d{\bf r}.
\end{eqnarray}
It can be written in an alternative form as follows. Substituting the expression
of the gravitational force
\begin{eqnarray}
\label{virial7b}
{\bf F}=-\nabla\Phi=-G\int\rho({\bf r}',t)\frac{{\bf r}-{\bf r}'}{|{\bf r}-{\bf r}'|^d}\, d{\bf r}'
\end{eqnarray}
into Eq. (\ref{virial7}), we obtain
\begin{eqnarray}
\label{virial7b2}
W_{ij}=-G\int\rho({\bf r},t)\rho({\bf r}',t)x_i\frac{x_j-x_j'}{|{\bf r}-{\bf r}'|^d}\, d{\bf r}d{\bf r}'.
\end{eqnarray}
Interchanging the prime and unprimed variables, we get
\begin{eqnarray}
\label{virial7b3}
W_{ij}=G\int\rho({\bf r},t)\rho({\bf r}',t)x'_i\frac{x_j-x_j'}{|{\bf r}-{\bf r}'|^d}\, d{\bf r}d{\bf r}'.
\end{eqnarray}
Taking the half-sum of the foregoing expressions, we find that
\begin{eqnarray}
\label{virial7b4}
W_{ij}=-\frac{G}{2}\int\rho({\bf r},t)\rho({\bf r}',t)\frac{(x_i-x_i')(x_j-x_j')}{|{\bf r}-{\bf r}'|^d}\, d{\bf r}d{\bf r}'.
\end{eqnarray}
Under this form, the gravitational potential energy tensor is manifestly
symmetric: $W_{ij}=W_{ji}$. The fourth term is
the external  potential energy tensor
\begin{eqnarray}
\label{virial9ext}
W^{\rm ext}_{ij}=-\int  \rho x_i \frac{\partial \Phi_{\rm ext}}{\partial x_j} \, d{\bf r}.
\end{eqnarray}
The fifth term is  the quantum potential energy tensor
\begin{eqnarray}
\label{virial8}
W^Q_{ij}=-\int   x_i \frac{\rho}{m} \frac{\partial Q}{\partial x_j} \, d{\bf r}.
\end{eqnarray}
Substituting Eq. (\ref{mad19}) into Eq. (\ref{virial8}), we get
\begin{eqnarray}
\label{virial8z}
W^Q_{ij}=-\int x_{i}\partial_{k}P_{jk}\, d{\bf r}=\int P_{ij}\, d{\bf r},
\end{eqnarray}
where $P_{ij}$ is the quantum pressure tensor defined by Eq. (\ref{mad20}).  Since $P_{ij}$ is symmetric, the quantum potential energy tensor is also symmetric: $W^Q_{ij}=W^Q_{ji}$. The sixth term is the frictional tensor
\begin{eqnarray}
\label{virial10}
- \int   x_i \xi\rho u_j  \, d{\bf r} =-\xi\int \rho x_i u_j\, d{\bf r}.
\end{eqnarray}
Substituting these results into Eq. (\ref{virial3}), we obtain the tensorial
virial theorem
\begin{eqnarray}
\label{virial12}
\frac{1}{2}\ddot I_{ij}+\frac{1}{2}\xi \dot I_{ij}=\int\rho u_iu_j\, d{\bf
r}+\delta_{ij}\int P\, d{\bf r}+W_{ij}^Q+W_{ij}+\frac{1}{2}(W^{\rm
ext}_{ij}+W^{\rm ext}_{ji}).
\end{eqnarray}
According to Eq. (\ref{virial8z}), it can also be written as
\begin{eqnarray}
\label{virial12b}
\frac{1}{2}\ddot I_{ij}+\frac{1}{2}\xi \dot I_{ij}=\int\rho u_iu_j\, d{\bf
r}+\int P_{ij}\, d{\bf r}+\delta_{ij}\int P\, d{\bf r}+W_{ij}+\frac{1}{2}(W^{\rm
ext}_{ij}+W^{\rm ext}_{ji}).
\end{eqnarray}
Contracting the indices, we obtain the scalar virial theorem
\begin{eqnarray}
\label{virial13}
\frac{1}{2}\ddot I+\frac{1}{2}\xi\dot I=2(\Theta_c+\Theta_Q)+d\int P\, d{\bf
r}+W_{ii}+W_{ii}^{\rm ext}.
\end{eqnarray}
This equation involves the virial of the quantum force
\begin{eqnarray}
\label{virial15}
W_{ii}^Q=-\int  \frac{\rho}{m} {\bf r}\cdot \nabla Q \, d{\bf r},
\end{eqnarray}
the virial of the gravitational force
\begin{equation}
\label{virial18}
W_{ii}=-\int \rho {\bf r}\cdot\nabla\Phi\, d{\bf r},
\end{equation}
and the virial of the external force
\begin{equation}
\label{virial17}
W_{ii}^{\rm ext}=-\int \rho {\bf r}\cdot\nabla\Phi_{\rm ext}\, d{\bf r}.
\end{equation}
According to Eqs. (\ref{virial8z}), (\ref{mad20b}) and (\ref{ef5}), the virial of the quantum force is equal to twice the quantum kinetic energy
\begin{eqnarray}
\label{virial16}
W_{ii}^Q=\int P_{ii}\, d{\bf r}=\frac{\hbar^2}{4m^2}\int\frac{(\nabla\rho)^2}{\rho}\, d{\bf r}=2\Theta_Q,
\end{eqnarray}
leading to the second term in the right hand side of Eq. (\ref{virial13}). From
Eq. (\ref{virial7b4}), we find that the virial of the gravitational
force is given by
\begin{eqnarray}
\label{virial19}
W_{ii}=-\frac{G}{2}\int\frac{\rho({\bf r},t)\rho({\bf r}',t)}{|{\bf r}-{\bf r}'|^{d-2}}\, d{\bf r}d{\bf r}'.
\end{eqnarray}
Therefore
\begin{equation}
\label{virial20}
W_{ii}=(d-2)W,\qquad (d\neq 2)
\end{equation}
\begin{equation}
\label{virial21}
W_{ii}=-\frac{GM^2}{2},\qquad (d=2)
\end{equation}
where $W$ is the gravitational potential energy
\begin{eqnarray}
\label{virial7w}
W=-\frac{G}{2(d-2)}\int\frac{\rho({\bf r},t)\rho({\bf r}',t)}{|{\bf r}-{\bf r}'|^{d-2}}\, d{\bf r}d{\bf r}'.
\end{eqnarray}

In the strong friction limit $\xi\rightarrow +\infty$ in which ${\bf
u}=O(1/\xi)$, the tensorial virial theorem (\ref{virial12})
becomes\footnote{This
expression can also be obtained directly from the generalized Smoluchowski
equation
(\ref{mad16}) by taking the time derivative of Eq. (\ref{virial1}), substituting
Eq. (\ref{mad16}) into the resulting expression, integrating by parts, and using
Eqs. (\ref{virial7}), (\ref{virial9ext}) and (\ref{virial8}).}
\begin{eqnarray}
\label{virial25}
\frac{1}{2}\xi \dot I_{ij}=\delta_{ij}\int P\, d{\bf
r}+W_{ij}^Q+W_{ij}+\frac{1}{2}(W^{\rm ext}_{ij}+W^{\rm ext}_{ji}).
\end{eqnarray}
According to Eq. (\ref{virial8z}), it can also be written as
\begin{eqnarray}
\label{virial26b}
\frac{1}{2}\xi \dot I_{ij}=\int P_{ij}\, d{\bf r}+\delta_{ij}\int P\, d{\bf
r}+W_{ij}+\frac{1}{2}(W^{\rm ext}_{ij}+W^{\rm ext}_{ji}).
\end{eqnarray}
Contracting the indices, we obtain the scalar virial theorem
\begin{eqnarray}
\label{virial27}
\frac{1}{2}\xi\dot I=2\Theta_Q+d\int P\, d{\bf r}+W_{ii}+W_{ii}^{\rm ext}.
\end{eqnarray}

At equilibrium ($\ddot I_{ij}=\dot I_{ij}=0$ and ${\bf u}={\bf 0}$), the tensorial virial theorem reduces to
\begin{eqnarray}
\label{virial22}
\delta_{ij}\int P\, d{\bf r}+W_{ij}^Q+W_{ij}+\frac{1}{2}(W^{\rm ext}_{ij}+W^{\rm
ext}_{ji})=0.
\end{eqnarray}
According to Eq. (\ref{virial8z}), it can also be written as
\begin{eqnarray}
\label{virial23}
\int P_{ij}\, d{\bf r}+\delta_{ij}\int P\, d{\bf r}+W_{ij}+\frac{1}{2}(W^{\rm
ext}_{ij}+W^{\rm ext}_{ji})=0.
\end{eqnarray}
Contracting the indices, we obtain the equilibrium scalar virial theorem
\begin{eqnarray}
\label{virial24}
2\Theta_Q+d\int P\, d{\bf r}+W_{ii}+W_{ii}^{\rm ext}=0.
\end{eqnarray}

For a spherically symmetric system, according to the Gauss theorem, we have
\begin{eqnarray}
\label{virial25q}
\nabla\Phi=\frac{GM(r,t)}{r^{d-1}}{\bf e}_r,
\end{eqnarray}
where
\begin{eqnarray}
\label{virial25b}
M(r,t)=\int_0^r \rho(r',t) S_d {r'}^{d-1}\, dr',
\end{eqnarray}
is the mass contained within the sphere of radius $r$. This is Newton's law in
$d$ dimensions. Using Eq. (\ref{virial25b}), the virial of the gravitational
force (\ref{virial18}) can be written as
\begin{equation}
\label{virial26}
W_{ii}=-S_dG\int \rho(r,t)M(r,t)r\, dr=-\int\frac{GM(r,t)}{r^{d-2}}\, dM(r,t).
\end{equation}
In $d=2$, we immediately recover Eq. (\ref{virial21}). In $d\neq 2$, using Eq.
(\ref{virial20}), we obtain the formula
\begin{equation}
\label{virial27b}
W=-\frac{1}{d-2}\int \rho(r,t)\frac{GM(r,t)}{r^{d-2}}S_d r^{d-1}\, dr,
\end{equation}
which is useful to calculate the gravitational potential energy of a spherically
symmetric distribution of matter (see Appendix \ref{sec_det}).

\subsection{Harmonic potential}

For the harmonic potential (\ref{mfgp8}), we have
\begin{eqnarray}
\label{virial28}
W^{\rm ext}_{ij} =-\omega_0^2 I_{ij},\qquad W_{ii}^{\rm ext}=-\omega_0^2
I=-2W_{\rm ext}.
\end{eqnarray}
The harmonic potential energy tensor is manifestly symmetric: $W_{ij}^{\rm
ext}=W_{ji}^{\rm ext}$.  The tensorial virial theorem can be written as
\begin{eqnarray}
\label{virial29}
\frac{1}{2}\ddot I_{ij}+\frac{1}{2}\xi \dot I_{ij}+\omega_0^2 I_{ij}=\int\rho
u_iu_j\, d{\bf r}+\delta_{ij}\int P\, d{\bf r}+W_{ij}^Q+W_{ij}
\end{eqnarray}
or, equivalently, as
\begin{eqnarray}
\label{virial30}
\frac{1}{2}\ddot I_{ij}+\frac{1}{2}\xi \dot I_{ij}+\omega_0^2 I_{ij}=\int\rho
u_iu_j\, d{\bf r}+\int P_{ij}\, d{\bf r}+\delta_{ij}\int P\, d{\bf r}+W_{ij}.
\end{eqnarray}
The scalar virial theorem can be written as
\begin{eqnarray}
\label{virial31}
\frac{1}{2}\ddot I+\frac{1}{2}\xi\dot I+\omega_0^2 I=2(\Theta_c+\Theta_Q)+d\int
P\, d{\bf r}+W_{ii}.
\end{eqnarray}

In the strong friction limit $\xi\rightarrow +\infty$, the tensorial virial theorem takes the form
\begin{eqnarray}
\label{virial35}
\frac{1}{2}\xi \dot I_{ij}+\omega_0^2 I_{ij}=\delta_{ij}\int P\, d{\bf
r}+W_{ij}^Q+W_{ij}
\end{eqnarray}
or, equivalently,
\begin{eqnarray}
\label{virial36}
\frac{1}{2}\xi \dot I_{ij}+\omega_0^2 I_{ij}=\int P_{ij}\, d{\bf
r}+\delta_{ij}\int P\, d{\bf r}+W_{ij},
\end{eqnarray}
and the scalar virial theorem takes the form
\begin{eqnarray}
\label{virial37}
\frac{1}{2}\xi\dot I+\omega_0^2 I=2\Theta_Q+d\int P\, d{\bf r}+W_{ii}.
\end{eqnarray}

The equilibrium tensorial virial theorem can be written as
\begin{eqnarray}
\label{virial32}
\delta_{ij}\int P\, d{\bf r}+W_{ij}^Q+W_{ij}-\omega_0^2 I_{ij}=0
\end{eqnarray}
or, equivalently, as
\begin{eqnarray}
\label{virial33}
\int P_{ij}\, d{\bf r}+\delta_{ij}\int P\, d{\bf r}+W_{ij}-\omega_0^2 I_{ij}=0.
\end{eqnarray}
The equilibrium scalar virial theorem can be written as
\begin{eqnarray}
\label{virial34}
2\Theta_Q+d\int P\, d{\bf r}+W_{ii}-\omega_0^2 I=0.
\end{eqnarray}

\section{Composite models}
\label{sec_com}

In Sec. \ref{sec_eos},  we
have considered particular classes of generalized GPP equations
associated with particular equations of state and particular entropies. In this
Appendix, we consider composite models. Specifically, we consider generalized
GPP
equations where the effective potential $h=\sum_i h_i$ is a sum of potentials
$h_i$ such as those studied in Sec. \ref{sec_eos}. In that case,
$P=\sum_i p_i$, $U=\sum_i U_i$, and $F=E_*-\sum_i T_i S_i$. For example, if
$P=K_1\rho^{\gamma_1}+K_2\rho^{\gamma_2}+A\ln\rho+\rho k_BT/m$, we get
$F=E_*-K_{1}S_{\gamma_1}-K_{2} S_{\gamma_2}-AS_L-TS_B$. We
emphasize, however, that the density $\rho$ associated with the equation of
state $P=\sum_i P_i$ (or the free energy $F=E_*-\sum_i T_i S_i$) is {\it not}
the sum of the densities $\rho_i$ 
associated with the equations of state $p_i$ (or the free energies
$F_i=E_*-T_i S_i$) taken individually:
$\rho\neq\sum_i\rho_i$.

Let us give some examples of physical interest in the context of dark
matter halos. The equations of state
\begin{eqnarray}
\label{com1}
P=K_1\rho^{\gamma_1}+K_2\rho^{\gamma_2},\qquad
\frac{1}{P}=\frac{1}{K_1\rho^{\gamma_1}}+\frac{1}{K_2\rho^{\gamma_2}}
\end{eqnarray}
describe a composite halo with a polytropic core and a polytropic halo. The
equation of state
\begin{eqnarray}
\label{com2}
P=K\rho^{\gamma}+\rho\frac{k_B T}{m}
\end{eqnarray}
with $\gamma>1$ (resp. $\gamma<1$) describes a composite halo with a
polytropic (resp. isothermal) core  and an isothermal (resp. polytropic)  
halo. Symmetrically, the equation of state 
\begin{eqnarray}
\label{com2b}
\frac{1}{P}=\frac{1}{K\rho^{\gamma}}+\frac{m}{\rho k_B T}
\end{eqnarray}
with $\gamma>1$ (resp. $\gamma<1$) describes a composite halo with an
isothermal (resp. polytropic)   core  and a polytropic (resp. isothermal)
halo.

More specifically, the equation of state 
\begin{eqnarray}
\label{com3}
P=\frac{2\pi a_s\hbar^2}{m^3}\rho^{2}+K\rho^{\gamma}
\end{eqnarray}
with $\gamma<2$ describes a composite halo with a BEC core and a polytropic
halo while the equation of state
\begin{eqnarray}
\label{com4}
P=\frac{2\pi a_s\hbar^2}{m^3}\rho^{2}+\rho\frac{k_B T}{m}
\end{eqnarray}
describes a composite halo with a BEC core and an isothermal halo.

Finally, the equation of state
\begin{eqnarray}
\label{com5}
\frac{1}{P}=\frac{1}{A\ln\rho}+\frac{m}{\rho k_B T}
\end{eqnarray}
describes a composite halo with a logotropic core and an isothermal halo while
the equation of state 
\begin{eqnarray}
\label{com6}
P=A\ln\rho+\rho\frac{k_B T}{m}
\end{eqnarray}
describes a composite halo with an isothermal core and a logotropic halo.
Similarly, the equations of state
\begin{eqnarray}
\label{com6b}
\frac{1}{P}=\frac{1}{A\ln\rho}+\frac{1}{K\rho^{\gamma}}\qquad {\rm and}\qquad
P=A\ln\rho+K\rho^{\gamma}
\end{eqnarray}
describe a composite halo with a logotropic core and a
polytropic halo, or the converse.

\section{A damped logarithmic Gross-Pitaevskii equation that conserves the
energy of the standard Gross-Pitaevskii equation}
\label{sec_ce}

The free energy associated with the generalized 
GPP equations (\ref{mi1}) and (\ref{mi2}) with a constant temperature $T$ can be
written as
$F=E_0-T S_B$ (see Eq. (\ref{mi9})). When $\xi=0$ and $T=0$, the
energy $E_0$ corresponds to the free energy of the standard GPP
equations (\ref{mfgp13}) and (\ref{mfgp14a}) and it is conserved. When $\xi\neq
0$ and $T\neq 0$, the
energy $E_0$ is not
conserved anymore. However, we can
consider a model in
which the temperature $T(t)$ varies with time in order to conserve $E_0$
exactly. This model can be relevant if, as advocated in the Introduction and in
Sec. \ref{sec_pi}, the generalized GPP equations (\ref{mi1}) and  (\ref{mi2})
provide a
coarse-grained representation of the GPP equations (\ref{mfgp13}) and
(\ref{mfgp14a}), and an effective model
of gravitational cooling. In that case, we can argue that the 
energy $E_0$ associated with the GPP equations (fine-grained)  should be
conserved by the generalized GPP equations (coarse-grained). This idea is
similar to the one developed in the context of the
theory of violent relaxation in Ref. \cite{csr}. Physically, the energy of
the core decreases until it reaches the ground state and, in
parallel, the 
energy lost by the core heats up the halo so that its  temperature $T(t)$
increases so
as to conserve the total (core + halo) energy $E_0=E_{\rm core}(t)+E_{\rm
halo}(t)$.

Taking the time derivative of Eq. (\ref{mi10}), and proceeding as in
Appendix \ref{sec_hth}, we find that
\begin{eqnarray}
\label{ce1}
\dot E_0=-\frac{k_B T(t)}{m}\int {\bf u}\cdot
\nabla\rho\, d{\bf r}-\xi\int\rho {\bf u}^2\, d{\bf r}. 
\end{eqnarray}
Taking the time derivative of the Boltzmann entropy (\ref{f9}) and using the
continuity equation (\ref{mad12}), we get
\begin{eqnarray}
\label{ce2}
\dot S_B=-\frac{k_B}{m}\int {\bf u}\cdot
\nabla\rho\, d{\bf r}. 
\end{eqnarray}
Recalling Eq. (\ref{ef3}), we find that Eq. (\ref{ce1}) can be rewritten as
\begin{eqnarray}
\label{ce3}
{\dot E}_0-T(t){\dot S}_B=-2\xi\Theta_c.
\end{eqnarray}
If we now impose that ${\dot E}_0=0$ according to the arguments given above, we
find that the temperature must evolve as
\begin{eqnarray}
\label{ce4}
T(t)=\frac{2\xi\Theta_c(t)}{{\dot S}_B(t)}.
\end{eqnarray}
In this manner, the  (coarse-grained) GPP equations
(\ref{mi1}) and (\ref{mi2}) with 
a time-dependent temperature $T(t)$ given by Eq. (\ref{ce4}) relax towards a
steady state with a
core-halo structure while
conserving the energy $E_0$ of the (fine-grained) GPP equations
(\ref{mfgp13}) and
(\ref{mfgp14a}). This model may provide an improved
coarse-grained parametrization of the process of gravitational cooling.

{\it Remark:} In the strong friction limit $\xi\rightarrow
+\infty$, the previous equations remain valid with ${\bf u}$ given by Eq.
(\ref{mad15}), where the pressure $P$ contains only the
contribution (\ref{f60}) arising from the self-interaction, not from the
temperature.

\section{Particular solution of the continuity equation}
\label{sec_velf}

In this Appendix, we derive 
a particular exact solution of the continuity equation (\ref{mad12}). We stress,
however, that this solution is generally not an exact solution of the Euler
equation (\ref{mad13}).

We consider a density profile of the form
\begin{eqnarray}
\label{velf02}
\rho({\bf r},t)=\frac{M}{R(t)^d}f\left \lbrack \frac{{\bf r}}{R(t)}\right
\rbrack
\end{eqnarray}
with $\int f({\bf x})\, d{\bf x}=1$ so as to ensure the normalization condition
(conservation of mass). We also consider a velocity profile of the form
\begin{eqnarray}
\label{velf01}
{\bf u}({\bf r},t)=H(t){\bf r}.
\end{eqnarray}
The continuity equation (\ref{mad12}) can be rewritten as
\begin{eqnarray}
\label{velf1}
\frac{\partial\ln\rho}{\partial t}+\nabla\cdot {\bf u}+ {\bf
u}\cdot \nabla\ln\rho=0.
\end{eqnarray}
From Eqs. (\ref{velf02}) and (\ref{velf01}), we obtain
\begin{eqnarray}
\label{velf2}
\frac{\partial\ln\rho}{\partial t}=-\frac{\dot R}{R}{\bf x}\cdot \nabla_{\bf
x}\ln f-d\frac{\dot R}{R},\qquad \nabla\ln\rho=\frac{1}{R}\nabla_{\bf x}\ln f,\qquad \nabla\cdot {\bf u}=dH.
\end{eqnarray}
Substituting the foregoing relations into Eq. (\ref{velf1}), we get
\begin{eqnarray}
\label{velf4}
\left (H-\frac{\dot R}{R}\right )\left (d+{\bf x}\cdot \nabla_{\bf x}\ln f\right
)=0.
\end{eqnarray}
This equation is satisfied for all ${\bf r}$ if, and only if,
\begin{eqnarray}
\label{velf5}
H(t)=\frac{\dot R}{R}.
\end{eqnarray}
Using Eqs.  (\ref{mad5}) and
(\ref{velf01}), we find that the action is given by
\begin{eqnarray}
\label{velf6}
S({\bf r},t)=\frac{1}{2}m H(t)r^2+S_0(t).
\end{eqnarray}
Therefore, the wave function is given by Eq. (\ref{mad1}) where $\rho({\bf r},t)$ is given by Eq. (\ref{velf02}) and $S({\bf r},t)$ is given by Eq. (\ref{velf6}). For a Gaussian density profile, we obtain  Eq. (\ref{a3}).

\section{Details of the calculation of the free energy with the Gaussian ansatz}
\label{sec_det}

We consider a Gaussian density profile of the form
\begin{eqnarray}
\label{det1}
\rho=A e^{-r^2/R^2}.
\end{eqnarray}
The associated  mass is
\begin{eqnarray}
\label{det2}
M=A\int_0^{+\infty} e^{-r^2/R^2} S_d r^{d-1}\, dr=\frac{1}{2}A S_d R^d \int_0^{+\infty} e^{-t}  t^{(d-2)/2}\, dt=\frac{1}{2}A S_d R^d \Gamma\left (\frac{d}{2}\right ),
\end{eqnarray}
where we have made the change of variables $t=r^2/R^2$. Equation (\ref{det2})
determines 
the normalization constant through the relation
\begin{eqnarray}
\label{det2v}
A=\frac{2M}{S_d R^d \Gamma\left
(\frac{d}{2}\right )}.
\end{eqnarray}

The moment of inertia (\ref{ef1d}) is given by
\begin{eqnarray}
\label{det3}
I=A\int_0^{+\infty} e^{-r^2/R^2} r^2 S_d r^{d-1}\, dr=\frac{1}{2}A S_d R^{d+2} \int_0^{+\infty} e^{-t}  t^{d/2}\, dt=\frac{1}{2}A S_d R^{d+2} \Gamma\left (\frac{d}{2}+1\right )=\frac{d}{2}MR^2,
\end{eqnarray}
where we have made the change of variables $t=r^2/R^2$ to get the second
equality, and used the identity $\Gamma(x+1)=x\Gamma(x)$ and  Eq. (\ref{det2v})
to get the fourth equality.

Substituting Eq. (\ref{a3x}) into Eq.  (\ref{ef3}),
we find that the classical kinetic energy can be written as
\begin{eqnarray}
\label{det4}
\Theta_c=\frac{I}{2R^2}\left (\frac{dR}{dt}\right )^2=\frac{d}{4}M\left (\frac{d R}{dt}\right )^2,
\end{eqnarray}
where we have used Eq. (\ref{det3}) to get the second equality.

Substituting Eq. (\ref{det1}) into Eq.  (\ref{ef5}), we find that the quantum
kinetic energy can be written as
\begin{eqnarray}
\label{det5}
\Theta_Q=\frac{\hbar^2 I}{2m^2R^4}=\frac{d\hbar^2 M}{4m^2R^2},
\end{eqnarray}
where we have used Eq. (\ref{det3}) to get the second equality. 

Substituting Eq. (\ref{det1}) into Eq.  (\ref{f31}), we find that the internal
energy of a fluid with a polytropic equation of state of index $\gamma>0$ is
given by
\begin{eqnarray}
\label{det6}
U=\frac{K}{\gamma-1}A^{\gamma}\int_0^{+\infty} e^{-\gamma r^2/R^2} S_d r^{d-1}\, dr=\frac{K}{\gamma-1}A^{\gamma-1}\frac{1}{\gamma^{d/2}}M,
\end{eqnarray}
where we have made the change of variables $\tilde r=\sqrt{\gamma}r$ to get the
second equality. Replacing $A$ by its expression given by Eq. (\ref{det2v}), we
obtain the result of Eq. (\ref{a5}). 

Substituting Eq. (\ref{det1}) into Eq. 
(\ref{f7}), we find that the internal energy of a fluid with an isothermal
equation of state is given by
\begin{eqnarray}
\label{det7}
U_{B}=\frac{k_B T}{m}\int \rho \left (\ln A-\frac{r^2}{R^2}-1\right )\, d{\bf r}=\frac{k_B T}{m}M(\ln A-1)-\frac{k_B T}{m}\frac{I}{R^2}.
\end{eqnarray}
Using the expression of $A$ given by Eq. (\ref{det2v}) and 
the expression of $I$ given by Eq. (\ref{det3}), we obtain the result of Eq.
(\ref{a6}). 

The potential energy of a harmonic potential is given by
\begin{eqnarray}
\label{det8}
W_{\rm ext}=\frac{1}{2}\omega_0^2 I=\frac{d}{4}\omega_0^2 M R^2,
\end{eqnarray}
where we have used Eq. (\ref{det3}) to get the second equality. 

To compute the gravitational potential energy in $d\neq 2$, we use Eq.
(\ref{virial27b}). For the Gaussian profile (\ref{det1}), we obtain
\begin{eqnarray}
\label{det9}
W=-\frac{1}{d-2}S_d G A\int_0^{+\infty} e^{-r^2/R^2} M(r,t) r\, dr.
\end{eqnarray}
Integrating by parts, we get
\begin{eqnarray}
\label{det10}
W=-\frac{1}{d-2}S_d G A \frac{R^2}{2}\int_0^{+\infty} e^{-r^2/R^2} M'(r) \, dr=-\frac{1}{d-2}S_d^2 G A^2 \frac{R^2}{2}\int_0^{+\infty} e^{-2r^2/R^2} r^{d-1} \, dr
=-\frac{1}{d-2}S_d G A \frac{MR^2}{2^{1+d/2}},\nonumber\\
\end{eqnarray}
where we have made the change of variables $\tilde r=\sqrt{2}r$ to get the third
expression and used Eq. (\ref{det2}). Replacing $A$ by its expression given by
Eq. (\ref{det2v}), we obtain the result of Eq. (\ref{a7}). In $d=2$, if we use
the formula [see Eqs. (\ref{pot2}) and (\ref{ney3})]
\begin{eqnarray}
\label{det11}
W=\frac{1}{2}G\int {\rho({\bf r})}{\rho({\bf r}')}\ln{|{\bf r}-{\bf r}'|} \, d{\bf r}d{\bf r}',
\end{eqnarray}
and make the change of variables ${\bf x}={\bf r}/R$, we obtain the result of Eq. (\ref{a7}) with
\begin{eqnarray}
\label{det12}
W_0=\frac{GM^2}{2\pi^2}\int e^{-(x^2+{x'}^2)}\ln|{\bf x}-{\bf x}'|\, d{\bf x}d{\bf x'}.
\end{eqnarray}
Using the identity
\begin{eqnarray}
\label{det13}
\ln|{\bf x}-{\bf x}'|=\ln x_{>}-\sum_{m=1}^{+\infty}\frac{1}{m}\left (\frac{x_<}{x_>}\right )^m\cos\left\lbrack m(\theta-\theta')\right\rbrack,
\end{eqnarray}
where $x_>$ (resp. $x_<$) denotes the largest (smallest) value of $x$ and $x'$,
we obtain 
\begin{eqnarray}
\label{det14}
\int e^{-(x^2+{x'}^2)}\ln|{\bf x}-{\bf x}'|\, d{\bf x}d{\bf x'}=4\pi^2\int_0^{+\infty}dx\, x\ln x e^{-x^2}\int_0^x dx'\, x' e^{-{x'}^2}+4\pi^2\int_0^{+\infty}dx\, x e^{-x^2}\int_x^{+\infty} dx'\, x'\ln x' e^{-x'^2}.\nonumber\\
\end{eqnarray}
Using simple integrations by parts, we get
\begin{eqnarray}
\label{det15}
\int e^{-(x^2+{x'}^2)}\ln|{\bf x}-{\bf x}'|\, d{\bf x}d{\bf x'}=4\pi^2
\int_0^{+\infty} e^{-x^2}\left (1-e^{-x^2}\right )x\ln x \,
dx=\frac{\pi^2}{2}(\ln 2-\gamma_E)=0.572099...
\end{eqnarray}
where $\gamma_E=0.57721566...$ is the Euler constant.
This leads to the result of Eq. (\ref{a9}).

\section{The generalized Jeans problem}

In this Appendix, we study the linear dynamical stability of an
infinite homogeneous self-gravitating BEC described by the
generalized GPP equations (\ref{mfgp9}) and (\ref{mfgp14}). This is a
generalization of the classical Jeans  problem \cite{jeans} to a quantum
dissipative system. We use the hydrodynamic representation
(\ref{mad12})-(\ref{mad14}) of the generalized GPP equations 
(\ref{mfgp9}) and (\ref{mfgp14}). We consider
an infinite homogeneous equilibrium state with $\rho_{\rm eq}({\bf r},t)=\rho$,
${\bf u}_{\rm eq}({\bf r},t)={\bf 0}$ and $S_{\rm eq}({\bf r},t)=-Et$. In the
presence of an external harmonic potential
(\ref{mfgp8}), the quantum Euler equation (\ref{mad13}) reduces to the condition
of
hydrostatic equilibrium $-\nabla\Phi-\omega_0^2{\bf r}={\bf 0}$. Considering the
Poisson
equation (\ref{mad14}), this condition can be satisfied only when
$\omega_0^2<0$, $\Phi=|\omega_0^2|r^2/2$, and $\rho=d|\omega_0^2|/S_dG$.
Then, we have $E=mh(\rho)$. In the absence of external potential, we make
the Jeans swindle \cite{bt} (see \cite{kiessling} for a mathematical
justification). 

Considering a small perturbation about the equilibrium state
and linearizing the hydrodynamic equations (\ref{mad12})-(\ref{mad14}), we
obtain
\begin{equation}
\label{j1}
\frac{\partial\delta}{\partial t}+\nabla\cdot {\bf u}=0,
\end{equation}
\begin{equation}
\label{j2}
\frac{\partial{\bf u}}{\partial t}=-c_s^2\nabla
\delta-\nabla\delta\Phi+\frac{\hbar^2}{4m^2}\nabla (\Delta\delta)-\xi{\bf u},
\end{equation}
\begin{equation}
\label{j3}
\Delta\delta\Phi=S_d G \rho\delta,
\end{equation}
where $c_s^2=P'(\rho)$ is the speed of sound and
$\delta({\bf r},t)=\delta\rho({\bf r},t)/\rho$ the density contrast. We
note that the external potential does not appear in the linearized
equations. Taking the time derivative of Eq. (\ref{j1}) and the divergence of
Eq. (\ref{j2}), we obtain a single equation for the density contrast
\begin{eqnarray}
\label{j4}
\frac{\partial^2\delta}{\partial t^2}+\xi\frac{\partial\delta}{\partial
t}=-\frac{\hbar^2}{4m^2}\Delta^2\delta+c_s^2\Delta\delta+S_d G\rho\delta.
\end{eqnarray} 
Expanding the solutions into plane waves of the form $\delta({\bf
r},t)\propto {\rm exp}\lbrack i({\bf k}\cdot{\bf r}-\omega t)\rbrack$, we
obtain the dispersion relation
\begin{eqnarray}
\label{j5}
\omega^2+i\xi \omega=\frac{\hbar^2k^4}{4m^2}+c_s^2k^2-S_d G\rho \qquad
\Rightarrow\qquad 
\omega=-i\frac{\xi}{2}\pm\sqrt{-\frac{\xi^2}{4}+\frac{\hbar^2k^4}{
4m^2}+c_s^2k^2-S_d G\rho}.
\end{eqnarray} 
For $\xi=\hbar=0$, we recover the celebrated Jeans dispersion relation
\cite{jeans}. For $\xi=G=0$, we recover the Bogoliubov
dispersion relation \cite{bogoliubov}. For $\xi=0$, we recover the
dispersion relation studied in Sec. V of \cite{prd1}. The general
dispersion relation (\ref{j5}) will be studied in detail in a forthcoming paper
(in preparation).

\end{document}